\documentclass{aa}  
\usepackage{arydshln} 
\usepackage{comment} 
\usepackage{xcolor}
\usepackage{graphicx}
\usepackage{multirow}
\usepackage{adjustbox}
\usepackage{txfonts}
\usepackage{tikz} 
\usepackage{fourier}
\DeclareMathAlphabet{\mathcal}{OMS}{cmsy}{m}{n}
\SetMathAlphabet{\mathcal}{bold}{OMS}{cmsy}{b}{n}
\newcommand{\STAB}[1]{\begin{tabular}{@{}c@{}}#1\end{tabular}}
\usepackage{natbib}
\bibpunct{(}{)}{;}{a}{}{,} 

\begin{document}

    \title{Unsupervised star, galaxy, qso classification}
   \subtitle{Application of HDBSCAN}

   \author{C. H. A. Logan
          \inst{1}
          \and
          S. Fotopoulou\inst{2}
          }

   \institute{H. H. Wills Physics Laboratory, University of Bristol \\
              \email{crispin.logan@bristol.ac.uk}
         \and
             Centre for Extragalactic Astronomy, Department of Physics, Durham University, Durham DH1 3LE, UK\\
             \email{sotiria.fotopoulou@durham.ac.uk}
             }

   \date{Received 06 September 2019; accepted 12 November 2019}

 
  \abstract
   {Classification will be an important first step for upcoming surveys that will detect billions of new sources such as LSST and Euclid, as well as DESI, 4MOST and MOONS. The application of traditional methods of model fitting and colour-colour selections will face significant computational constraints, while machine-learning methods offer a viable approach to tackle datasets of that volume.}
   {While supervised learning methods can perform very well for classification tasks, the creation of representative and accurate training sets is a resource and time consuming task. We present a viable alternative using an unsupervised machine learning method to separate stars, galaxies and QSOs using photometric data.}
   {The heart of our work uses Hierarchical Density-Based Spatial Clustering of Applications with Noise (\textsc{hdbscan}) to find the star, galaxy and QSO clusters in a multidimensional colour space. We optimized the hyperparameters and input attributes of three separate \textsc{hdbscan} runs, each to select a particular object class, and thus treat the output of each separate run as a binary classifier. We subsequently consolidate the output to give our final classifications, optimized on the basis of their F1 scores. We explore the use of Random Forest and PCA as part of the pre-processing stage for feature selection and dimensionality reduction.}
   {Using our dataset of $\sim$ 50,000 spectroscopically labelled objects we obtain an F1 score of 98.9, 98.9 and 93.13 respectively for star, galaxy and QSO selection using our unsupervised learning method. We find that careful attribute selection is a vital part of accurate classification with \textsc{hdbscan}. We applied our classification to a subset of the SDSS spectroscopic catalogue and demonstrate the potential of our approach in correcting misclassified spectra useful for DESI and 4MOST. Finally, we created a multiwavelength catalogue of 2.7 million sources using the KiDS, VIKING and ALLWISE surveys and publish corresponding classifications and photometric redshifts.}
   {}


   \maketitle
%

\section{Introduction} \label{section.intro}

The identification and classification of astronomical objects is at the forefront of any astronomical analysis. \citet{Hubble1926} introduced the eponymous tuning-fork diagram of galaxy morphologies, while exotic objects such as quasi-stellar radio objects (quasars) and later quasi-stellar objects (QSO) were identified as sources with point-source appearance on photographic plates but with non-stellar spectra \citep{Schmidt1963}.  Modern multiwavelength extragalactic surveys have created sophisticated colour selection criteria to isolate stars, galaxies, and active galactic nuclei (AGN) including QSO. \citep[e.g.][]{BPT1981,Daddi2004,Stern2005, Patel2012,Stern2012, Assef2018}. 

Machine-learning methods can be broadly categorized according to the strategy during the training phase, split into supervised or unsupervised learning. Supervised learning requires a training set used to learn the underlying correlations between the input features and the target. 
Neural networks have been embraced in their ability to classify point-like versus extended sources in several early works \citep{Odewahn1992, Storrie-Lombardi1992, Lahav1995,Weir1995}, with very notable application the stellarity parameter in Sextractor \citep{Bertin1996}. More recent works have taken advantage of the excellent imaging capabilities of the Hubble space telescope to create accurate training samples \citep[e.g.][]{Huertas-Company2015} and engagement with the public through citizen-science projects such as Galaxy Zoo \citep{Lintott:2008a,Lintott:2011a,Willett:2013a} deployed to classify galaxies observed with the Sloan Digital Sky Survey \citep[SDSS,][]{York2000}. Another popular supervised approach is the exploitation of decision trees.  Methods based on decision trees have also been widely applied on imaging data to separate stars and galaxies \citep{Weir1995,Vasconcellos:2011a,Bai2019}.

Another approach for machine-learning classification is unsupervised learning, which searches for clusters in the input feature space with minimal tuning. Distance based algorithms such as k-means (also Voronoi tesselation) require an a priori definition of the expected number of clusters, while density based algorithms, such as \textsc{DBSCAN} \citep{Ester:1996a}, require only the definition of the minimum number of objects belonging in a cluster and a distance between points. Other applications of unsupervised learning include source classification and physical parameter estimation with self-organizing maps \citep{Geach2012, Masters2017ApJ...841..111M, Masters2019ApJ...877...81M, Hemmati2019ApJ...881L..14H}.

Upcoming large extragalactic surveys will produce an unprecedented volume of data which will require new, robust, and automated processing methods \citep{Dubath2017}. Machine-learning methods offer viable solutions for a category of problems, including source classification which is the focus of this work. Indeed, a number of recent machine-learning methods have been applied on morphological classification \citep{Huertas-Company2015, Barchi2019}, point-source identification \citep{Kuntzer2016,VafaeiSadr2019}, QSO detection \citep{Jin2019}, star classification \citep{Torres2019}, and novelty and anomaly detection \citep{Gieseke2017}.

In this work, we apply Hierarchical Density-Based Spatial Clustering of Applications with Noise, \citep[HDBSCAN,][]{Campello:2013a} on the problem of star, galaxy, and QSO separation. In \S \ref{section.dataset} we describe the dataset used in this work to build the classification model. In \S \ref{section.metrics} we define the performance metrics used throughout the paper. In \S \ref{section.attribute_selection} we present our method for feature selection and dimensionality reduction used as input in the final classifier. In \S \ref{section.classification_model} we describe \textsc{hdbscan}, and the optimization and final construction of the classifier. In \S \ref{section.results} we discuss the impact of the input attributes, including colours and half light radii, as well as the photometric depth and photometric uncertainties. In \S \ref{section.discussion_and_application} we compare our results to the automatic labels of the Sloan Digital Sky Survey (SDSS DR14), the refined labels of the SDSS Quasars catalogue of \citet{Paris2018}, and the KIDS DR3 quasar catalogue of \citet{Nakoneczny2019}. Finally, we apply and publicly release our classifications on the Kilo Degree Survey, matched with the VIKING and ALLWISE surveys ($\sim$3 million sources). 

Due to its speed and scalability, we motivate the use of our classifications and photometric redshift estimates to separate stars, galaxies and QSO, particularly useful to assess the optimal spectroscopic templates for spectroscopic redshift estimation, applicable for future all-sky spectroscopic surveys such as 4MOST and DESI. Throughout the paper use the AB magnitude system.

\section{Dataset} \label{section.dataset}
To explore the capabilities of unsupervised learning in classification we use the sample of \citet[][hereafter FP18]{Fotopoulou:2018a}. This sample was constructed to be a representative population of spectroscopically observed stars, galaxies, and QSO selected on the basis of their complete photometric coverage in the optical, near infrared and mid-infrared wavelengths. In this work, we are using the uk -- ir sample, a total of 49,181 sources with continuous photometric coverage from $u$ -- $W2$ without any missing photometric data.

\subsection{Catalogues}
Briefly, the FP18 sample is a collection of spectroscopically observed sources from public spectroscopic surveys ( SDSS/DR12\footnote{\url{http://www.sdss.org/dr12/data_access/bulk/}} \citep{Alam2015}, GAMA/DR2\footnote{\url{http://www.gama-survey.org/dr2/data/cat/SpecCat/v08/}} \citep{Liske2015},  VIPERS/DR1\footnote{\url{http://vipers.inaf.it/rel-pdr1.html}} \citep{Garilli2014},  VVDS/DR2 \citep{LeFevre2013}, 
PRIMUS/DR1 \citep{Coil2011,Cool2013}, 6df/DR3 \citep{Jones2004,Jones2009} ) matched with associated z, Y, J, H, and K photometry from the ESO near-infrared Public VISTA surveys \footnote{\url{https://www.eso.org/sci/observing/PublicSurveys/sciencePublicSurveys.html}} \citep{Arnaboldi2007} (VIKING ($\rm{J_{lim, AB}=22.1}$, PI W. Sutherland) and VIDEO ($\rm{J_{lim, AB}=24.5}$, PI M. Jarvis). 

The optical filters (u, g, r, i, z) originate from the SDSS survey \citep[DR12, $\rm{i_{lim, AB}=21.3}$,][]{Alam2015}, CFHTLS \citep[T0007, $\rm{i_{lim, AB}=24.8}$,][]{Hudelot2012} and KiDS \citep[DR2, $\rm{i_{lim, AB}=24.2}$,][]{deJong2015} surveys. 
The mid-infrared observations in the W1 and W2 filters of the WISE satellite \citep[ALLWISE\footnote{\url{http://wise2.ipac.caltech.edu/docs/release/allwise/}}, $\rm{W1_{lim,AB}=20.3}$,][]{Wright2010,Mainzer2011}. All photometric measurements were corrected according to the Schlegel maps of Galactic absorption \citep{Schlegel1998} and the Cardelli law for the Milky way \citep{Cardelli1989}.

\subsection{Spectroscopic labels} \label{section.dataset_speclabels}
FP18 created a framework for optimal photometric redshift estimation, suited for large area surveys. To that end, they created a labelling system tailored to selecting the best photometric redshift library setup and performed supervised learning using Random Forest. 
In this work, we use unsupervised learning methods to identify the nature of the sources without any prior knowledge. Labels are only needed to judge the performance of the classifier and they are not used at any point during the training stage of the algorithm\footnote{Since we use spectroscopic classifications to judge the performance of the classifier after the clustering is performed, our approach could also be described as semi-supervised}. Contrary to FP18, we use the labelling assigned by each spectroscopic survey.

However, the spectroscopic labels are assigned by a non-homogeneous process, either automatic as in the case of SDSS spectra or manual as is the case for example in VIPERS and VVDS. Table \ref{tab:labels} shows the breakdown of the spectral labels. About 52\% of the spectra unfortunately carry no label (UNKNOWN). We label the 42 sources with UNKNOWN label at z<0.0015 (vertical dashed line) as stars, while all remaining 25,372 sources are labelled as galaxies, assuming that the lack of information means that there was nothing special in the spectrum to report.

\begin{figure*}
    \centering
    \begin{tabular}{cc}
    \includegraphics[width=\columnwidth]{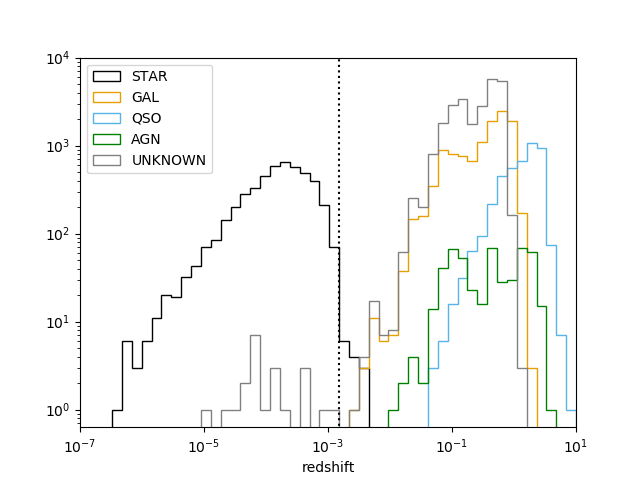}&
    \includegraphics[width=\columnwidth]{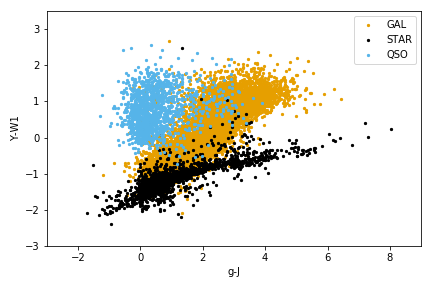} 
    \end{tabular}
\caption{Left: redshift distributions of the spectroscopic classes, as defined in each survey. The category AGN is mixture of QSO and AGN. Part of the subsample lacking categorization, labelled as UNKNOWN are evidently stars (z<0.0015). Right: colour-colour plot colour coded according to spectroscopic labels.}
    \label{fig:hclass}
\end{figure*}

The left hand side of Figure \ref{fig:hclass} shows the spectroscopic redshift distribution of our sample split according to the spectroscopic labels. As expected, sources classified as stars lie at very low redshift values, while sources classified as QSO tend to be at higher redshifts compared to the GAL category. Only four of the spectroscopic surveys assigned an AGN category (Primus, SDSS, VIPERS, VVDS). Given the small amount of spectra in this category, and not having access to all 1D spectra to assess their overlap with the QSO class, we decided to exclude these objects from our sample, reducing the sample that we use to 48,686 sources.

The right-hand side of Figure \ref{fig:hclass} shows the loci of the spectroscopically assigned labels on a colour-colour space known to separate well between the populations. We note that about 10\% of sources with spectroscopically assigned QSO label are located in the galaxy locus. These sources include reddened quasars, either due to intrinsic absorption or due to the intergalactic medium, and AGN. We will discuss further the impact of this in \S \ref{section:discussion-labels}.

\begin{table}
    \centering
    \scalebox{0.95}{
    \begin{tabular}{ccr@{,}l|ccr@{,}l}\hline\hline
    \multicolumn{4}{c|}{Initial labels} & \multicolumn{4}{c}{Final labels}\\\hline
       hclass  &  label & \multicolumn{2}{c|}{N sources} & hclass  &  label &  \multicolumn{2}{c}{N sources} \\ \hline 
        STAR  &  0  & 7&689 & STAR & 0 & 7&731 \\
        GAL & 1 & 11&391 & GAL & 1 & 36&763 \\
        AGN & 2 &  \multicolumn{2}{c|}{495} &  AGN & -- &  \multicolumn{2}{c}{--}\\
        QSO & 3 & 4&192 &QSO & 2 & 4&192\\
        unknown & -1 & 25&414 & unknown & -- &  \multicolumn{2}{c}{--}\\ \hline
        Total & & 49&181 & Total & & 48&686\\\hline\hline
    \end{tabular}}
    \caption{Breakdown of spectroscopic redshift labels, as given by the surveys considered in this work and the rectified number of sources. We remove all AGN sources from our sample, and attribute `unknown' labelled sources as either stars (z < 0.0015) or galaxies (z > 0.0015).}
    \label{tab:labels}
\end{table}

\section{Performance metrics} \label{section.metrics}
In this section, we describe the metrics that we use in this work to measure the performance of the star/galaxy/QSO classification model that we build, and define some of the terms that we use in this work.

We refer to the input features (colours, half light radii) to the machine learning algorithm as attributes. We refer to a parameter of the machine learning model, whose value is set before the training process and can be changed to improve model performance, as a hyperparameter. We refer to a binary classifier, or a model with binary labels, when we separate only two populations - OBJECT and NON-OBJECT (where OBJECT is one of star/galaxy/QSO); whereas when we refer to a multi-label classifier, we are separating all three populations simultaneously. It is possible to create a multi-label classifier with a number of binary classifiers, and combine their output labels, which is ultimately the setup that we choose to use in this work.

We use various metrics to evaluate the success of our classifications. They utilize the following information, which we obtain by comparing the predicted labels from our model with the true (spectroscopic) labels. We will explain them using the example of a binary star classifier.
\begin{itemize}
    \item TP - true positive: an object with true label star is classified as a star
    \item TN - true negative: an object with true label non-star is classified as a non-star
    \item FP - false positive: an object with true label non-star is classified as a star
    \item FN - false negative: an object with true label star is classified as a non-star
\end{itemize}

The metrics that we use in this paper are:

\paragraph{Accuracy} Fraction of correct predictions:
\begin{equation}
\rm{ACC = \frac{TP+TN}{TP+TN+FP+FN}}
\end{equation}
\paragraph{Precision} Fraction of correct positive predictions. In astronomy it is common to refer to precision as purity, and the words precision and purity will be used interchangeably in this work:
\begin{equation}
\rm{P = \frac{TP}{TP+FP}}
\end{equation}
\paragraph{Recall} Fraction of truly positive predictions. In astronomy it is common to refer to recall as completeness, and the words recall and completeness will be used interchangeably in this work:
\begin{equation} \label{eqn.recall}
\rm{R = \frac{TP}{TP+FN}}
\end{equation}	
\paragraph{F1 score} Harmonic mean of precision and recall:
\begin{equation}
\rm{F1 = 2\cdot\frac{P\cdot R}{P+R}}
\end{equation}	
\paragraph{Fall-out} Fraction of FP over negative condition: 
\begin{equation}
\rm{F = \frac{FP}{TN+FP}}
\end{equation}

When building a classifier, one tries to obtain the highest accuracy, precision and recall (and therefore F1 score), and the lowest fall-out. Throughout this work, we focus on the F1 score, as it is a trade off between precision and recall. However, if a higher precision may be favoured over a higher recall, then it is straightforward to optimize the final classification model that we present in this work to suit those needs.

Additionally, we also use the area under the curve (AUC) of the Receiver Operating Characteristic curve (ROC, \citealp{Bradley:1997a}) to assess classifier performance. The ROC curve is a plot that shows how the True Positive rate (TPR, also referred to as recall which we define in equation \ref{eqn.recall}) and False Positive Rate (FPR, defined as TN/(TN+FP)) change for a binary classifier as the classification threshold is varied. As our algorithm, \textsc{hdbscan}, does not provide a useful probability for each classification (except for the `highest-probability' method presented in \S \ref{subsection.consolidation}) we calculate AUC scores for the spectroscopic labels as true labels, and the predicted labels. The AUC gives us the probability that the classifier will rank a randomly chosen positively classified object higher than a randomly chosen negatively classified object, and is often used as a summary statistic of a ROC curve or more generally of a classifier's performance.

\section{Pre-processing and feature selection} \label{section.attribute_selection}
Most machine learning algorithms are, in general, sensitive to the selection of input attributes and the presence of correlations between attributes. In this section we explain how we create the sets of attributes that we used as input into the \textsc{hdbscan} classifier (the \textsc{hdbscan} algorithm is outlined in \ref{subsection.hdbscan}). 

\subsection{Pre-processing} \label{subsection.preprocessing}
Our sample contains 48,686 data points with measured total and 3$''$ aperture magnitudes in the u, g, r, i, z, Y, H, J, K bands and total magnitude in the W1 and W2 bands. We created all unique colour combinations from these photometric bands, creating 190 colours which we use to identify the optimal input attributes for \textsc{hdbscan}.

In addition, we explore the use of the half light radius (HLR) values and their ratios for the Y -- K bands. Whenever we use the Y -- K HLR values as part of our attributes, we remove all data points that have missing data for any of their Y -- K HLR values, or those that have unrealistic values (0 $''$ < HLR < 20 $''$), giving 43,348 data points. We do not use the u -- z HLR values, as there are too many missing values in our data set to use these without drastically reducing the size of our data set.

Before inputting the chosen attributes to any of the machine learning methods used, we use \texttt{scikit-learn} \citep{Pedregosa:2011a} to normalize and whiten the data to give zero mean and a variance of one for the attributes. It is crucial that the same scaling transformation that is applied on the training data is also applied on any new data.


\subsection{Dimensionality reduction and feature selection} \label{section.dim_reduction_and_feature_selection}
Dimensionality reduction and attribute selection is often the most important part of a machine learning pipeline. We want to reduce our high number of attributes that we have from the colour (and HLR) information to a lower number of attributes, as machine learning methods often struggle when faced with a large number of attributes. We do this by either converting all of our attributes into a lower number of dimensions (\S \ref{subsubsection.PCA}) using Principal Component Analysis (PCA), or by selecting the most informative attributes for our needs (\S \ref{subsubsection.RF}) using Random Forest (RF, \citealp{Breiman:2001a}) or we use a combination of RF followed by PCA. For both algorithms we use the \texttt{scikit-learn} implementations in Python \citep{Pedregosa:2011a}.

\subsubsection{Principal Component Analysis} \label{subsubsection.PCA}
Principal Component Analysis (PCA) is a dimensionality reduction technique, and has been widely applied to problems in astronomy (e.g. \citealp{Francis:1999a,McGurk:2010a,Paraficz:2016a}). \citet{Shlens:2014a} gives a nice introduction to PCA. In short, PCA works by creating an optimal rotation of a new coordinate system aligned with the maximum variance of the data. PCA then outputs new vectors for each data point by converting the data points in the new coordinate system, encompassing the maximum information for each data point from the high-dimensional parameter space. The new attributes are then used as input to  \textsc{hdbscan}. It is crucial that the same PCA transformation that is applied on the training data is also applied on any new data.

\subsubsection{Random Forest} \label{subsubsection.RF}

\begin{table}
    \centering
    \begin{tabular}{cccccc}\hline
     STAR     &  GAL & QSO & ALL  \\\hline 
     J$_3$-W1 & K-Y$_3$   & z-u$_3$   &  K-Y$_3$  \\
     K-J$_3$  & K-J$_3$   & i-u$_3$   &  K-J$_3$  \\
     Y$_3$-W1 & K-Z$_3$   & r-g$_3$   &  K-H$_3$  \\
     K-H$_3$  & K-H$_3$   & u$_3$-z$_3$  &  J$_3$-W1  \\
     J$_3$-K$_3$ & J$_3$-K$_3$ & u$_3$-i$_3$ &  J$_3$-K$_3$  \\
     H$_3$-W1 & Y$_3$-K$_3$   & Y-u$_3$   &  Y$_3$-W1  \\
     K-Y$_3$  & J$_3$-W1   & u-z   &  H$_3$-W1  \\
     H$_3$-K$_3$ & Y$_3$-W1 & z-g$_3$ &  H$_3$-K$_3$  \\
     Y$_3$-W2 & J-K   & r-u$_3$   &  J-K  \\
     J-K      & H$_3$-K$_3$ & u-i    &  Y$_3$-K$_3$  \\ \hline
    \end{tabular}
    \caption{Top 10 attributes from the output of RF. The first three columns are the top 10 attributes for when the labels were binary for STAR/non-STAR, GAL/non-GAL, QSO/non-QSO and the fourth column is for the multi-label setup.}
    \label{tab:RFtopatts}
\end{table}

Random Forest (RF, \citealp{Breiman:2001a}) is a supervised learning method that can be used to classify data. An RF is an ensemble learning method, and consists of a collection of decision trees. Each decision tree is trained using a different random sub-sample of the total input attributes. To use the RF as a classifier, each decision tree's final output is used as a `vote' to assign a class for each input data point: the RF can either return a probability\footnote{The probability is equal to the number of `votes' that the object is in that class divided by the total number of decision trees.} for each data point to belong to a certain class, or return a single class prediction for each data point by returning the class with the highest number of `votes' for each data point. The RF's use of multiple decision trees, each with different attribute sets, ensures that it is robust to overfitting, in contrast to a single decision tree. In addition to functioning as a supervised learning classifier, the RF algorithm can return an importance for each of the input attributes, which represents predictive power of each attribute. When using the RF to calculate importances in this work, we use Gini impurity based importances.

We use the RF to obtain ranked lists of attributes for four different label setups (see Table \ref{tab:RFtopatts} for the top 10 from each). We keep the original multi-label setup (referred to in Table \ref{tab:RFtopatts} as ALL), with labels for STAR/GAL/QSO, and run the RF on three other labels setups, where we convert the labels to be binary for STAR/non-STAR, GAL/non-GAL and QSO/non-QSO (referred to in Table \ref{tab:RFtopatts} as STAR, GAL and QSO respectively). We produce the list of importances by running the RF 1000 times for each of the three binary label setups and also the multi-label setup, and rank the attributes by average importance over all 1000 runs for each different setup. This is necessary as RF is non-deterministic, due to the fact that the random sub-samples chosen for each decision tree in the training phase of the RF will be different each time the RF is trained. 

\subsubsection{Lists of Colour Attributes from FP18} \label{subsubsection.ABC_atts}
In addition to the four ranked lists of important attributes from the RF runs that we detail above, we also use the ranked lists of important attributes from the RF classifiers A, B and C from FP18 (see paper for more detail on these classifiers), and the two colours g-J and Y-W1 as used in Figure 4 in FP18. Although classifiers B and C were constructed for different purposes compared to the work we are doing, we still expect the output colours from them to be significantly more informative than just choosing colours at random. We also find that the attributes in classifiers B and C are generally not highly correlated (see \S \ref{subsubsection.correlatedatts} and right panel in Figure \ref{fig:corrABC_and_SGQA}), further validating our choice to use them in the construction of attribute sets.

\subsubsection{Correlated attributes} \label{subsubsection.correlatedatts}

\begin{figure*}
    \centering
    \begin{tabular}{cc}
    \includegraphics[width=0.5\linewidth]{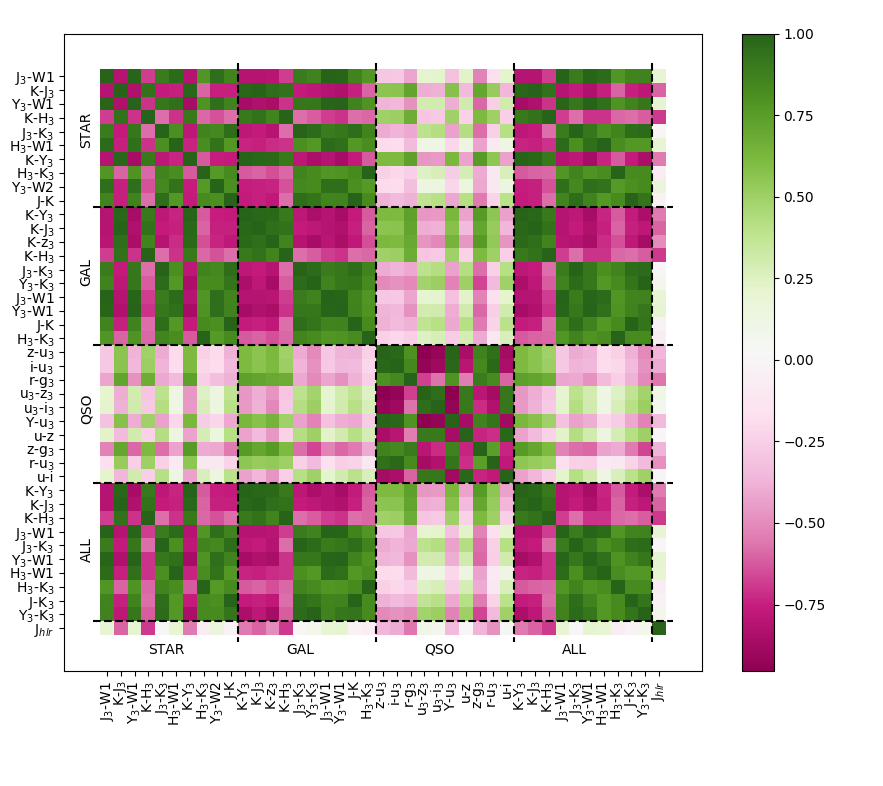}
    &
    \includegraphics[width=0.5\linewidth]{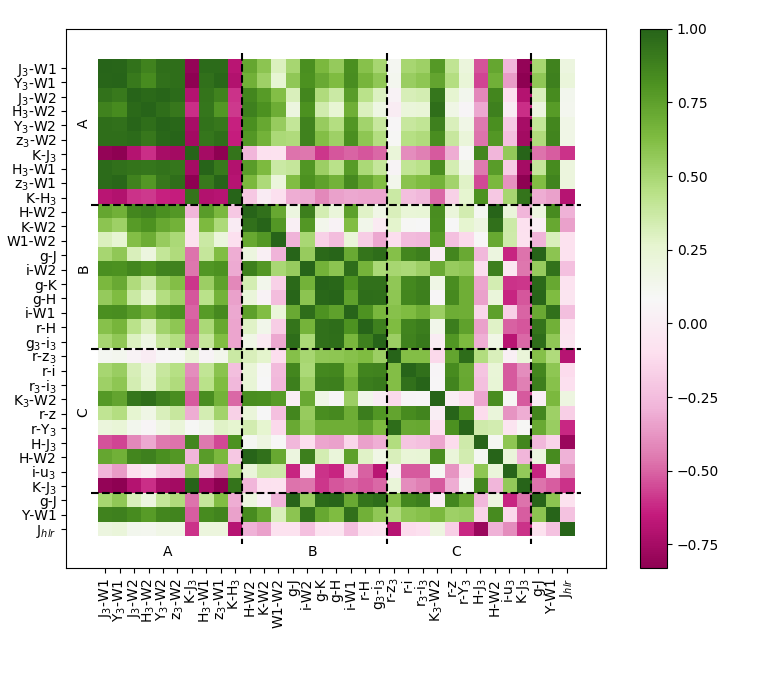}
    \end{tabular}
    \caption{Left: Correlation matrix for the top 10 attributes from the three RF binary label and one multi-label output, and the $J_{hlr}$ attribute. The STAR, GAL, QSO, ALL labels show the top 10 attributes grouped together for the different RF setups (explained in \S \ref{subsubsection.RF}).
    Right: Correlation matrix for the top 10 attributes from the three classifiers A, B and C from FP18, the colours g-J and Y-W1, and the $J_{hlr}$ attribute. The A, B, C labels show the top 10 attributes grouped together for the classifiers A, B and C from FP18.}
    \label{fig:corrABC_and_SGQA}
\end{figure*}

Figure \ref{fig:corrABC_and_SGQA} (left) shows that there is a high degree of correlation among the input attributes. This is also seen in Table \ref{tab:RFtopatts} where it is clear that many similar attributes are in the top 10 attributes from RF - for example both J$_3$-W1 and Y$_3$-W1 are in the top three most important attributes for the STAR RF binary classifier. This is because RF looks only at a single attribute at each decision point in each decision tree. This is in contrast to \textsc{hdbscan}, where all attributes are used simultaneously. This means that if we use e.g. the top 10 attributes from the list of important attributes from the RF star classifier, some are likely to be highly correlated with each other, so when using the combination of these 10 attributes as an input to \textsc{hdbscan}, it might not actually be an informative selection of attributes for \textsc{hdbscan}; in fact, by adding highly correlated attributes to the input attributes, we may simply be adding noise. 

To mitigate against this issue of inputting highly correlated attributes to \textsc{hdbscan}, we apply the following approaches: 
\begin{itemize}
    \item We combine the top attributes as given by each of the three binary and the one multi-label RF classifiers and also combine those from the classifiers A, B and C from FP18, 
    as the combinations are expected to be less correlated. This is confirmed by inspecting a correlation matrix (see Figure \ref{fig:corrABC_and_SGQA}). 
    \item We add the colours g-J and Y-W1 to the top attributes from the classifiers A, B and C from FP18, as these are uncorrelated with certain attributes from these classifiers' ranked lists of attributes. For example, g-J and the top 10 attributes from classifier A in Figure \ref{fig:corrABC_and_SGQA} are not highly correlated, so adding g-J to this set of attributes will likely improve the performance of \textsc{hdbscan} when using the list of attributes from classifier A with the colour g-J added as opposed to without it. 
    \item We `trim' the top 10 important attributes from the RF output by removing the most correlated attributes. 
    \item We investigate the impact of adding the HLR information to the colour data, as the HLR is not expected to be highly correlated with our colour input attributes (this is clear in Figure \ref{fig:corrABC_and_SGQA}). We would expect the HLR information for each data point to be useful, as it gives an indicator of how point-like the object is. 
    In addition, ratios between HLRs can be used as extra attributes, and are also useful in separating galaxies from point sources. As mentioned in \S \ref{subsection.preprocessing}, for this work we use the Y -- K HLR values and their ratios as attributes.
\end{itemize}

\subsubsection{Construction of Attribute Sets} \label{subsection.att_sets}

Using the RF's lists of importances for different classifier setups as explained above we created a large number of different sets of attributes, by selecting different numbers of attributes from each list: the top 3 -- 20 attributes at a time were chosen from the RF classifiers' lists of attributes by importance (top 10 of each are shown in Table \ref{tab:RFtopatts}), and the top 3 -- 10 for classifiers A, B and C from FP18 to create different attribute sets. We also used the methods as discussed in \S \ref{subsubsection.correlatedatts} to create further attribute sets with reduced correlation between attributes in each attribute set. We also added the Y -- K HLR values and their ratios to these attribute sets - however, we also keep the attribute sets with just colour information, as we can compare the classification results of using just colour data versus colour data \textit{and} HLR values, to investigate the impact of adding HLR values to the colour information (see \S \ref{subsection.adding_HLR_impact}). 

We then performed PCA on these sets of attributes to reduce the number of input attributes further. From an initial exploration of \textsc{hdbscan}, we found that the best performance on our data set was obtained when our parameter space was reduced from $\sim$200 attributes to $\sim$5 -- 30 attributes and PCA was then performed to reduce those attributes to $\sim$ 2 -- 5 dimensions, so our sets of attributes are constructed loosely following these rules. In total, we constructed $\sim$ 4,000 different attribute combinations. Figure \ref{fig:tikz_gridsearch} shows a summary of the method we used for creating our attribute sets 

\section{Classification model} \label{section.classification_model}

\subsection{Algorithm description}

\textsc{hdbscan} (Hierarchical Density-Based Spatial Clustering of Applications with Noise, \citealp{Campello:2013a}) is an unsupervised clustering algorithm, based on \textsc{dbscan} (Density-based Spatial Clustering of Applications with Noise, \citealp{Ester:1996a}).
The \textsc{hdbscan} algorithm is first presented in \citep{Campello:2013a}.
In this work, we are using the implementation\footnote{\url{https://pypi.org/project/hdbscan/}} of \citet{McInnes:2017a}. 

\subsubsection{DBSCAN}

\begin{figure}
    \centering
    \includegraphics[width=\columnwidth]{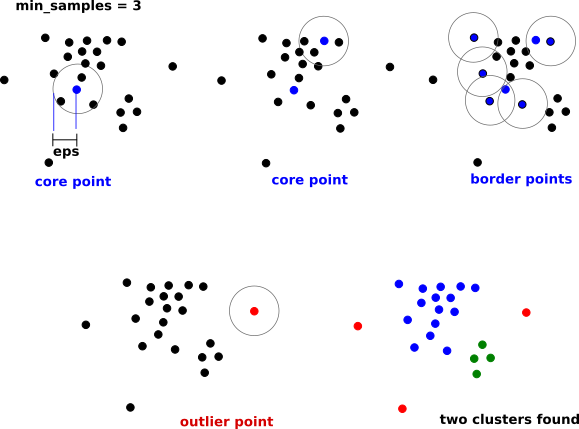}
    \caption{The main steps in the \textsc{dbscan} algorithm are shown, for an example where the \texttt{min\_samples} hyperparameter is set to 3. }
    \label{fig:dbscan_algo}
\end{figure}

\begin{figure}
    \centering
    \includegraphics[width=\columnwidth]{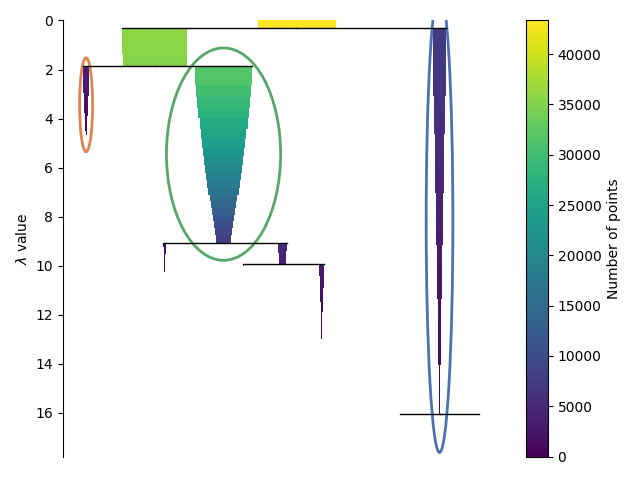}
    \caption{An example of a dendrogram from an \textsc{hdbscan} run is shown. In this example, three clusters are found, and are circled. The $\lambda$ value is 1/distance.}
    \label{fig:hdbscan_dendro}
\end{figure}

\textsc{dbscan} \citep{Ester:1996a}
is a density-based clustering algorithm. Density-based clustering algorithms work by finding clusters of points by looking for regions in the parameter space of the data where there is a high density of points, surrounded by a region where there is a lower density of points. In addition to being an intuitive way of searching for clusters, it also works well in practice.

\textsc{dbscan} requires two hyperparameter values to be specified: a neighbourhood distance (\texttt{eps} in the \texttt{scikit-learn} implementation) which is used to define the maximum distance between two points for them to be considered in the same cluster, and minimum points (\texttt{min\_samples} in the \texttt{scikit-learn} implementation) which defines the value at which a neighbourhood of points is considered dense. The algorithm works by defining core points, border points and noise points. A sketch of the \textsc{dbscan} algorithm's key steps are shown in Figure \ref{fig:dbscan_algo}. Briefly:

\begin{itemize}
    \item A core point is defined such that it has at least the value of \texttt{min\_samples} of points within a distance \texttt{eps}. 
    \item A border point is defined such that it has at least one core point in its neighbourhood and isn't a core point itself.
    \item A noise or outlier point is any other point that is not a core or a border point.
    \item Clusters are formed by picking a core point in the data set and searching for any other points that are within a distance \texttt{eps} of this core point, assigning them to the same cluster as this initial core point. The algorithm repeats this process until all core points have been assigned to a cluster.
    \item Clusters consist of core and border points. Noise points are considered outliers, and are not part of any cluster.
\end{itemize}

Advantages of using \textsc{dbscan} include: it can discover any number of clusters; it can find clusters of varying size and shape; it can detect and ignore outliers. Disadvantages of using \textsc{dbscan} include: it is sensitive to the choice of the hyperparameter \texttt{eps} - if it is too small, then a sparse cluster will be labelled as noise, and if it is too large then dense clusters will be merged together as one cluster. 

\subsubsection{HDBSCAN} \label{subsection.hdbscan}

The main difference between \textsc{dbscan} and \textsc{hdbscan} is that instead of counting points within a fixed radius \texttt{eps} to define core, boundary and noise points, \textsc{hdbscan} effectively does this using an expanding radius, such that the only hyperparameter of importance is the \texttt{min\_cluster\_size} (the minimum size that a cluster can be). By effectively running \textsc{dbscan} with every possible value of \texttt{eps}, \textsc{hdbscan} takes advantage of the fact that in the \textsc{dbscan} algorithm, when decreasing \texttt{eps}, clusters will only fragment into smaller clusters (or remain the same). This means that a hierarchical tree or dendrogram can be produced that shows the clustering output (for a given value of \textsc{hdbscan}'s hyperparameter \texttt{min\_cluster\_size}). By making a cut through the tree we can produce an equivalent outcome as running \textsc{dbscan} for a particular value of \texttt{eps}. \textsc{hdbscan} finds the optimal cut through the dendrogram by returning the most persistent clusters. An example of a dendrogram output from a run of \textsc{hdbscan} is shown in Figure \ref{fig:hdbscan_dendro}. From the hierarchy of potential clusters, \textsc{hdbscan} returns a flat clustering, which corresponds to the cluster labels. 

One advantage 
of \textsc{hdbscan} over \textsc{dbscan} is that it can identify clusters of varying density. For our data, the density of points in individual clusters varies significantly, so being able to find clusters of varying density is an important requirement. \textsc{hdbscan} is also faster than \textsc{dbscan} by a factor of $\sim$ 2 for the $\sim$ 50,000 data points that we are training on; this discrepancy in speed only increases with more data points. In addition, \textsc{hdbscan} is more robust to the hyperparameter selection of the model, whereas \textsc{dbscan} is very sensitive to the choice of hyperparameters. 

We explored hyperparameters of \textsc{hdbscan} to assess their impact on the performance of the classifier. We found that 1) the distance metric `Euclidean' and `Manhattan' are the best choices, giving similar results and speed. We chose to use `Euclidean' distance in this work. 2) the minimum number of points in a neighbourhood for a point to be recognized as a core point (\texttt{min\_samples}), did not impact the results. Therefore, we used the default, where it is automatically set to the same value as \texttt{min\_cluster\_size}. 3) finally, we explored the \texttt{cluster\_selection\_method} and found that the default Excess of Mass (`eom') worked best. In summary, changing the other hyperparameters (aside from \texttt{min\_cluster\_size}) never improved the performance of \textsc{hdbscan} with regards to our needs. 

Being an unsupervised learning method, \textsc{hdbscan} simply returns a number of clusters and their members, and does not assign an object class (STAR/GAL/QSO) to each cluster. It is possible to visually inspect the clustering output, and assign clusters to an object - however this becomes inefficient over a large number of runs. Instead, we automate this step, by assigning each cluster to the most frequently occurring object in that cluster, in order to then be able to calculate the metric scores for each setup for \textsc{hdbscan} that we explore (see \S \ref{section:model_construction}). In the case where there are more than three clusters returned by \textsc{hdbscan}, the cluster with the largest number of each object is the cluster we assign to that object. In some cases, more than three clusters are found, and in these cases we do not ascribe a class to these other clusters that are found.

In addition to returning a number of clusters and their members, \textsc{hdbscan} also returns an outlier class for points that do not clearly belong to any other cluster, and returns an outlier score, which tells us how `strong' of an outlier each outlier point is. Additionally, for each data point (including those in the outlier class) \textsc{hdbscan} can return the probability that it belongs to a certain cluster; however this is a simple probability that is based on the distance of a point from the centre of a cluster - we present a method to produce more informative probabilities in \S \ref{subsection.consolidation}.

\usetikzlibrary{shapes.geometric, arrows}
\tikzstyle{startstop} = [rectangle, rounded corners, minimum width=3cm, minimum height=1cm,text centered, draw=black]
\tikzstyle{arrow} = [thick,->,>=stealth]
\tikzstyle{arrow_dashed} = [ thick , ->, densely dashed ]

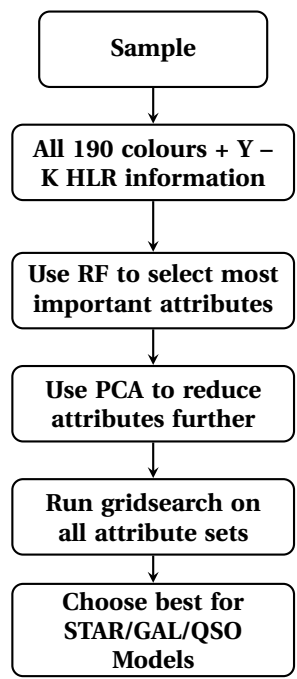
\begin{figure}
\centering
\scalebox{1.0}{
\begin{tikzpicture}[node distance=2.5cm,thick]
\node (sample) [startstop] {\textbf{Sample}};
\node (colours) [startstop, below of=sample,yshift=1cm,text width=3.5cm] {\textbf{All 190 colours + Y -- K HLR information}};
\node (RF) [startstop, below of=colours,xshift=0cm,yshift=0.8cm,text width=3.5cm] {\textbf{Use RF to select most important attributes}};
\node (PCA) [startstop, below of=RF,xshift=0cm,yshift=1cm,text width=3.5cm] {\textbf{Use PCA to reduce attributes further}};
\node (gridsearch) [startstop, below of=PCA,xshift=0cm,yshift=1cm,text width=3.5cm] {\textbf{Run gridsearch on all attribute sets}};
\node (choose) [startstop, below of=gridsearch, yshift=1cm,text width=3.5cm] {\textbf{Choose best for STAR/GAL/QSO Models}};
\draw [arrow] (sample) -- (colours);
\draw [arrow] (colours) -- (RF);
\draw [arrow] (RF) -- (PCA);
\draw [arrow] (PCA) -- (gridsearch);
\draw [arrow] (gridsearch) -- (choose);
\end{tikzpicture}
}
\caption{Flowchart to show the method we used to select the best attributes for our \textsc{hdbscan} classifier. The first four steps are detailed in \S \ref{section.dim_reduction_and_feature_selection} and the final two steps are detailed in \S \ref{section:model_construction}}
\label{fig:tikz_gridsearch}
\end{figure}

\begin{figure}
\centering
\scalebox{0.8}{
\begin{tikzpicture}[node distance=2.5cm,thick]
\node (starmodel) [startstop, xshift=-4cm,yshift=0cm,text width=3cm] {\textbf{STAR Model (output interpreted as STAR/non-STAR)}};
\node (galmodel) [startstop, xshift=0cm,yshift=0cm,text width=3cm] {\textbf{GAL Model (output interpreted as GAL/non-GAL)}};
\node (qsomodel) [startstop, xshift=4cm,yshift=0cm,text width=3cm] {\textbf{QSO Model (output interpreted as QSO/non-QSO)}};
\node (consol) [startstop, below of=galmodel, yshift=0.5cm] {\textbf{Consolidation}};
\node (classifications) [startstop, below of=consol, yshift=0.5cm] {\textbf{Final Object Classifications}};
\draw [arrow] (starmodel) -- (consol);
\draw [arrow] (galmodel) -- (consol);
\draw [arrow] (qsomodel) -- (consol);
\draw [arrow] (consol) -- (classifications);
\end{tikzpicture}
}
\caption{Flowchart to show the final model setup. The setups for the STAR, GAL and QSO models are detailed in Tables \ref{tab:finalmodelsetups_nohlr} and \ref{tab:finalmodelsetups} respectively for when just colour and for when both colour and HLR information are used as attributes. The `Consolidation' step is described in \S \ref{subsection.consolidation}}
\label{fig:tikz_model_setup}
\end{figure}
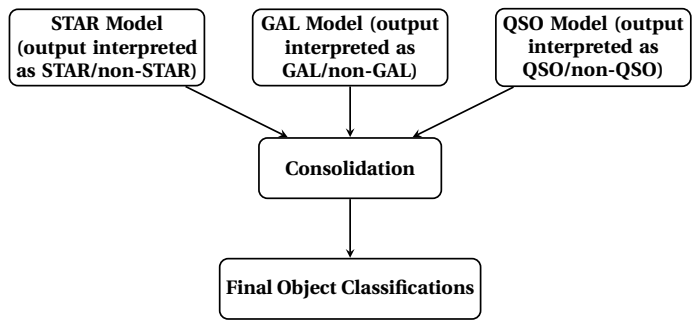

\subsection{Model construction} \label{section:model_construction}
In this section we describe the method that we use to find the optimal setup for our \textsc{hdbscan} classifier. This section describes the final two steps in Figure \ref{fig:tikz_gridsearch}. The goal of the attribute selection step is to select the attributes (and corresponding \texttt{min\_cluster\_size} value) that give optimal performance for \textsc{hdbscan} for the classification of stars, galaxies and QSOs.

From our large list of potential attribute sets (see \S \ref{subsection.att_sets}), we search for the optimal set of attributes to use in our final model setup by trying each attribute set as input into \textsc{hdbscan}, and varying the \texttt{min\_cluster\_size} hyperparameter of the \textsc{hdbscan} algorithm over a range of values for each attribute set. We use the metrics as described in \S \ref{section.metrics} (with the focus on F1 score for the purposes of this work) to evaluate the performance of \textsc{hdbscan} for when each of the attribute sets were used as input. The values of the \texttt{min\_cluster\_size} hyperparameter that we iterate over are:  2 -- 70 (1), 70 -- 100 (2), 100 -- 120 (5), 120 -- 200 (20), 200 -- 500 (50), 500 -- 1000 (100), 1000 -- 2000 (200), 200 - 5000 (500), with the ranges shown and their step size in parentheses. In total, there are 114 different \texttt{min\_cluster\_size} values that we test. We refer to this step as a gridsearch (over the different attribute sets and over the different \texttt{min\_cluster\_size} values to find which attribute set and \texttt{min\_cluster\_size} combination gives the best classification performance). This step is a brute force method; however, we constructed our different attribute sets in an informed manner (see \S \ref{subsubsection.correlatedatts} and \ref{subsection.att_sets}). This gridsearch provides us with predicted labels for all points, from which we can calculate the associated metrics (defined in \S \ref{section.metrics}) for the STAR, GAL and QSO class, for each set of input attributes, and for each of the \texttt{min\_cluster\_size} values that we iterated over. 

Inspecting these output metrics, it is clear that the best metric scores for a specific object (e.g. the STAR class), are obtained with a specific attribute setup that is different to the attribute setup with which we can obtain the highest metric scores for a different object (e.g. the QSO class). Thus, for our final model of classifying STAR/GAL/QSO, we opt to use the output of three \textsc{hdbscan}  runs, enabling us to use different attribute and \texttt{min\_cluster\_size} value setups for the classification of each object class. Although \textsc{hdbscan} returns the number of clusters it finds automatically, we convert the clustering the output to a binary one e.g. STAR, non-STAR allowing us to use the output as a binary classification. 
The metric that we use to select our best classifier setups (attribute set and \texttt{min\_cluster\_size}) is the F1 score, though this is flexible - if for a specific science case purity was considered more important, then it is straightforward to choose the setup that maximizes purity, while keeping the completeness above a certain threshold, for example. We choose the setup that maximizes the F1 score for the STAR class, the GAL class, and the QSO class separately, giving us three binary classifiers in our final classifier model.

Table \ref{tab:finalmodelsetups_nohlr} shows the optimal setups of our three binary object classifiers, and their respective metric scores at this point in the method (i.e. before consolidation - see \S \ref{subsection.consolidation}), using just the colour data as input attributes and using all data points in our dataset. Table \ref{tab:finalmodelsetups} shows the optimal setups of our three binary classifiers, and their respective metric scores at this point in the method (i.e. before consolidation - see \S \ref{subsection.consolidation}), using the Y -- K HLR information in addition to the colour data as input attributes, with the reduced dataset where data points with missing Y -- K HLR values have been removed. Although when inputting just colour attributes to \textsc{hdbscan} we do use the full sample (with results reported in Table \ref{tab:finalmodelsetups_nohlr}), we also input to \textsc{hdbscan} the same colour attributes with the reduced dataset where data points with missing (or unrealistic) Y -- K HLR values have been removed, and the results using this reduced dataset and just colour attributes are also shown in Table \ref{tab:finalmodelsetups}, so as to be able to make a fair comparison between the case where our attribute sets include the Y -- K HLR values and their ratios to the case where we are using only the colour attributes (see \S \ref{subsection.adding_HLR_impact}).

The time taken to train \textsc{hdbscan} depends on the number of attributes fed to the algorithm, and also the \texttt{min\_cluster\_size} that we use. Most of the attribute combinations that we use as input to \textsc{hdbscan} have 2 -- 5 attributes, and take at most a few seconds to train (apart from when we use very high values of \texttt{min\_cluster\_size} values, where \textsc{hdbscan} can take up to 30 seconds to train). The longest time to train \textsc{hdbscan} is when we use all attributes that we have together ($\sim$200), when it takes $\sim$ 10 minutes to train with little dependence on the \texttt{min\_cluster\_size}. We used one core to train each model on an Intel(R) Xeon(R) CPU E5-2680 v4.

\subsection{Consolidation} \label{subsection.consolidation}
During the consolidation phase (see Figure \ref{fig:tikz_model_setup}), we combine the outputs of these three binary classifiers to give the final classifications of each data point. For the final classifications of all of the data points, the objects that are classified positively by the STAR binary classifier are considered STAR, and so on for GAL and QSO. 

We consider three consolidation methods `optimal', `alternative', and the `highest-probability' defined as follows:
\paragraph{Optimal method} For data points that are classified positively by two binary classifiers separately, we assign the point to the rarest object class, as follows:
\begin{itemize}
    \item sources classified as both STAR and GAL, we call STAR
    \item sources classified as both QSO and GAL, we call QSO.
    \item sources classified as both STAR and QSO, we call QSO. 
\end{itemize}
We also define an outlier class, (different to the \textsc{hdbscan} outlier class) that we refer to as the `post-consolidation outlier' class. These post-consolidation outliers are those that are not classified positively by any of the binary classifiers. 
\paragraph{Alternative method} The second consolidation method we investigated, which we refer to as the `alternative' method, we simply reassign all doubly positively classified objects to the post-consolidation outlier class.

\paragraph{Highest-probability method}
Finally, we created 100 catalogue realizations based on the CPz sample using a Gaussian distribution centred on that data point's value with a sigma equal to the uncertainty value. The maximum value for sigma that we allow is 1, so for any error values above this we use a sigma of 1; 8,175 data points in the u -- W2 photometric bands have errors greater than 1, with the majority of these being in the u or u3 bands. The Y -- K HLR values have no associated errors, so we use the same value in all new realizations.

From the predicted labels of the 100 realizations based on the CPz catalogue, we estimated a probability for each data point to belong to a certain class, calculated as the number of times it was classified (using the `optimal' consolidation method) as a certain class, divided by the number of realizations.
From the probability information we can also predict a final label for each data point, by finding the object class (including post-consolidation outlier) that has the highest probability value for each data point. In the cases where two classes are equally probable, we assign the data point to the rarest object class (as in the `optimal' consolidation method). We call these new final labels the `highest-probability' labels.

A comparison of the F1 scores obtained using the three different consolidation methods is shown in Tables \ref{tab:consolidation_method_comp} and \ref{tab:consolidation_method_comp_hlrs} when using just the colour information as input data, and when using both the colour and Y -- K HLR information as attributes respectively. Comparing the `alternative' method to the  `optimal' method, the GAL classifier metric scores do not change, but for the STAR and QSO classifiers the precision increases (and fall-out decreases) for the `alternative' method, however all the other metric scores decrease for the STAR and QSO classifiers, leading to the F1 scores for the STAR and QSO classes decreasing. However, if precision was the desired outcome, the `alternative' method would be the best choice for combining labels.

We find that the `highest-probability' method produces fewer outliers compared to the other two methods, while the F1 scores change by a negligible amount. Since we find that the `highest-probability' method gives similar results to running our \textsc{hdbscan} model once on the observed data, we do not choose to use the `highest-probability' labelling method over the original method; however, the probabilities for each point given by this method are clearly useful and more informative than the simple \textsc{hdbscan} probabilities, and we present these in the catalogues as well as the `highest-probability' labels (see \S \ref{section.appendix.catalogue_descriptions}).

\subsection{F1 uncertainties} \label{section.uncertainties}
In order to obtain an estimate of the level of uncertainty on the metric values that we present in Tables \ref{tab:consolidation_method_comp} and \ref{tab:consolidation_method_comp_hlrs} we predicted the labels and metric scores for the 100 catalogue realizations described in \S \ref{subsection.consolidation}, using the optimal \textsc{hdbscan} model trained on the original CPz catalogue. When training on colour data, we find the mean F1 scores for the star, galaxy and QSO class to be 98.52$\pm$0.04, 98.42$\pm$0.03 and 89.19$\pm$0.18 (error given is the standard deviation). When training on both colour and HLR data, we find the mean F1 scores for the star, galaxy and QSO class to be 98.62$\pm$0.05, 98.71$\pm$0.02 and 91.85$\pm$0.14. We note that the F1 scores are slightly lower than those in Tables  \ref{tab:consolidation_method_comp} and \ref{tab:consolidation_method_comp_hlrs}; this is expected as we are testing on data different to what the \textsc{hdbscan} model was optimized on. However, by using a predefined model and a perturbed dataset we can assess the generalization of the model and use it with confidence on previously unseen data. 

\subsection{Summary of model setup} \label{section.final_model_setup}

Figures \ref{fig:tikz_gridsearch} and \ref{fig:tikz_model_setup} summarize the process followed for the classification of STAR, GAL and QSO using \textsc{hdbscan}. Briefly, starting from the initial parameter space of about 200 attributes, we identified the highest importance attributes using Random Forest and further reduced that parameter space to three components using PCA. Finally, we ran a gridsearch to identify the best hyperparameters on the basis of the F1 score, estimated using the spectroscopic labels as the truth.

By selecting favorable input attributes and \textsc{hdbscan} hyperparameters per object class we constructed three classifiers. The output of the binary classifiers can then be consolidated for overall optimal F1 score or optimal precision. The specific setup of the STAR, GAL and QSO model is shown in Table \ref{tab:finalmodelsetups_nohlr} and in Table \ref{tab:finalmodelsetups} respectively for when we use just colour attributes and for when we use both colour attributes and HLR information as input. The exact attribute lists are given in Table \ref{tab:appendix_colours_conversion}.


\section{HDBSCAN performance} \label{section.results}
In the following section we discuss the performance of our classifier under various selections of the input attributes, assuming that the spectroscopic labels are correct. We will revisit this assumption in \S \ref{section:discussion-labels}. 

\begin{table*}[ht]
    \centering
    \scalebox{1}{
    \begin{tabular}{c|cccccccccc}
    \hline
         Attribute list & \multicolumn{10}{c}{colours} \\ \hline
         \textit{best\_star\_colours} &  K-Y$_3$ &  K-J$_3$ &  K-z$_3$ &  K-H$_3$ &  J$_3$-K$_3$ 
          &                              Y$_3$-K$_3$ &  J$_3$-W1 &  Y$_3$-W1 & J-K &  H$_3$-K$_3$ \\
          &                              H$_3$-W1  & Y-K &  H-Y$_3$ &  Y$_3$-W2 &  J$_3$-W2 
          &                              i-g$_3$ & z$_3$-W1 &  z$_3$-K$_3$ &  z-u$_3$ &  H-J$_3$ \\ \hline
         \textit{best\_gal\_colours} & g-J &  Y-W1 &  J$_3$-W1 &  Y$_3$-W1 &  J$_3$-W2
         &                              H$_3$-W2 &  Y$_3$-W2 &  z$_3$-W2 & K-J$_3$ &  H$_3$-W1 \\
         &                              z$_3$-W1 &  K-H$_3$ &H-W2 &  K-W2 &  W1-W2 
         &                              i-W2 & g-K &  g-H &  i-W1 &  r-H\\
         &                              g$_3$-i$_3$ &  r-z$_3$ &  r-i &  r$_3$-i$_3$ &  K$_3$-W2 
         &                              r-z &  r-Y$_3$ &  H-J$_3$ &  i-u$_3$ &\\ \hline
         \textit{best\_qso\_colours} & J$_3$-W1 &  Y$_3$-W1 &  J$_3$-W2 &  H$_3$-W2 &  Y$_3$-W2
         &                          z$_3$-W2 &  K-J$_3$ &  H$_3$-W1 & z$_3$-W1 &  K-H$_3$\\
         &                          H-W2 &  K-W2 &  W1-W2 &  g-J &  i-W2 
         &                          g-K &  g-H & i-W1 &  r-H &  g$_3$-i$_3$ \\
         &                          r-z$_3$ &  r-i &  r$_3$-i$_3$ &  K$_3$-W2 &  r-z 
         &                          r-Y$_3$ & H-J$_3$ &  i-u$_3$ &&\\\hline
         best\_star\_colours\_comp & g-J &  Y-W1 &  J$_3$-W1 &  Y$_3$-W1 &  J$_3$-W2 
         &                          H$_3$-W2 &  Y$_3$-W2 &  H-W2 & K-W2 & W1-W2 \\
         &                          i-W2 &   r-z$_3$ & r-i & r$_3$-i$_3$ & K$_3$-W2 
         &                          r-z & &&&\\\hline
         best\_gal\_colours\_comp & g-J &  Y-W1 &  J$_3$-W1 &  Y$_3$-W1 &  J$_3$-W2 
         &                          H$_3$-W2 &  Y$_3$-W2 &  z$_3$-W2 &  K-J$_3$ &  H$_3$-W1 \\
         &                          z$_3$-W1 &  K-H$_3$ & & &\\\hline
         best\_qso\_colours\_comp & J$_3$-W1 &  Y$_3$-W1 &  J$_3$-W2 &  H$_3$-W2 &  Y$_3$-W2 
         &                          z$_3$-W2 &  K-J$_3$ &  H$_3$-W1 &   z$_3$-W1 &  K-H$_3$ \\
         &                          H-W2 & K-W2 & W1-W2 & g-J & i-W2 
         &                          g-K & g-H & i-W1 &  r-H & g$_3$-i$_3$ \\
         &                          r-z$_3$ & r-i & r$_3$-i$_3$ & K$_3$-W2 & r-z 
         &                          r-Y$_3$ & H-J$_3$ & H-W2 & i-u$_3$ & K-J$_3$ \\\hline
         \textbf{best\_star\_atts} & J$_3$-W1 &  Y$_3$-W1 &  J$_3$-W2 &  H-W2 &  K-W2 
         &                          W1-W2 &  r-z$_3$ &  r-i & r$_3$-i$_3$ &  K$_{HLR}$ \\\hline
         \textbf{best\_gal\_atts} & g-J &  Y-W1 &  J$_3$-W1 &  Y$_3$-W1 &  J$_3$-W2 
         &                          H-W2 &  K-W2 &  W1-W2 & r-z$_3$ &  r-i \\
         &                          r$_3$-i$_3$ &  K$_{HLR}$ & & &\\\hline
         \textbf{best\_qso\_atts} & H-W2 &  K-W2 &  W1-W2 &  g-J &  i-W2 
         &                          i-W1 &  r-H &  g$_3$-i$_3$ & Y$_{HLR}$ &\\
         \hline
    \end{tabular}
    }
    \caption{Attribute lists of binary model setups in Tables \ref{tab:finalmodelsetups_nohlr} and \ref{tab:finalmodelsetups}. The attributes in italics are the colours that are part of the attributes that are used as part of the final best model setup when just colour data are used as attributes. The attributes in bold are the colours and HLR values or ratios that are part of the attributes that are used as part of the final best model setup for when HLR data are used in addition to the colour data as attributes. }
    \label{tab:appendix_colours_conversion}
\end{table*}

\begin{table*}
    \centering
    \scalebox{1}{
    \begin{tabular}{cc|cccc:cccc}\hline
     &Model     &  Attributes & PCA & min\_cluster\_size & F1 & ACC & P & R & F \\\hline 
     &STAR & \textit{best\_star\_colours}             & 3 & 33 & 98.64 & 99.57 & 99.51 & 97.79 & 0.09 \\ 
     \textsc{hdbscan}&GAL  & \textit{best\_gal\_colours}            & 3 & 59 & 98.65 & 97.96 & 98.25 & 99.06 & 5.43 \\  
     48,686 data points&QSO  & \textit{best\_qso\_colours}             & 3 & 49 & 91.07 & 98.51 & 94.44 & 87.93 & 0.49 \\ \hline 
    \end{tabular}
    }
    \caption{Pre-consolidation colour-only classification setup and performance. The full list of colours from the Attributes column are shown in Table \ref{tab:appendix_colours_conversion}.}
    \label{tab:finalmodelsetups_nohlr}
\end{table*}

\begin{table*}
    \centering
    \scalebox{1}{
    \begin{tabular}{cc|cccc:cccc}\hline
     &Model     &  Attributes & PCA & min\_cluster\_size & F1 & ACC & P & R & F \\\hline 
     \textsc{hdbscan}&STAR & best\_star\_colours\_comp            & 3 & 61 & 98.84 & 99.61 & 99.52 & 98.17 & 0.1 \\ 
     43,348 data points&GAL  & best\_gal\_colours\_comp     & 3 & 62 & 98.62 & 97.97 & 98.29 & 98.96 & 4.82 \\ 
     (no HLR data used)&QSO  & best\_qso\_colours\_comp  &  3 & 56 & 92.28 & 98.61 & 94.14 & 90.49 & 0.57\\ \hline 
     \textsc{hdbscan}&STAR & \textbf{best\_star\_atts}            & 3 & 78 & 98.92 & 99.63 & 99.48 & 98.36 & 0.11 \\ 
     
     
     43,348 data points&GAL  & \textbf{best\_gal\_atts}     & 3 & 36 & 98.83 & 98.28 & 98.53 & 99.14 & 4.13 \\ 
     HLR data used&QSO  & \textbf{best\_qso\_atts}  &  3 & 34 & 93.13 & 98.77 & 95.12 & 91.22 & 0.47 \\ \hline 
    \end{tabular}
    }
    \caption{Same as Table \ref{tab:finalmodelsetups_nohlr} for colour and HLR input attributes. Sources with missing HLR have been removed from this sample.
        }
    \label{tab:finalmodelsetups}
\end{table*}

\begin{table}
    \centering
    \scalebox{0.75}{
    \begin{tabular}{cc|ccccccc}\hline
consolidation  & \multirow{2}{*}{class}  &  \multirow{2}{*}{F1} & \multirow{2}{*}{ACC} & \multirow{2}{*}{P} & \multirow{2}{*}{R} & \multirow{2}{*}{F} & \multirow{2}{*}{AUC} & \multirow{2}{*}{Nsources} \\ 
          method  &  &   &  &  &  &  &  \\ \hline 
          &STAR  &  98.64 & 99.57 & 99.51 & 97.78 & 0.09 & 0.988 & 7,596 \\
        optimal &GAL   & 98.7 & 98.04 & 98.43 & 98.98 & 4.86 & 0.971 & 36,967   \\
        &QSO   & 91.07 & 98.51 & 94.44 & 87.93 & 0.49  & 0.937 & 3,903 \\ \hline
        &STAR  & 98.45 & 99.51 & 99.64 & 97.28 & 0.07 & 0.986 &  7,548\\
        alternative &GAL   & 98.7 & 98.04 & 98.43 & 98.98 & 4.86 & 0.971 & 36,967 \\
        &QSO   & 90.91 & 98.5 & 94.94 & 87.21 & 0.44 & 0.934 &  3,851 \\ \hline
        highest-&STAR  &  98.62 & 99.56 & 99.46 & 97.79 & 0.1 & 0.988 & 7,601  \\
        probability &GAL   & 98.72 & 98.06 & 98.42 & 99.02 & 4.91 & 0.971 & 36,990   \\
        &QSO   & 91.11 & 98.52 & 94.33 & 88.1 & 0.5  & 0.938 & 3,915 \\ \hline
    \end{tabular}
    }
    \caption{Post-consolidation colour-only performance. Comparison of the optimal and alternative consolidation methods, defined in \S \ref{subsection.hdbscan}. The `optimal', `alternative', and `highest-probability' consolidation methods give 220, 320 and 180 post-consolidation outliers respectively.}
    \label{tab:consolidation_method_comp}
\end{table}

\begin{table}
    \centering
    \scalebox{0.75}{
    \begin{tabular}{cc|ccccccc}\hline
         consolidation  & \multirow{2}{*}{class}  &  \multirow{2}{*}{F1} & \multirow{2}{*}{ACC} & \multirow{2}{*}{P} & \multirow{2}{*}{R} & \multirow{2}{*}{F} & \multirow{2}{*}{AUC} & \multirow{2}{*}{Nsources} \\ 
          method  &  &   &  &  &  &  &  \\ \hline 
         &STAR  &  98.9 & 99.62 & 99.5 & 98.31 & 0.1 & 0.991 & 7,359  \\
        optimal &GAL   & 98.9 & 98.39 & 98.71 & 99.1 & 3.61 & 0.983 & 32,051   \\
        &QSO   & 93.13 & 98.77 & 95.12 & 91.22 & 0.47  & 0.925 & 3,811 \\ \hline
        &STAR  & 98.83 & 99.6 & 99.54 & 98.13 & 0.09 & 0.990 &  7,343\\
        alternative  &GAL   & 98.9 & 98.39 & 98.71 & 99.1 & 3.61 & 0.983 & 32,051 \\
        &QSO   & 92.7 & 98.7 & 95.49 & 90.06 & 0.43 & 0.920 &  3,748 \\ \hline
        highest-&STAR  &  98.94 & 99.64 & 99.5 & 98.39 & 0.1 & 0.991 & 7,365  \\
        probability &GAL   & 98.92 & 98.41 & 98.7 & 99.14 & 3.64 & 0.983 & 32,067   \\
        &QSO   & 93.01 & 98.74 & 94.87 & 91.22 & 0.5  & 0.927 & 3,821 \\ \hline
   \end{tabular}
    }
    \caption{Same as Table \ref{tab:consolidation_method_comp} for colour and HLR input attributes. The `optimal', `alternative', and `highest-probability' consolidation methods give 127, 206 and 95 post-consolidation outliers respectively.}
    \label{tab:consolidation_method_comp_hlrs}
\end{table}

\subsection{Classification performance}
 For context, a very bad classification scenario can be assessed if we assume all of the data points to belong to one cluster, i.e. all points classified as STAR, or GAL, or QSO. In this extreme scenario, the recall and fall-out would be 100\% (see definitions in \S \ref{section.metrics}), the accuracy and precision would be equal to each other, being 15.88\% for stars, 75.71\% for galaxies and 8.61\% for QSO. Respectively, the F1 score would be 27.41\% for stars, 86.05\% for galaxies, and 15.85\% for QSO.
 
 \subsubsection{Colour data as attributes}
We present our final model setup and performance metrics in Table \ref{tab:finalmodelsetups_nohlr} for the case when we use just colour data as attributes. The optimal setups are shown for the STAR, GAL and QSO binary classifier models, specifically detailing the optimal \texttt{min\_cluster\_size} value and input attributes, and the number of dimensions to which those attributes are then reduced using PCA. The colours in each of these top attributes are shown in Table \ref{tab:appendix_colours_conversion}. 

Using the `optimal' consolidation method, the F1 scores that we obtain when using just colour attributes and the full data set are $\mathrm{F1_{STAR}=98.64}$, $\mathrm{F1_{GAL}=98.7}$ and $\mathrm{F1_{QSO}=91.07}$ (Table \ref{tab:consolidation_method_comp}). On the other hand, the `alternative' consolidation method that favours precision over completeness gives $\mathrm{F1_{STAR}=98.45}$, $\mathrm{F1_{GAL}=98.7}$ and $\mathrm{F1_{QSO}=90.91}$.
The `optimal' method produces 220 post-consolidation outliers, and the `alternative' method produces 320 post-consolidation outliers. The `highest-probability' method gives $\mathrm{F1_{STAR}=98.62}$, $\mathrm{F1_{GAL}=98.72}$ and $\mathrm{F1_{QSO}=91.11}$ and 180 outliers. In addition, we present the AUC scores for all consolidation methods in Table \ref{tab:consolidation_method_comp}. We find similar AUC scores for all consolidation methods, between 0.986 and 0.988 for stars, 0.971 for galaxies and between 0.934 and 0.938 for QSOs.

 \subsubsection{Colour and HLR data as attributes}

In Table \ref{tab:finalmodelsetups} we present the final model setup and performance metrics when using the Y -- K HLR data in addition to the colour data as the attributes. Similarly to the colour-only case we show the optimal setups for the STAR, GAL and QSO binary classifier models including the optimal \texttt{min\_cluster\_size} values, input attributes, and the number of PCA components. The colours and HLR data in each of these top attributes are shown in Table \ref{tab:appendix_colours_conversion}. 

After the consolidation step using the `optimal' consolidation method, the F1 scores that we obtain when using the Y -- K HLR data in addition to the colour data, and a reduced data set with no missing Y -- K HLR values (43,348 sources), are $\mathrm{F1_{STAR}=98.9}$, $\mathrm{F1_{GAL}=98.9}$ and $\mathrm{F1_{QSO}=93.13}$ with 127 post-consolidation outliers (Table \ref{tab:consolidation_method_comp_hlrs}). The `alternative' consolidation method shows $\mathrm{F1_{STAR}=98.83}$, $\mathrm{F1_{GAL}=98.9}$ and $\mathrm{F1_{QSO}=92.7}$ with 206 outliers. The `highest-probability' method gives $\mathrm{F1_{STAR}=98.94}$, $\mathrm{F1_{GAL}=98.92}$ and $\mathrm{F1_{QSO}=93.01}$ and 95 outliers. In addition, we present the AUC scores for all consolidation methods in Table \ref{tab:consolidation_method_comp_hlrs}. We find the AUC scores to be between 0.990 and 0.991 for stars, 0.983 for galaxies and between 0.920 and 0.927 for QSOs.

\begin{figure*}
\centering
\begin{tabular}{ccc}
\includegraphics[width=0.5\linewidth]{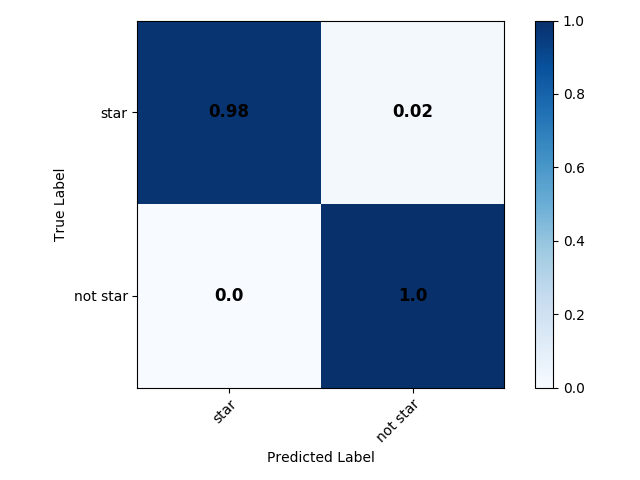} & \includegraphics[width=0.5\linewidth]{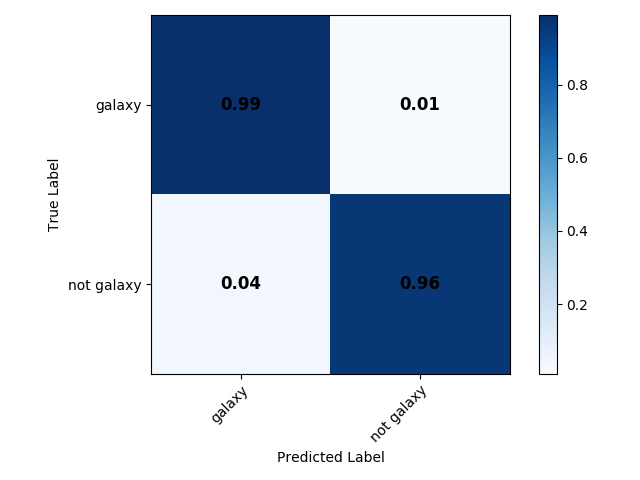} \\
(a) & (b)  \\
\includegraphics[width=0.5\linewidth]{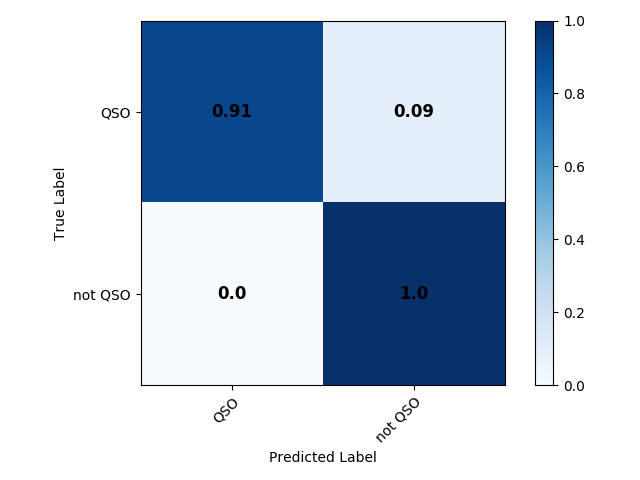} & \includegraphics[width=0.5\linewidth]{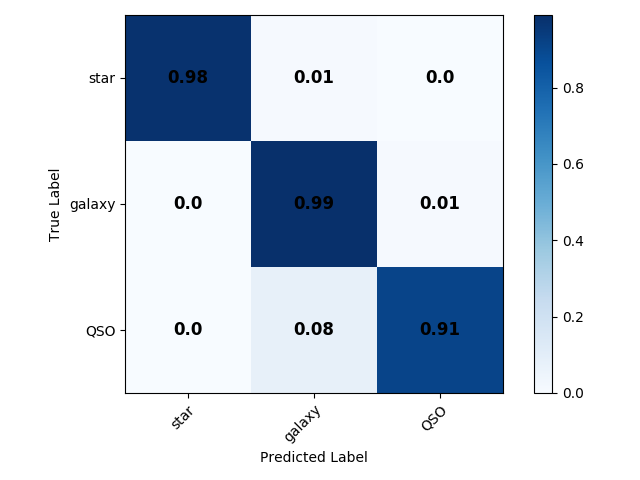} \\
 (c) & (d) \\
\end{tabular}
\caption{ Confusion matrices of post-consolidation predicted labels using colour and HLR input attributes. (a) STAR (b) GAL, (c) QSO and (d) the combined output. We note that for panel (d), the predicted label 'post-consolidation outlier' is not shown, hence the star and QSO rows' values do not sum to 1.}
\label{fig:confusion_plots_hlr}
\end{figure*}

The confusion matrices for each of the binary classifiers are shown in Figure \ref{fig:confusion_plots_hlr}. In panel (d) it is clear that the reason QSOs have poorer F1 scores than the galaxy and star class because about $\sim$8\% are classified as galaxies, even with the inclusion of the Y -- K HLR data. This is due to the overlap between the QSO and galaxy clusters in the colour space as discussed in \S \ref{section.dataset_speclabels} (see Fig. \ref{fig:hclass}), and more specifically the broad definition of QSO which can include both blue, unobscured objects as well as broad-absorption line systems.

\subsection{Optimal Attribute Lists, PCA components and Feature Importances} \label{section.results_att_lists_comments}

\begin{figure*}[ht]
\centering
\begin{tabular}{ccc}
\includegraphics[width=0.325\linewidth]{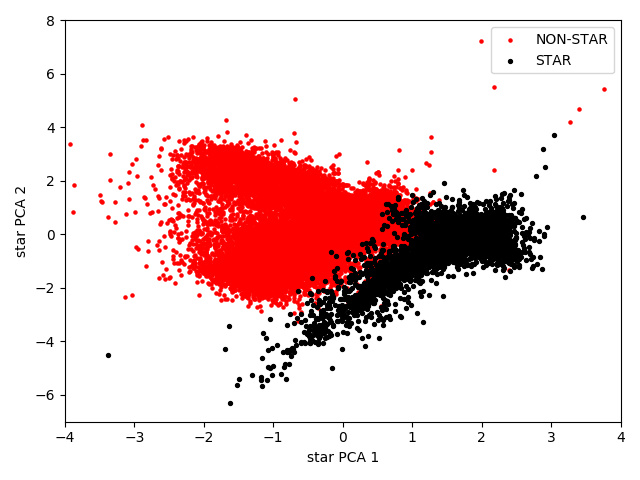} &  \includegraphics[width=0.325\linewidth]{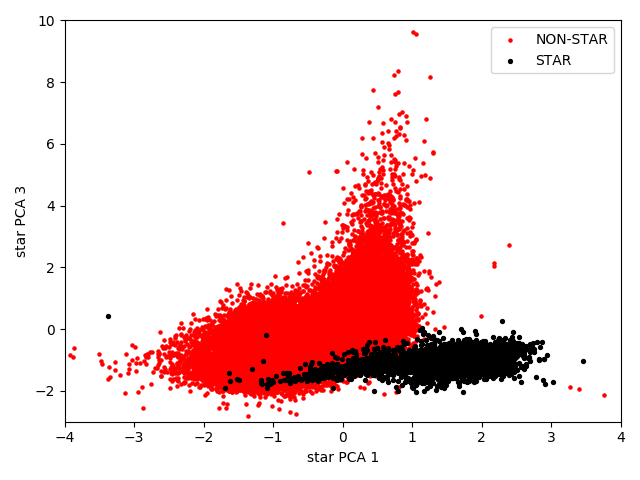} & \includegraphics[width=0.325\linewidth]{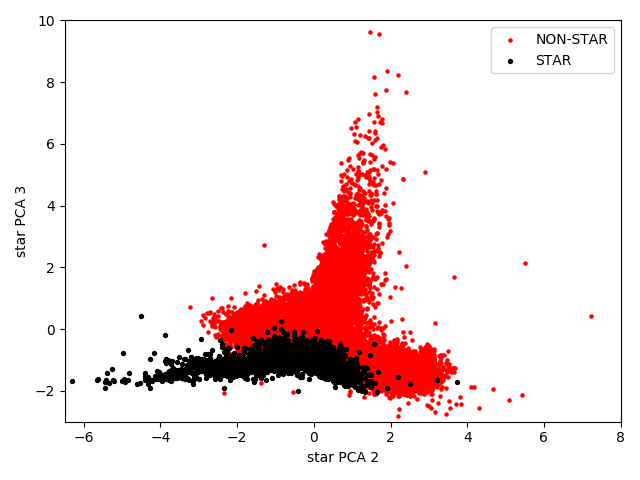} \\
\includegraphics[width=0.325\linewidth]{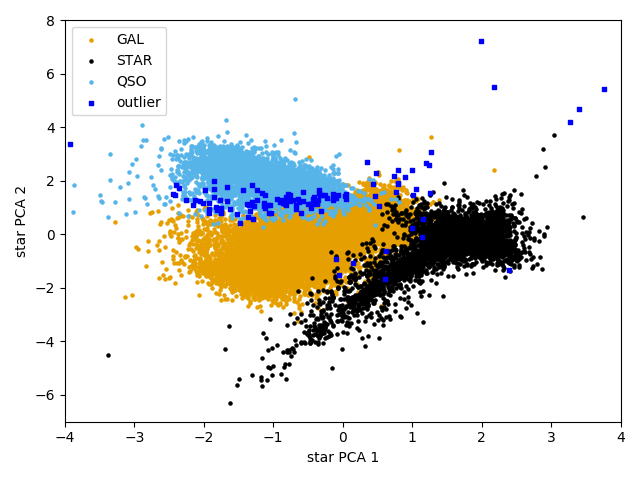}  &
\includegraphics[width=0.325\linewidth]{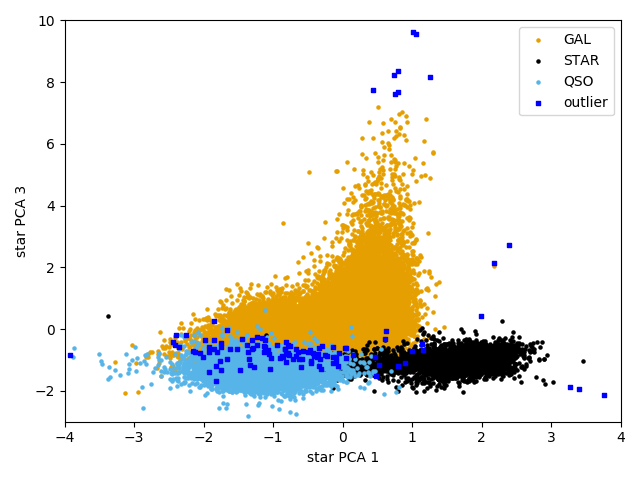}  &
\includegraphics[width=0.325\linewidth]{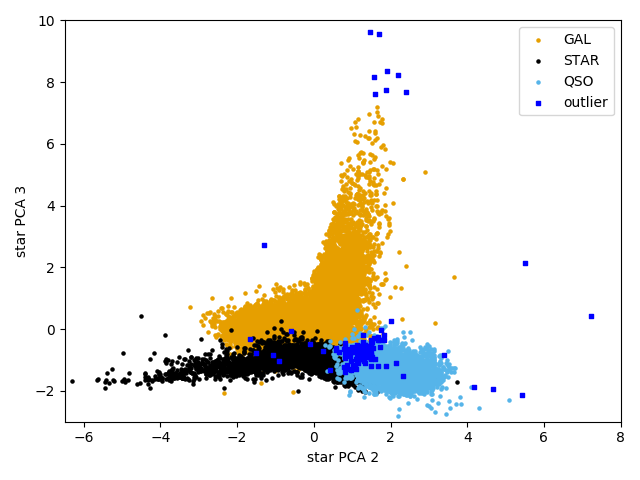}\\ 
\end{tabular}
\caption{PCA components for the STAR classifier using colour and HLR data as input attributes. The left column 
shows the labels that we obtain from the \textsc{hdbscan} STAR classifier. The right column shows the post-consolidation labels including outliers.}
\label{fig:star_PCA}
\end{figure*}

\begin{figure*}[ht]
\centering
\begin{tabular}{ccc}
\includegraphics[width=0.325\linewidth]{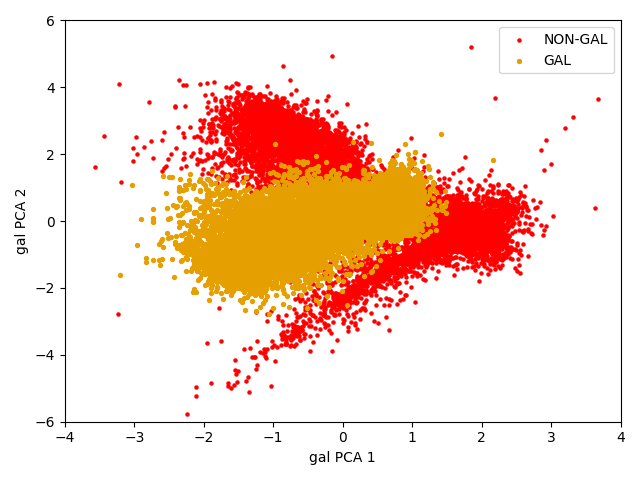} &  \includegraphics[width=0.325\linewidth]{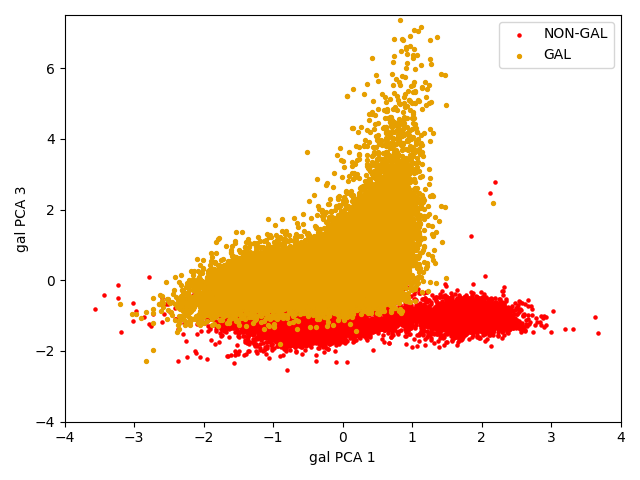} & \includegraphics[width=0.325\linewidth]{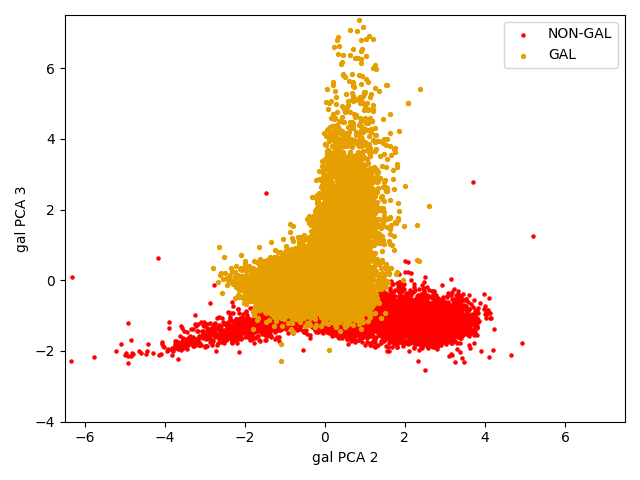} \\
\includegraphics[width=0.325\linewidth]{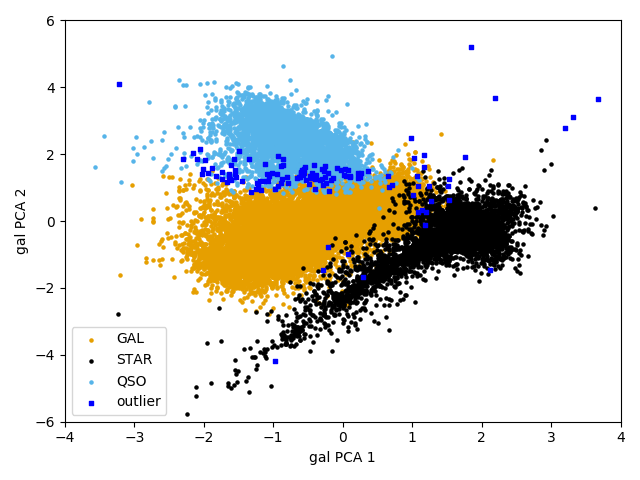}  &
\includegraphics[width=0.325\linewidth]{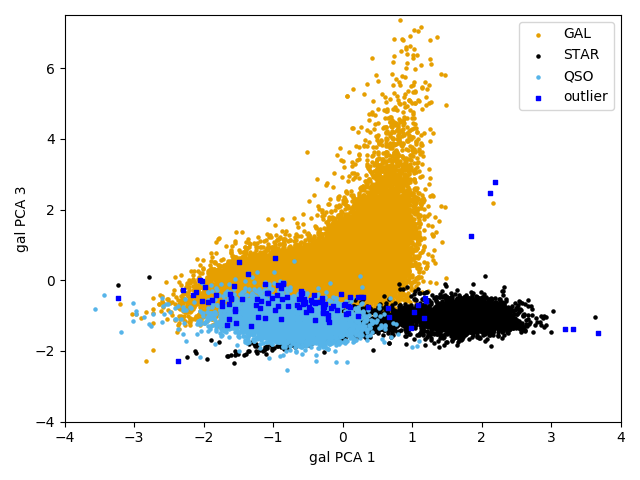}  &
\includegraphics[width=0.325\linewidth]{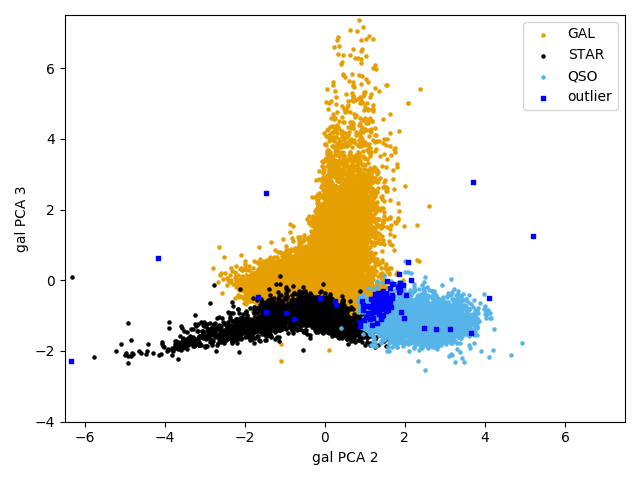}\\ 
\end{tabular}
\caption{Same as Figure \ref{fig:star_PCA} for the GAL classifier.
}
\label{fig:gal_PCA}
\end{figure*}

\begin{figure*}[ht]
\centering
\begin{tabular}{ccc}
\includegraphics[width=0.325\linewidth]{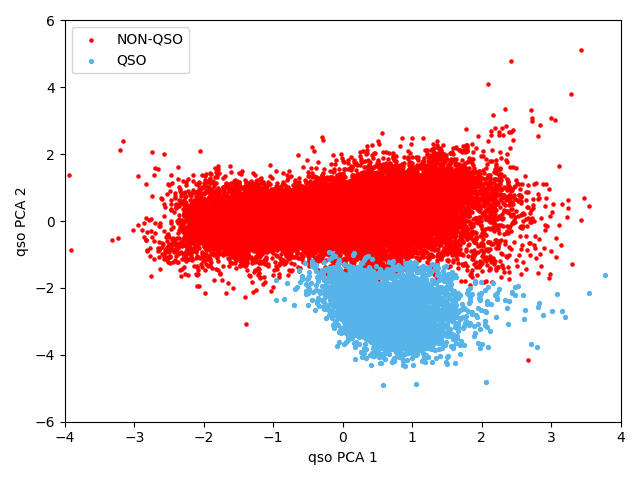} &  \includegraphics[width=0.325\linewidth]{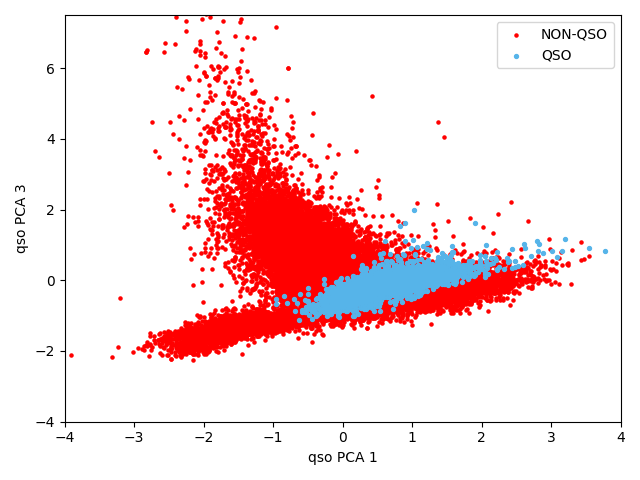} & \includegraphics[width=0.325\linewidth]{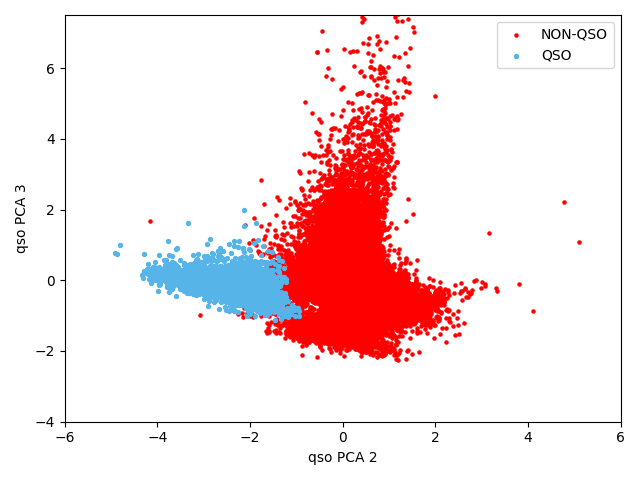} \\
\includegraphics[width=0.325\linewidth]{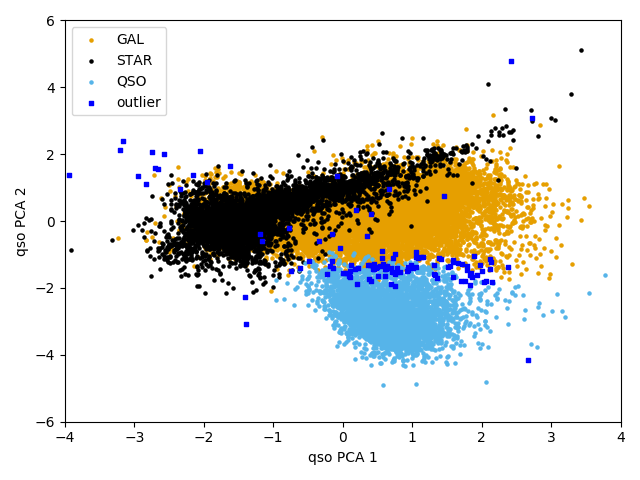}  &
\includegraphics[width=0.325\linewidth]{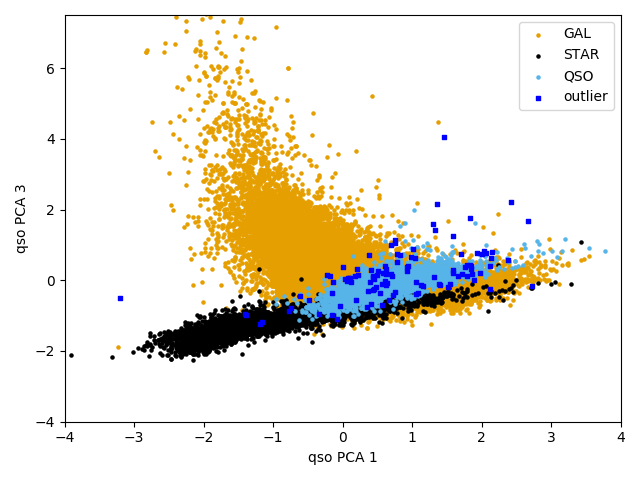}  &
\includegraphics[width=0.325\linewidth]{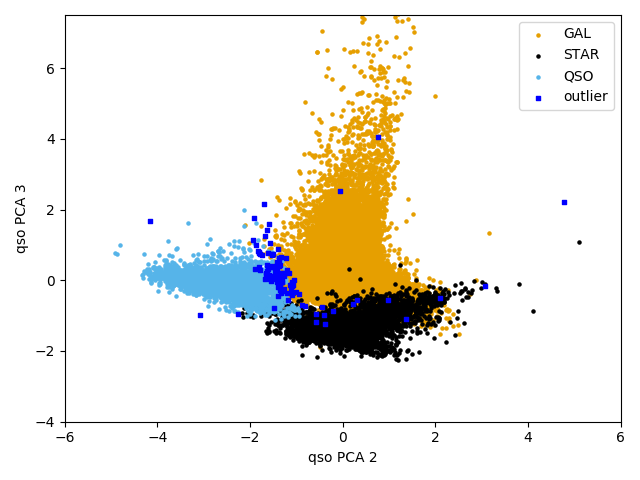}\\ 
\end{tabular}
\caption{Same as Figure \ref{fig:star_PCA} for the QSO classifier.
}
\label{fig:qso_PCA}
\end{figure*}

Using all of the 190 colours that are available from the u -- W2 photometric bands as input attributes to \textsc{hdbscan}, we can obtain an F1 scores of $\mathrm{F1_{STAR}=87.92}$, $\mathrm{F1_{GAL}=86.05}$ and $\mathrm{F1_{QSO}=45.13}$ when running the \textsc{hdbscan} in binary mode. Even with the use of PCA to reduce this large number of attributes to 3 dimensions, we only obtain F1 scores of $\mathrm{F1_{STAR}=94.42}$, $\mathrm{F1_{GAL}=93.5}$ and $\mathrm{F1_{QSO}=54.76}$. Thus, successful attribute selection is critical to optimize the performance of our model. We find that for our optimal setups for our \textsc{hdbscan} final classifier, we generally reduce $\sim$ 10 - 30 attributes to 3 attributes using PCA for input into \textsc{hdbscan} (see Tables \ref{tab:finalmodelsetups_nohlr}, \ref{tab:finalmodelsetups} and \ref{tab:appendix_colours_conversion}). 

In Figures \ref{fig:star_PCA}, \ref{fig:gal_PCA}, \ref{fig:qso_PCA} we show the PCA components of the binary STAR, GAL and QSO classifiers respectively, when using both colour and Y -- K HLR data in our attributes for the \textsc{hdbscan} model setup. It is clear the each different component combination is providing some extra information for \textsc{hdbscan} to be able to separate different object types. The left column in these figures shows the binary class assignment, while the right-hand side shows the post-consolidation class assigned in each respective PCA space projection, including the post-consolidation outliers. 

In Table \ref{tab:appendix_colours_conversion}, we show the list of attributes for the optimal model setups for each of the STAR, GAL and QSO binary classifiers. The best attribute list names are shown in italics and bold respectively when using just the colour data as attributes, and when HLR data are included. It can be seen that in all of these top attributes, the WISE bands (i.e. W1 and W2) appear frequently, suggesting that the inclusion of these WISE bands is critical in order to classify STAR, GAL and QSO to a high level of precision and accuracy. Interestingly, the RF's list of top 10 attributes for the binary QSO classifier (shown in Table \ref{tab:RFtopatts}) setup does not include any of the WISE bands in the colours, whereas the optimal attribute list for the \textsc{hdbscan} QSO binary classifier model includes many colours that use the WISE band photometry (see Table \ref{tab:appendix_colours_conversion}). This confirms that different attributes are more important for \textsc{hdbscan} compared to RF, which validates our choice to use the top attributes from classifiers A, B and C from FP18 when creating the potential attribute lists for \textsc{hdbscan} which we tested as input to our \textsc{hdbscan} binary classifiers. Even though classifiers B and C were used for different purposes as to what we are doing in this work (as explained in \S \ref{subsubsection.ABC_atts}), their attributes were still useful for \textsc{hdbscan}.

It is also clear that combining information from the different apertures is important - a mix of 3 $''$ aperture magnitudes and total magnitudes carries information particularly for stars and QSOs, providing information on the morphology. This was also the case for the RF top attributes output, in Table \ref{tab:RFtopatts}.

\subsection{Impact of HLR information} \label{subsection.adding_HLR_impact}
When removing missing Y -- K HLR data, the number of faint sources decreases hence the immediate increase in classification performance compared to using only colours. In Table \ref{tab:finalmodelsetups} we show the best gridsearch result (as described in \S \ref{section:model_construction}) run over all attribute sets using just colour attributes using the dataset where the data points with missing Y -- K  HLR are removed, so as to be able to make a fair comparison. The STAR classification performance increases slightly (from 98.84 to 98.92), the GAL performance also increases slightly (from 98.62 to 98.83), but it is the QSO performance that increases the most (from 92.28 to 93.13).

Additionally, in Table \ref{tab:appendix_colours_conversion} we see that the optimal attribute lists (for the case where we use both colour data and HLR data as attributes) for each of the STAR, GAL and QSO binary \textsc{hdbscan} classifiers in our final model include the $K_{HLR}$ attribute and also the $Y_{HLR}$ attribute. However, if only one HLR value is available (for example the $K_{HLR}$ data), good performance can still be achieved. In the case of using the optimal attributes that include just the $K_{HLR}$ value from the HLR data for the QSO classifier, the F1 scores changes from 93.13 (achieved using an attribute list that includes the $K_{HLR}$ attribute) to 93.0. There is also a benefit of using just one HLR value, especially clear in the case of our dataset; if there are missing (or unrealistically high - HLR > 20 $''$) Y -- K HLR values, then it is only necessary to drop those objects that have missing e.g. $J_{HLR}$ values. In the case of our dataset, we could then use 47,652 points from our data set if just removing those points that have missing $K_{HLR}$ values, compared with 43,348 when we remove all of data points from our data set that have missing Y -- K HLR values.

\subsection{Dependence of performance on photometric depth and redshift}
\begin{figure}
    \centering
    \includegraphics[width=80mm]{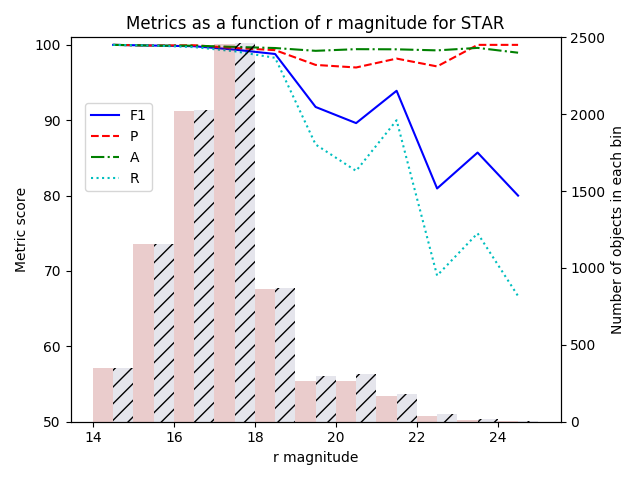} \\ 
    \includegraphics[width=80mm]{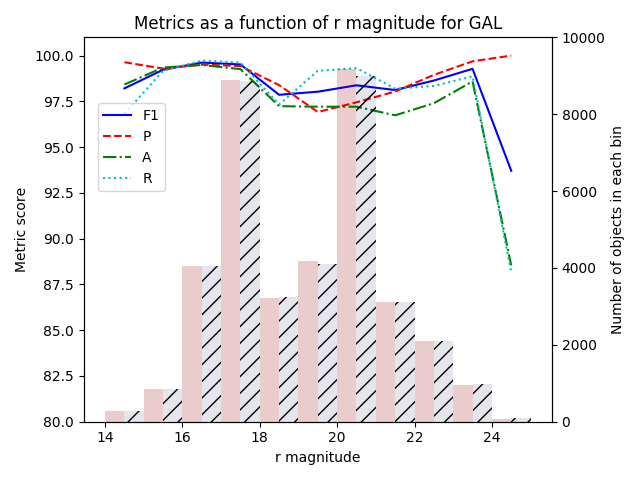} \\ 
    \includegraphics[width=80mm]{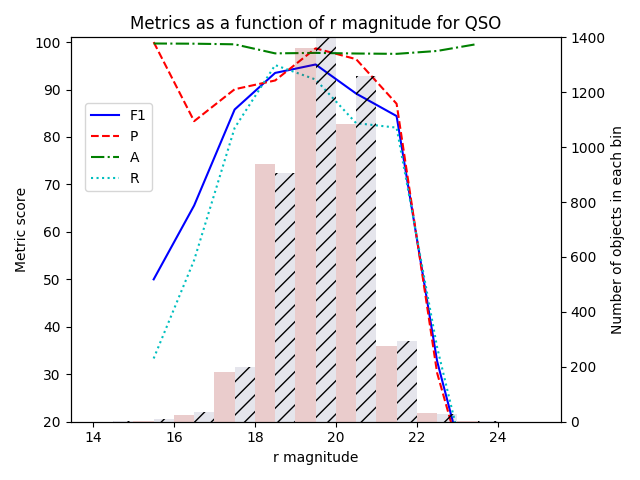} \\ 
    \caption{Post-consolidation colour-only classification performance as a function of magnitude.
    The number of objects in each magnitude bin are shown as red bars for the final output of the \textsc{hdbscan} model, and the hatched grey bars are for the spectroscopic labels. 
    }
    \label{fig:perf_as_fn_of_mag_nohlr}
\end{figure}

\begin{figure}
    \centering
    \includegraphics[width=80mm]{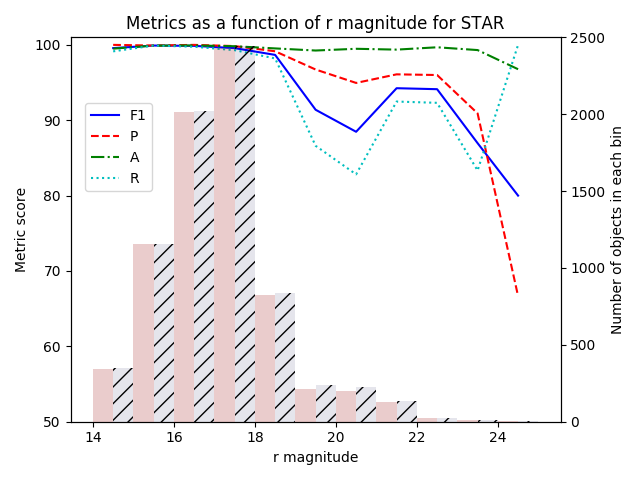} \\ 
    \includegraphics[width=80mm]{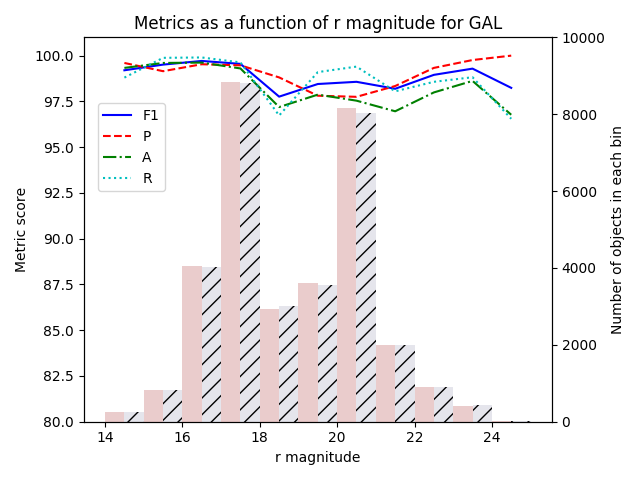} \\ 
    \includegraphics[width=80mm]{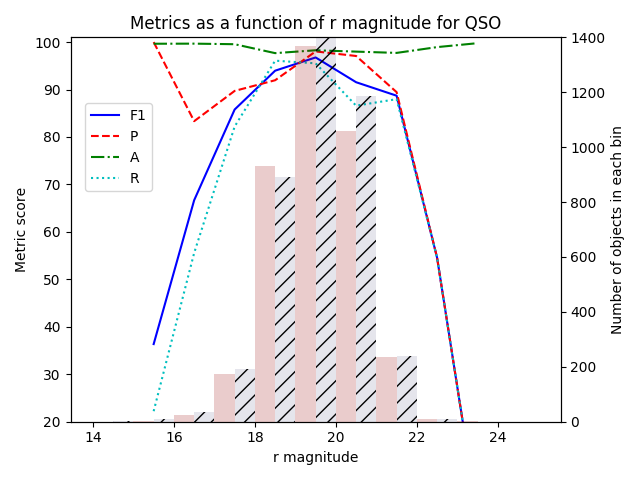} \\ 
    \caption{Same as Figure \ref{fig:perf_as_fn_of_mag_nohlr} for colour and HLR classification performance.}
    \label{fig:perf_as_fn_of_mag}
\end{figure}

Figure \ref{fig:perf_as_fn_of_mag_nohlr} shows the performance of the colour-only \textsc{hdbscan} model as  function of the $r$ magnitude\footnote{The double peaked magnitude distribution in the galaxy histograms is due to the combination of shallow (SDSS DR12) and deep surveys (CFHTLS, KiDS) surveys in the CPz sample of FP18.} for each of the object classes. Figure \ref{fig:perf_as_fn_of_mag} shows the same distributions for the colour and HLR classifiers. The number of points in each magnitude bin is also shown on the right hand axes.

In general, all classifiers achieve very high accuracy across all magnitudes (dot-dashed green line). The star classifier shows excellent performance at least up to magnitude r=19. At fainter magnitudes, the accuracy remains very high, however the completeness seems to decrease. At the same time those bins suffer from very low number of objects and this performance at those magnitudes should be re-examined with deeper data.

\begin{figure*}
    \centering
    \begin{tabular}{cc}
    \includegraphics[width=\columnwidth]{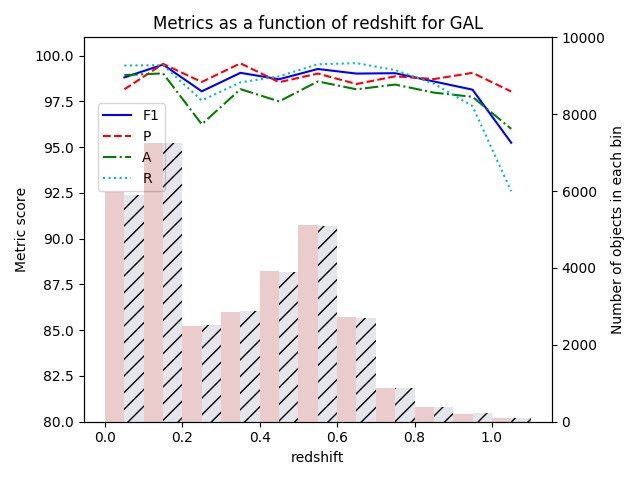} &    \includegraphics[width=\columnwidth]{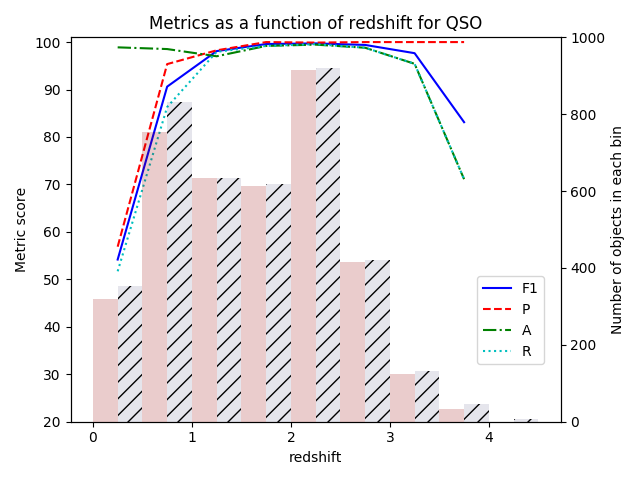} \\
    \end{tabular}
    \caption{Post-consolidation colour and HLR classification as a function of redshift is shown for galaxies (left panel) and QSOs (right panel). The number of objects in each magnitude bin are shown as red bars for the final output of the \textsc{hdbscan} model (after the consolidation step), and the hatched grey bars are for the spectroscopic labels.    }
    \label{fig:perf_as_fn_of_redshift}
\end{figure*}

The galaxy classifier shows overall great performance and stability, apart from the faintest magnitude bin (r>24) which suffers from low number statistics. The stable performance is retained also as a function of redshift as seen in the left-hand side of Fig. \ref{fig:perf_as_fn_of_redshift}.

The QSO classifier shows excellent accuracy performance across all magnitudes, in line with the two other classifiers. The difference appears in the completeness (recall curve, dotted cyan line). As mentioned in \S \ref{section:discussion-labels}, the definition of QSO based on the spectroscopic sample contains a variety of objects, including unobscured quasars, AGN, and BALs. The missed objects that reduce the completeness of this classifier are the 10\% of the labelled data that overlap with the galaxy cloud. Given that \textsc{hdbscan} performs unsupervised clustering, this population is not expected to be identified as QSO. The impact of this is also seen in the right-hand side of Fig. \ref{fig:perf_as_fn_of_redshift}. At the low redshift range, AGNs with colours not dominated by the quasar have been spectroscopically classified as QSO. These sources typically lie in the transition region between the QSO and GAL clouds. However, at the high redshift regime the QSO sources are impacted by the presence of BALs and intergalactic absorption hence extinguishing the blue part of the spectrum and moving these objects to galaxy-like appearance in the optical magnitudes. 

\begin{figure*}
\centering
\begin{tabular}{cc}
\includegraphics[width=0.5\linewidth]{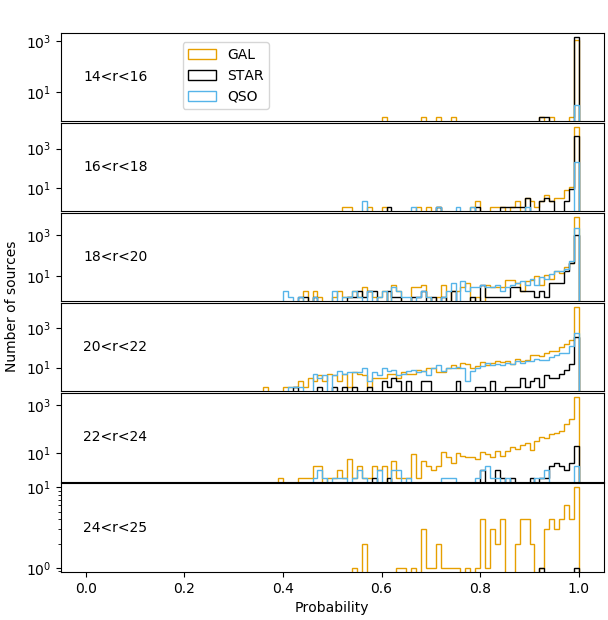}&
\includegraphics[width=0.5\linewidth]{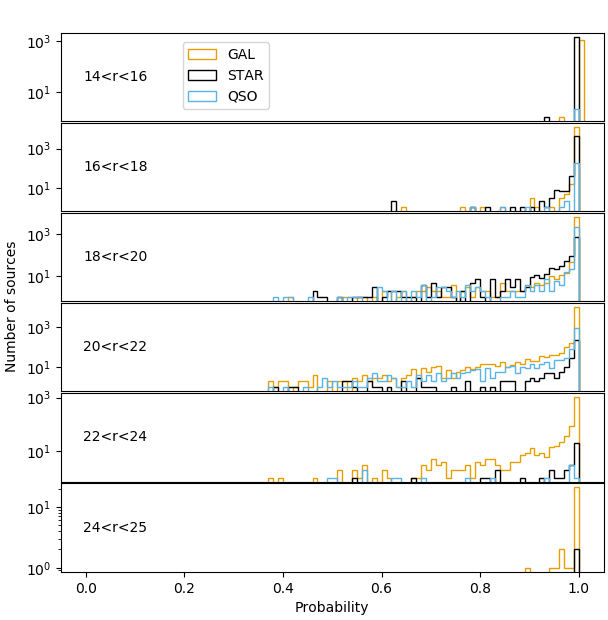}\\
\end{tabular}
\caption{The probability distributions for the CPz sample for six different r magnitude bins are shown. The classification probability of each object is obtained using the `highest-probability' method (see \S \ref{subsection.consolidation}). The left panel is for when just colours were used as attributes, and the right panel is for when both colour and HLR information were used as attributes. The objects are in classes according to their `highest-probability' final label. }
\label{figure.av_prob_of_classification_for_highest_prob_consolidation_CPz_colours_and_colours_and_HLR}
\end{figure*}

We also plot the probability distributions given by the ‘highest-probability’ consolidation method (see \S \ref{subsection.consolidation}) in different r magnitude bins in Figure \ref{figure.av_prob_of_classification_for_highest_prob_consolidation_CPz_colours_and_colours_and_HLR}. The left panel shows the colour-only CPz sample while the right panel shows the colour and HLR CPz performance.

\section{Application to KiDS, VIKING and ALLWISE (KiDS-VW)} \label{section.app_to_KiDS-VW}

\begin{table}
    \centering
    \scalebox{0.75}{
    \begin{tabular}{cc|ccccccc}\hline
consolidation  & \multirow{2}{*}{class}  &  \multirow{2}{*}{F1} & \multirow{2}{*}{ACC} & \multirow{2}{*}{P} & \multirow{2}{*}{R} & \multirow{2}{*}{F} & \multirow{2}{*}{AUC} & \multirow{2}{*}{Nsources} \\ 
          method  &  &   &  &  &  &  &  \\ \hline 
        &STAR  &  96.04 & 98.58 & 98.08 & 94.08 & 0.41 & 0.979 & 5,352  \\
        optimal &GAL   & 97.42 & 96.55 & 96.05 & 98.82 & 7.82 & 0.963 & 20,659   \\
        &QSO   & 90.47 & 97.17 & 97.57 & 84.33 & 0.4  & 0.925 & 4,196 \\ \hline
        &STAR  &  95.82 & 98.51 & 98.1 & 93.65 & 0.41 & 0.978 & 5,326  \\
        alternative &GAL   & 97.42 & 96.55 & 96.05 & 98.82 & 7.82 & 0.963 & 20,659   \\
        &QSO   & 89.96 & 97.04 & 97.75 & 83.32 & 0.36  & 0.920 & 4,138 \\ \hline
        highest-&STAR  &  96.13 & 98.61 & 98.08 & 94.25 & 0.41 & 0.979 & 5,361  \\
        probability &GAL   & 97.37 & 96.49 & 95.98 & 98.81 & 7.97 &  0.963& 20,673   \\
        &QSO   & 90.69 & 97.23 & 97.56 & 84.72 & 0.4  & 0.927 & 4,216 \\ \hline
    \end{tabular}
    }
    \caption{Performance metrics for the KiDS-VW sample, excluding sources already in the CPz sample. The `optimal', `alternative', and `highest-probability' consolidation methods give 307, 391 and 264 post-consolidation outliers respectively.}
    \label{tab:highest_prob_label_comp_KiDS-VW}
\end{table}

\subsection{Catalogue creation}

We updated the CPz sample of FP18 by cross matching the KiDS DR4v3 catalogue\footnote{Downloaded single detection band catalogues from \url{http://kids.strw.leidenuniv.nl/DR4/format.php#cols1}}, with the VIKING DR3v3 survey\footnote{Downloaded in its entirety using the ESO Catalogue Facility \url{https://www.eso.org/qi/}}, and the ALLWISE catalogue\footnote{Downloaded from \url{https://irsa.ipac.caltech.edu/data/download/wise-allwise/}}. The overlap area among these surveys is approximately 200 $\mathrm{deg^2}$ and about $\mathrm{2.7\cdot10^6}$ sources have no missing magnitudes in any of the filters u--W2. Similarly to FP18, we are using 2.8'' aperture magnitudes and total (auto or petrosian magnitudes) and model magnitdes for WISE. A comparison between aperture and total magnitudes revealed offsets $\mathrm{\Delta m= m_{aper}-m_{auto}=zp}$. We estimated and applied a general aperture correction by taking the offset in the magnitude range 16-18 in each filter (u--K). We found for the KiDS survey an offset of $\mathrm{zp=0.12}$ is needed in each band while for VIKING, $\mathrm{zp=-0.02}$ for the $z$ band and $\mathrm{zp=-0.04}$ for the $J$, $H$, and $Ks$ bands. Hereafter, we refer to this catalogue as the KiDS-VW catalogue.

\subsection{Classification}
Applying the methodology described in \S \ref{section.classification_model} and Fig. \ref{fig:tikz_model_setup}, we classified all sources without missing magnitudes into star, galaxy and QSO. We used the \textsc{hdbscan} model setup that uses only colour information as attributes (see Table \ref{tab:finalmodelsetups_nohlr}), as no HLR data was available for the Y -- K HLR bands. We applied the \textsc{hdbscan} model trained on the CPz dataset of 48,686 data points to the 2,728,329 data points in the KiDS-VW sample. We then combined the three sets of predicted labels from each of the binary classifers in the consolidation phase (as described in \S \ref{subsection.consolidation}; we used the `optimal' consolidation method). Crucially, the normalization and scaling that was applied to the new data had to be the same that was applied on the training data, and this was also true for the PCA transformation. \footnote{Each of the three binary classifiers in the \textsc{hdbscan} model took $\sim$ 10 minutes to predict the 3 million labels, so overall the total prediction stage took $\sim$ 30 minutes. This was run on one core on an Intel(R) Xeon(R) CPU E5-2680 v4. The memory usage was dictated by the size of the input data for prediction, however this potential bottleneck can be easily mitigated by predicting smaller batches of objects.}

The results we obtained from the prediction stage on the KiDS-VW sample was as follows: 1,184,222 objects classified as STAR, 1,379,850 as GAL and 123,084 as QSO, with 41,173 post-consolidation outliers. The distribution of sources within different classes is different to that of the CPz sample due to the construction of the catalogues (e.g. higher fraction of QSOs in the CPz catalogue as sources had to have spectroscopic classifications to be included, and QSOs are more likely to be spectroscopically followed up than other objects). In Figures \ref{fig:star_PCA_newcat}, \ref{fig:gal_PCA_newcat}, \ref{fig:qso_PCA_newcat}, we show the pre-consolidation and post-consolidation labels in PCA component space for the STAR, GAL and QSO \textsc{hdbscan} binary classifiers respectively for when predicting on the KiDS-VW catalogue having trained on the CPz sample.

\subsection{Photometric redshift estimation} \label{subsection:photometric_redshifts}

\begin{figure*}
    \centering
    \begin{tabular}{cc}
    \includegraphics[width=\columnwidth]{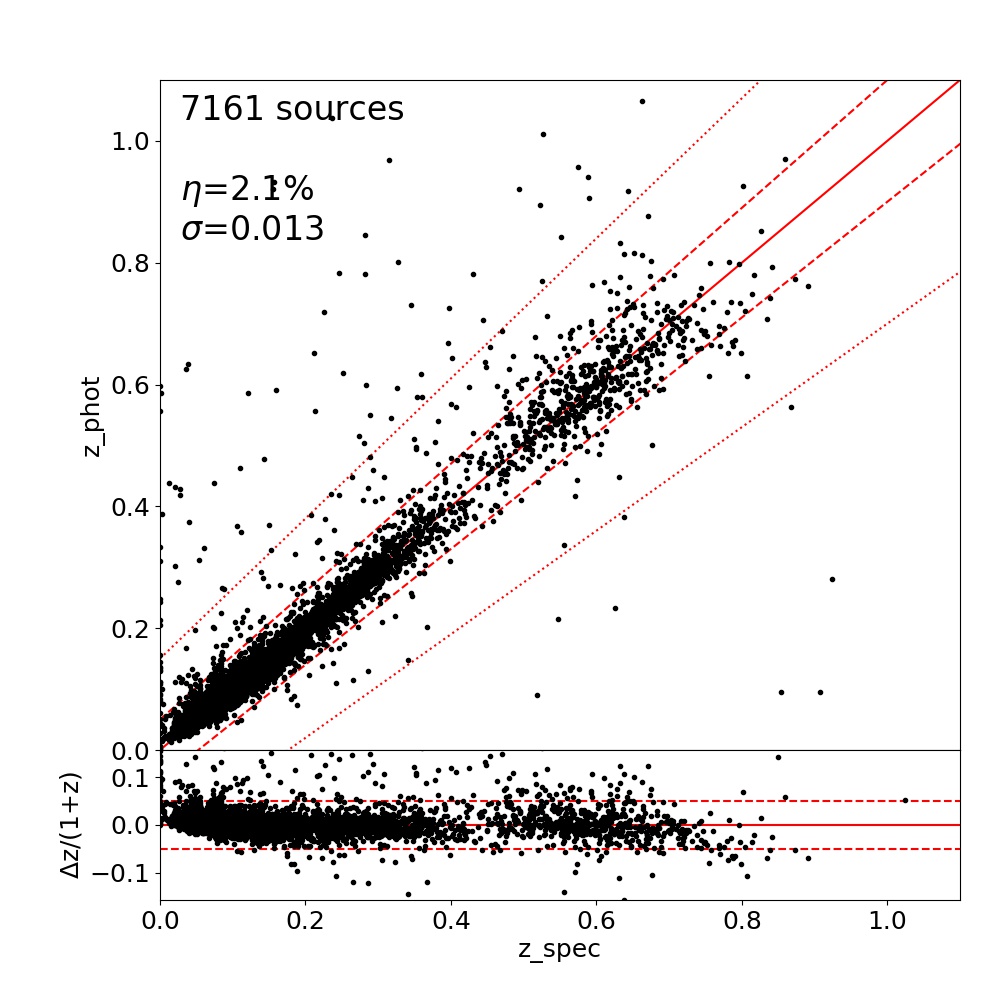} &
    \includegraphics[width=\columnwidth]{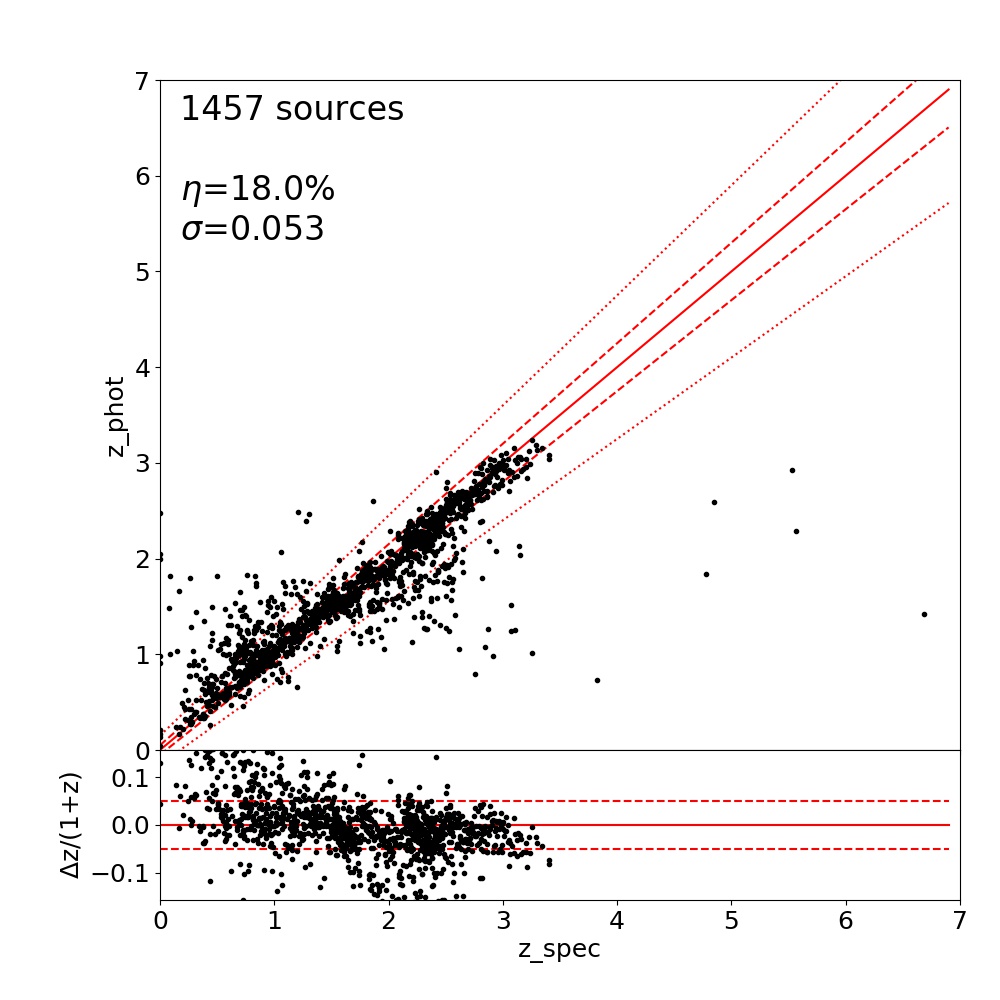} \\
    (a) & (b) \\
    \end{tabular}
    \caption{Comparison of the predicted photometric redshifts vs the spectroscopic redshifts for the test sample (20\%) of the SDSS DR14 sources in KiDS-VW. (a) Train the GAL\_PHOTOZ\_PREDICTOR and test on galaxies. (b) Train the QSO\_PHOTOZ\_PREDICTOR and test on QSOs. 
    The metric scores are also shown. For all plots and metric, we select the galaxies and QSOs according to their classification by our \textsc{hdbscan} model, using the `optimal' method for consolidation, and just the colour attributes.    }
    \label{fig:photoz_specz_comp_with_metrics}
\end{figure*}

For the photometric redshift estimation of the 2,728,329 sources in the KiDS-VW sample, we train and test a RF (the RF algorithm is explained in \S \ref{subsubsection.RF}), using the 55,383 sources in KiDS-VW that are also in SDSS DR14 and therefore have a spectroscopic redshift. We then use the trained RF to predict photometric redshifts for the rest of the sample.

\citet{Salvato2009, Salvato2011} demonstrated that photometric redshifts for X-ray AGN and QSO should be estimated with tailored templates and absolute magnitude prior selection. The classification-aided photometric redshift estimation (CPz) of FP18, generalized this approach by pre-classifying all sources according to the best photometric library setup, in the absence of X-ray information. Following the CPz approach, we trained two different RFs for this purpose in the following ways:
\begin{itemize}
    \item GAL\_PHOTOZ\_PREDICTOR - trained on galaxies. 
    \item QSO\_PHOTOZ\_PREDICTOR - trained on QSOs. 
\end{itemize}
For the methods above, we select the galaxies and QSOs according to their classification by our \textsc{hdbscan} model, using the `optimal' method for consolidation, and just the colour attributes.

To find the optimal hyperparameter setups for each of the two RFs, we split the 55,383 SDSS DR14 sources into a train, validation and test sample (60\%, 20\% and  20\% respectively). Using the train and validation samples, we iterated over the \texttt{n\_estimators} (10, 20, 40, 60, 100, 200) and \texttt{max\_depth} (3, 5, 10, 20, 50, None = no limit) hyperparameters of the RF, training on the train sample and calculating metric scores using the trained RF's predictions for the validation sample and the spectroscopic redshifts as the truth. We used two metrics (accuracy, $\sigma_{NMAD}=1.48|\frac{z_{phot}-z_{spec}}{1+z_{spec}}|$, and catastrophic outliers, $\eta$, ($N(|\frac{z_{phot}-z_{spec}}{1+z_{spec}}|)>0.15$) as well as visually inspecting each plot for any systematic issues. We found that \texttt{n\_estimators} = 200 and \texttt{max\_depth} = None provided the best overall performance for both of the RFs. 

Having found the best hyperparameter setup for our RFs, we then tested the performance on the test set. We present the metric values and plots of the predicted photometric redshift compared to the spectroscopic redshift for the test sample in Figure \ref{fig:photoz_specz_comp_with_metrics} for both of the RFs.

We also predicted the photometric redshifts for the whole KiDS-VW sample, using the trained RFs. We present these predicted photometric redshifts in the final KiDS-VW catalogue (\S \ref{section.appendix.catalogue_descriptions} in columns 65-66). Even though the RFs are optimized for certain objects, we still predict redshifts for the whole sample, and if a user of the catalogue just wants to select galaxies with redshifts calculated using the GAL\_PHOTOZ\_PREDICTOR, they can, using the information in the catalogue (i.e. the hdbscanclass columns in the final catalogue).

\section{Discussion} \label{section.discussion_and_application}
In this section we compare our best classifier to the literature, looking into more depth on the quasar identification of spectroscopic and machine-learning methods.

\subsection{Classifications}\label{section:discussion-labels}
\begin{table*}
    \centering
    \scalebox{0.9}{
        \begin{tabular}{cc|ccc}\hline
            &   & SDSS DR14        &   \\
            
            &          & STAR     & GALAXY     & QSO   \\ \hline 
            
            \STAB{\rotatebox[origin=c]{90}{HDBSCAN}}&       \STAB{\rotatebox[origin=bottom]{90}{STAR}}   &  \raisebox{-.5\height}{\includegraphics[width=0.32\textwidth]{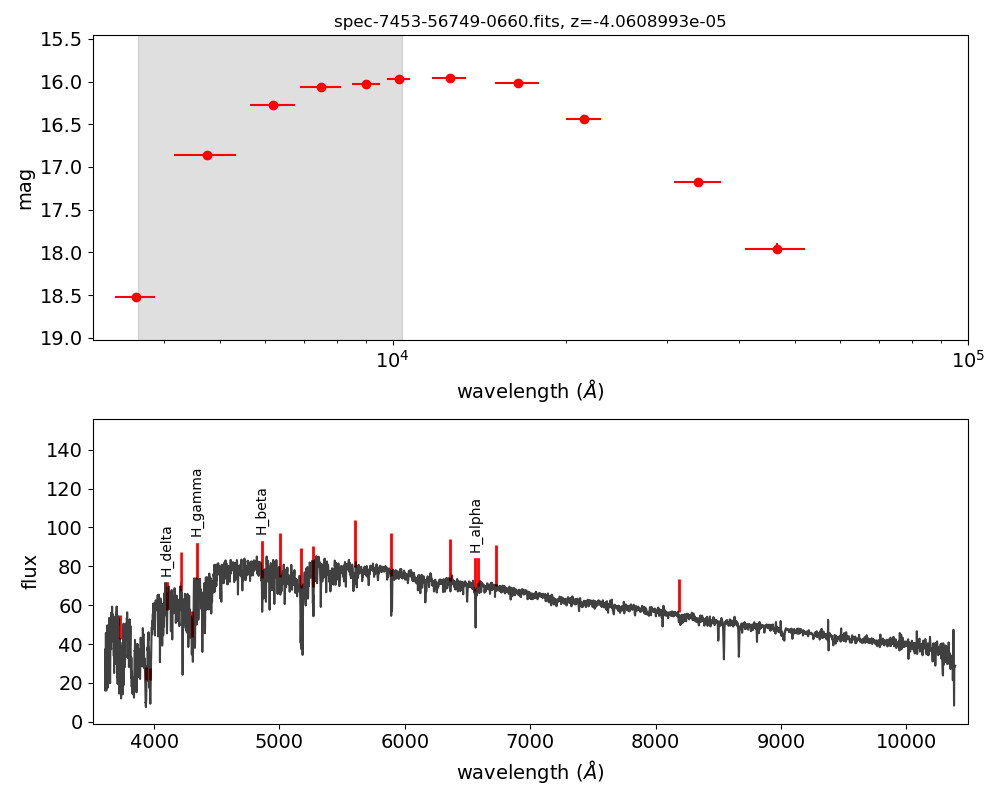}}&  \raisebox{-.5\height}{\includegraphics[width=0.32\textwidth]{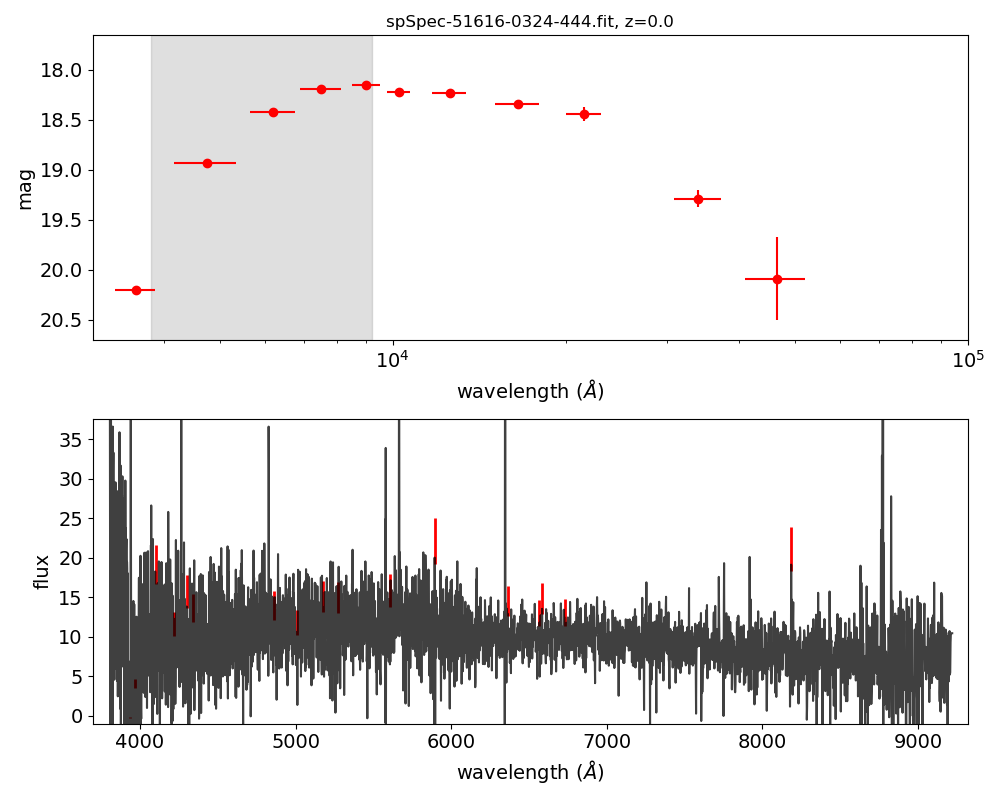}}& \raisebox{-.5\height}{\includegraphics[width=0.32\textwidth]{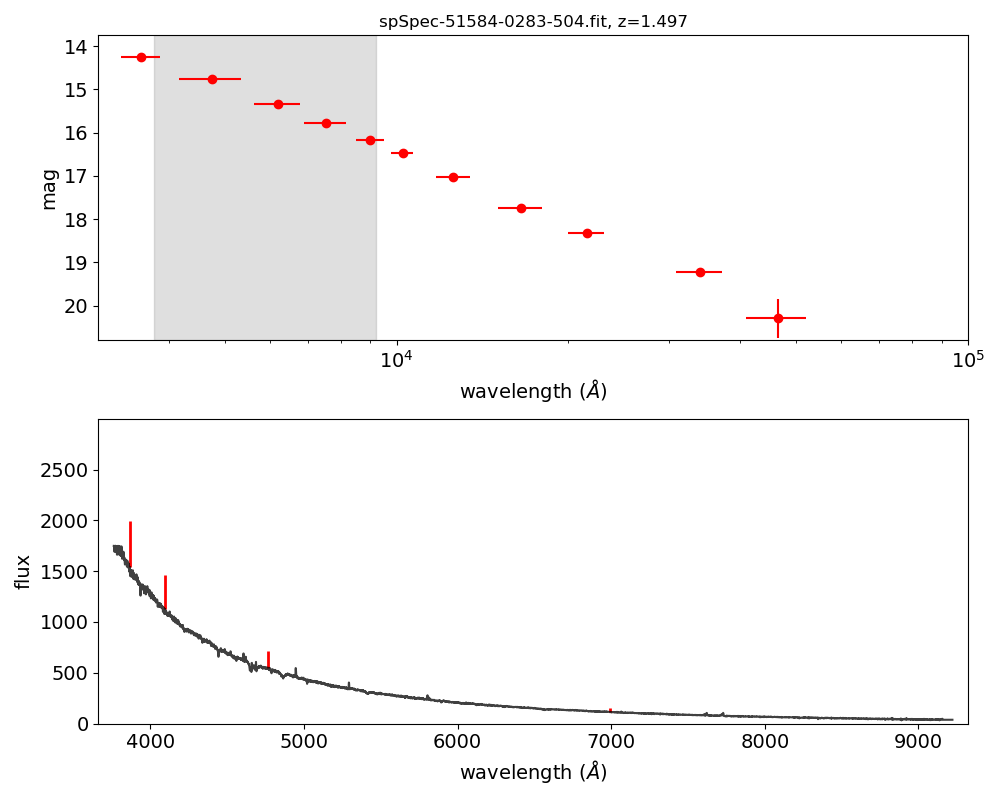}}\\ \hline
            
            &       \STAB{\rotatebox[origin=c]{90}{GALAXY}}   &        \raisebox{-.5\height}{\includegraphics[width=0.32\textwidth]{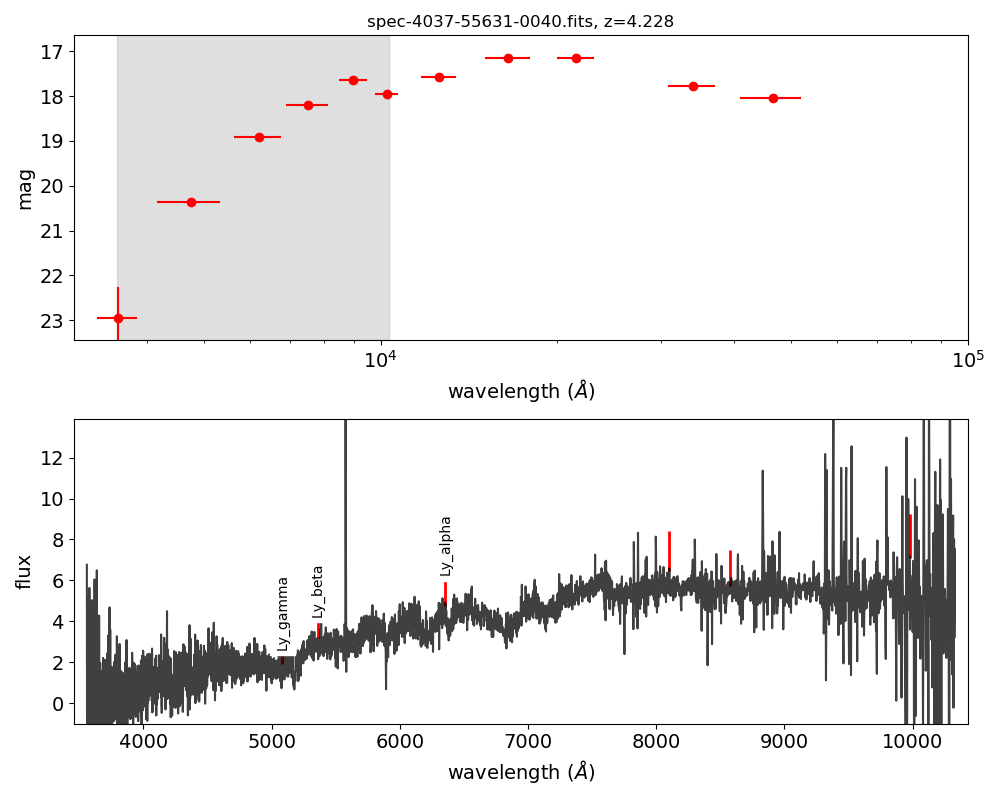}} & \raisebox{-.5\height}{\includegraphics[width=0.32\textwidth]{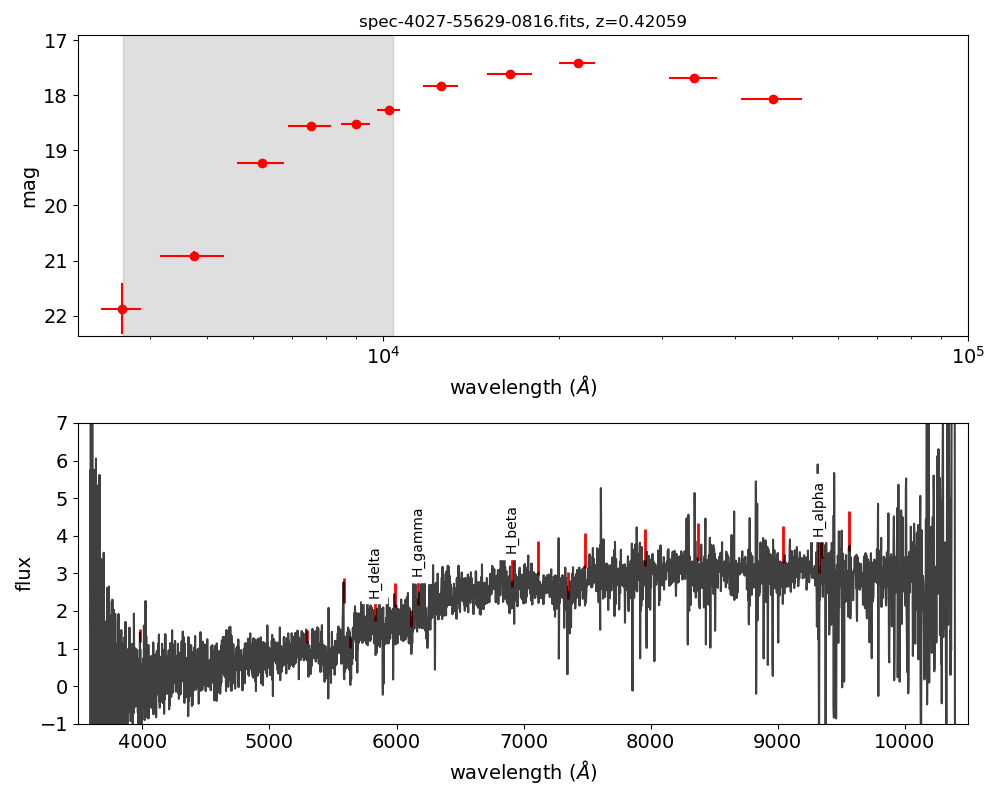}} & \raisebox{-.5\height}{\includegraphics[width=0.32\textwidth]{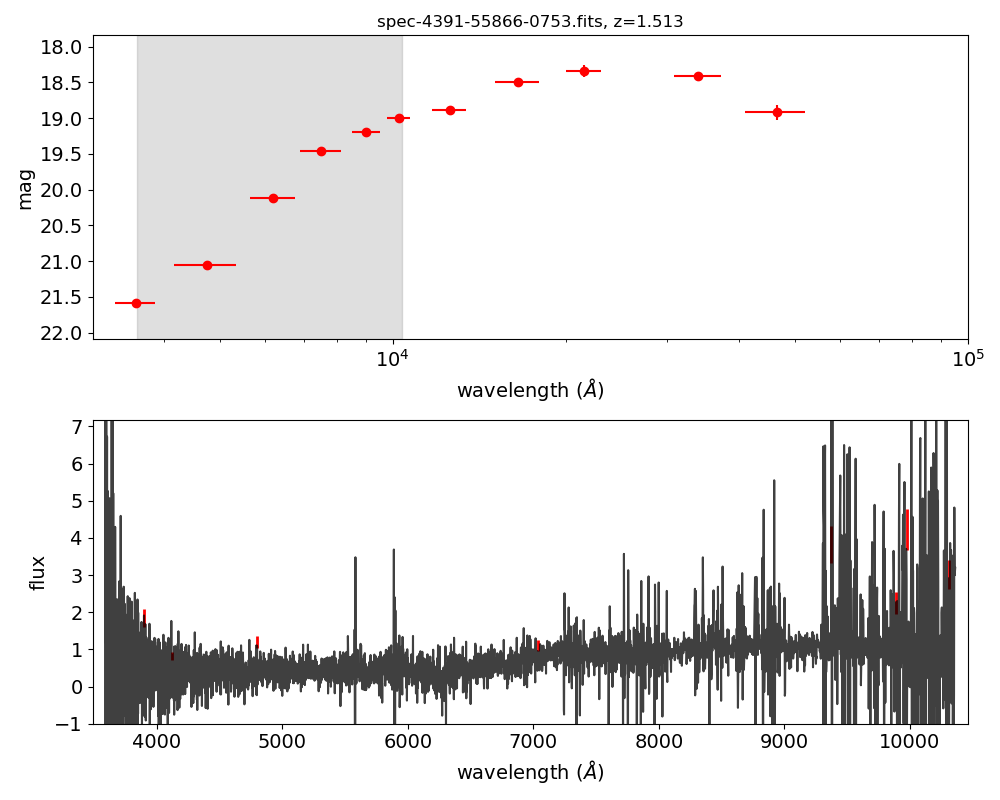}}\\ \hline
            
            &      \STAB{\rotatebox[origin=c]{90}{QSO}}   &       \raisebox{-.5\height}{\includegraphics[width=0.32\textwidth]{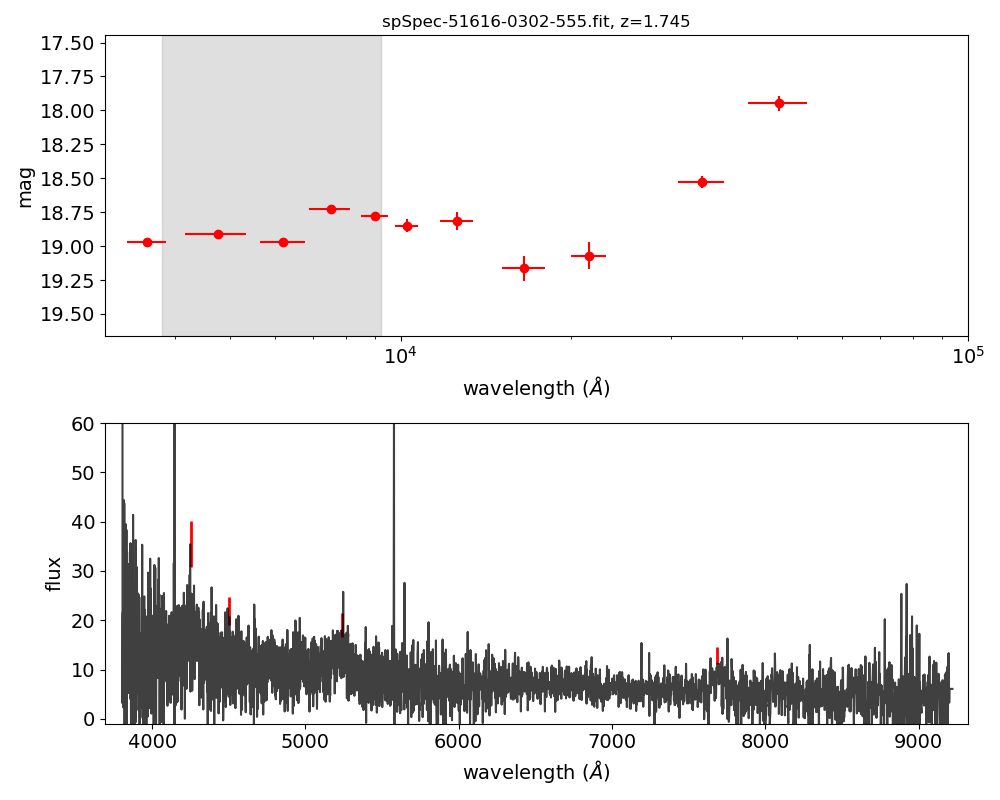}} & \raisebox{-.5\height}{\includegraphics[width=0.32\textwidth]{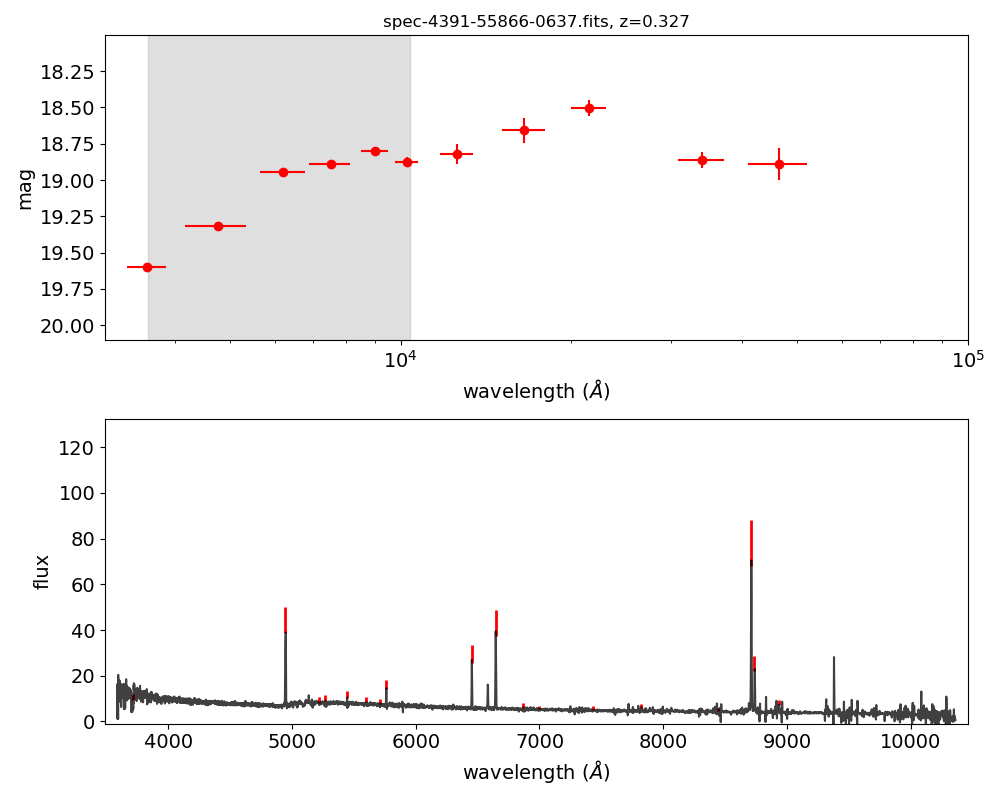}} &\raisebox{-.5\height}{ \includegraphics[width=0.32\textwidth]{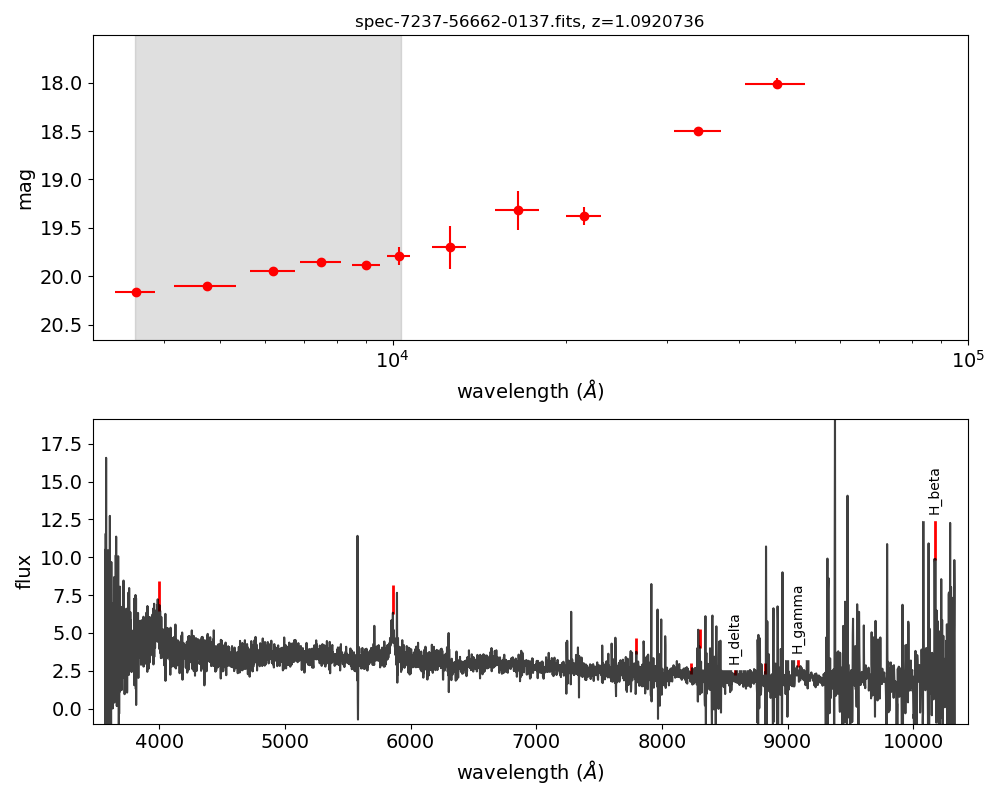}} \\ \hline
            
            &       \STAB{\rotatebox[origin=c]{90}{OUTLIER}}   &       \raisebox{-.5\height}{\includegraphics[width=0.32\textwidth]{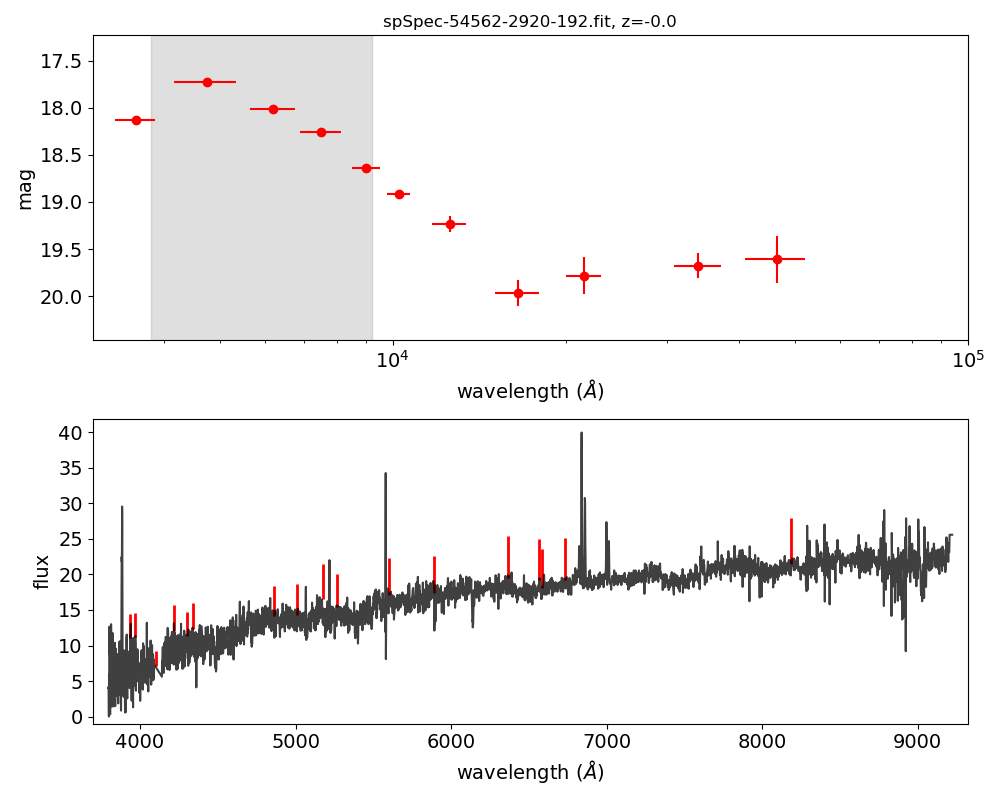}} & \raisebox{-.5\height}{\includegraphics[width=0.32\textwidth]{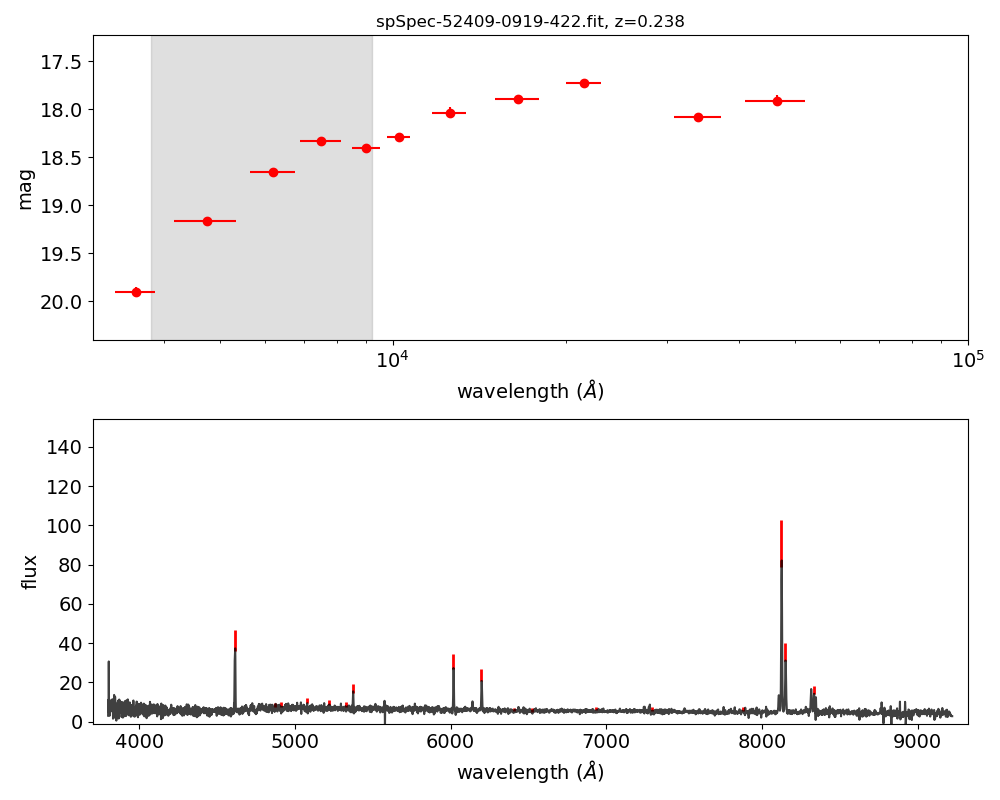}} & \raisebox{-.5\height}{\includegraphics[width=0.32\textwidth]{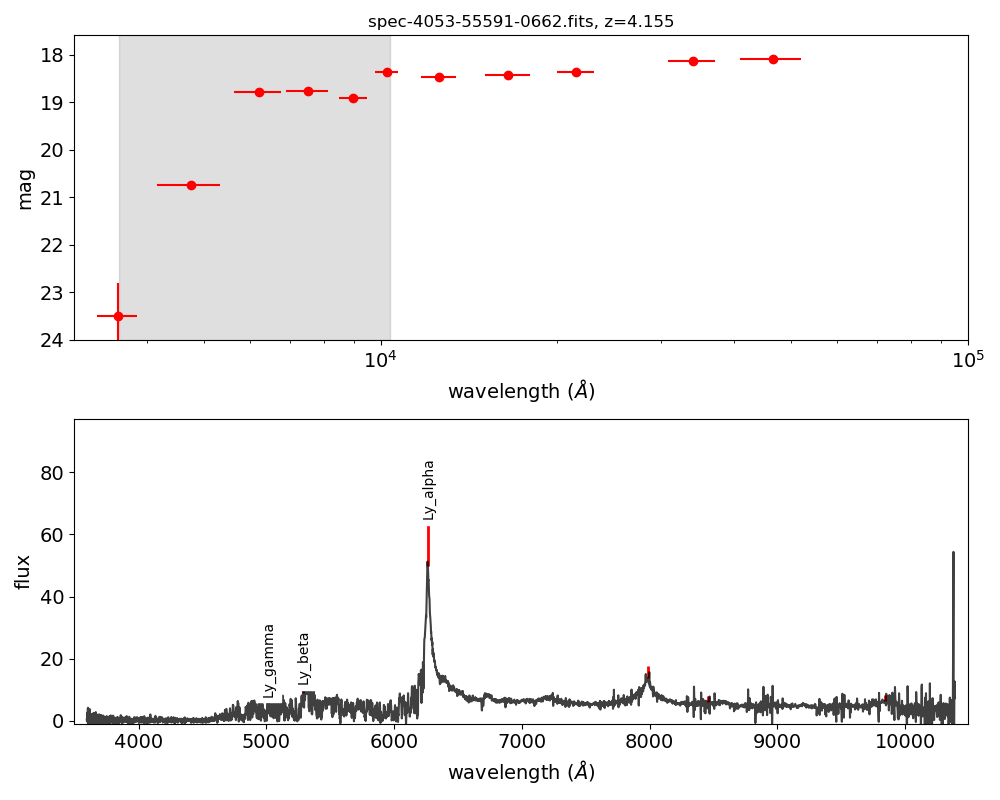}} \\
            \hline
        \end{tabular}
    }
    \caption{Example of classification overlap among HDBSCAN, SDSS DR14 and DR14Q for the CPz/SDSS-DR14 matched sample of 39,447 objects.}
    \label{tab:cpz_sdss_examples}
\end{table*}

Quasar identification and spectroscopic follow up has been one of they main focus points of SDSS \citep[e.g.][]{Richards2001,Schneider2010,Bovy2011}. \citet{Paris2018} presented an updated spectroscopic QSO catalogue for SDSS-IV/eBOSS (DR14Q) which includes target selection based on all previous SDSS quasar catalogues, WISE and the Palomar Transient Factory. Briefly, the authors created a superset of confirmed quasars from SDSS-I/II \citep[79,487 sources][]{Schneider2010} and quasar candidates for SDSS-III/IV \citep[819,611, following ][]{Paris2017} for a total of 899,098 quasar candidates. Given the large number of candidates, they partly rely on automated classification by the SDSS pipeline classifying the sources as star, QSO, or galaxy and partly on visual inspection of the spectra. From the superset of 899,098, the authors report that only 42,729 (4.5\%) of the sources did not have a secure star, QSO, or galaxy designation. Of those 32,621 (3.6\%) were visually inspected during the construction of previous versions of the SDSS quasar catalogue, leaving only 10,108 (1\%) candidates to be inspected for DR14Q. The final quasar catalogue of \citet{Paris2018} contains 526,356 quasars with 73.5\% selected through visual assessment and 26.5\% though an automated classification.

\begin{figure}
    \centering
    \begin{tabular}{c}
    \includegraphics[width=\columnwidth]{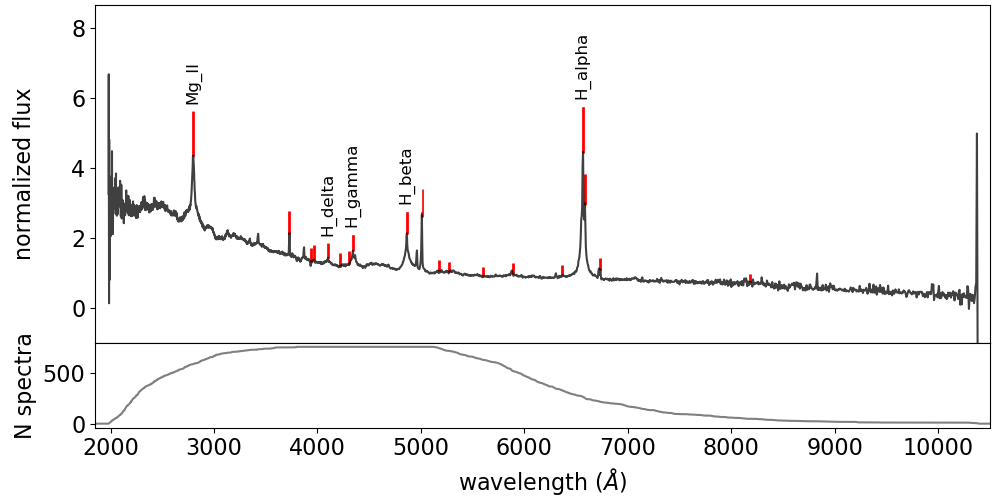}\\
    \includegraphics[width=\columnwidth]{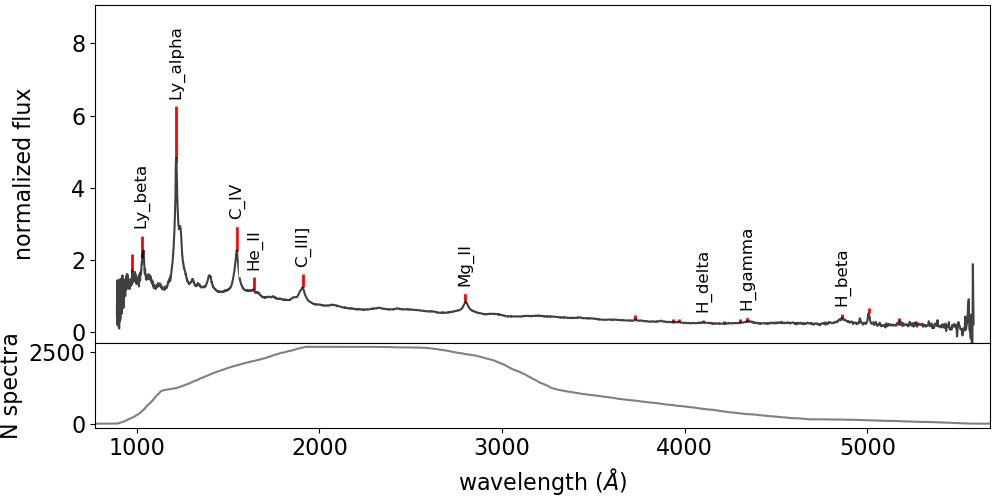}\\
    \end{tabular}
    \caption{SDSS DR14 median stacked spectrum of \textsc{hdbscan} identified QSO. Top: sources at 0<z<0.8, normalized at rest-frame 5,100 \AA. Bottom: sources at 0.8<z<3 normalized at 1,925 \AA. The lower subplot shows the number of stacked spectra per pixel. }
    \label{fig:median_qso_spec}
\end{figure}

We first matched the CPz dataset (43,348 sources) with SDSS DR14 (39,447 matches) within 1'' match radius. In Table \ref{tab:sdss_cross_matched} we show the breakdown of the classifications between CPz and SDSS DR14. Overall, the fraction of stars, galaxies and QSO is roughly comparable (stars: $\sim$18\%, galaxies: $\sim$72\%, $\sim$QSO: 10\%). However, there are variations on the actual objects per class as revealed by the confusion matrix between HDBSCAN and SDSS DR14.

We also matched the CPz dataset (43,348 sources) with the \citet{Paris2018} DR14Q catalogue (3,666 matches) within 1'' match radius. In Table \ref{tab:dr14q_cross_matched} we show the breakdown of the classification of HDBSCAN and SDSS DR14 using the 3,666 matched quasars of the DR14Q quasar catalogue of \citet{Paris2018}. It is tempting to interpret the 99.5\% quasar sample of SDSS DR14 as a very high quasar accuracy. However, there are two effects that contribute largely to this. First, the procedure of \citet{Paris2018} trusts the automatic classification of SDSS under certain quality conditions. Secondly and more importantly, a naive selection of QSO in the SDSS DR14 catalogue according to the pipeline assigned class returns 996,936 quasars. \citet{Paris2018} demonstrated that the number of reliable quasars is 526,356 (53\%), meaning that the number of SDSS DR14 QSO is overestimated. 

A visual inspection of the disagreement between the DR14Q catalogue and HDBSCAN shows that in many cases the spectra show prominent Lyman breaks and Broad Absorption Lines (BALs). Both cases are naturally more common at $\mathrm{z>2}$. This contributes to the decreasing HDBSCAN performace as a function of redshift shown in Fig. \ref{fig:perf_as_fn_of_redshift}. Other cases of confusion are AGN and obscured quasars at high redshift which appear similar to normal galaxies in the optical and near infrared observed frame wavelengths, but have prominent QSO mid infrared emission. Table \ref{tab:cpz_sdss_examples} shows a visualization of the confusion matrix highlighting selected cases of disagreement between HDBSCAN and SDSS DR14. A complete characterization of the galaxy/AGN parameters in the overlap regions between these high dimensional clusters is outside the scope of this work, as it will be only possible with flux limited redshift complete samples of future generation spectroscopic surveys.

Our unsupervised classification of a photometric dataset has recovered spectroscopically confirmed quasar populations with very high accuracy compared to DR14Q ($\mathrm{>97\%}$) and with comparable quality to the visual inspection of \citet{Paris2018} thus, our classifier can be used to automate and scale source identification to arbitrarily large datasets, particularly useful for all future photometric and spectroscopic large area surveys. 

\begin{table}
    \centering
    \scalebox{0.85}{
    \begin{tabular}{cc|cccc|ccc}\hline
 &           & \multicolumn{4}{c|}{HDBSCAN} & \multicolumn{3}{c}{SDSS DR14} \\
 &          & S     & G & Q & O & S     & G     & Q   \\ 
 &          & 18.0\%  & 72.1\% & 9.6\% & 0.2\% & 18.3\%     & 71.2\%     & 10.5\%   \\ \hline 
      \multirow{4}{*}{\STAB{\rotatebox[origin=c]{90}{\tiny{HDBSCAN}}}}&       S   & 7,114  &  -- & --  & --  &  7,065 & 44    & 5     \\
 &       G   &       &   28,58 & --  & -- & 113 & 27,938 & 407\\
 &       Q   &       &        & 3,779 & -- & 18 & 65 & 3,696 \\
 &       O   &       &       &       & 96 & 12 & 42 & 42 \\ \hline
        
     \multirow{3}{*}{\STAB{\rotatebox[origin=c]{90}{\tiny{SDSS}}}} &       S    &       &       &       &  & 7,208 & -- & -- \\
 &       G    &       &       &       &  &       & 28,089 & -- \\
 &       Q    &       &       &       &  &       &       & 4,150 \\ \hline
        
     \end{tabular}
    }
    \caption{Confusion matrix between HDBSCAN and SDSS DR14 for the CPz/SDSS-DR14 matched sample of 39,447 objects. The `O' refers to the post-consolidation outliers.}
    \label{tab:sdss_cross_matched}
\end{table}

\begin{table}
    \centering
\scalebox{0.85}{
    \begin{tabular}{cc|cccc|ccc}\hline
 &           & \multicolumn{4}{c|}{HDBSCAN} & \multicolumn{3}{c}{SDSS DR14}\\
 &          & S     & G & Q & O & S     & G     & Q \\ 
  &          & 0.03\%  & 1.9\% & 97.6\% & 0.5\% & 0.05\%     & 0.4\%     & 99.5\% \\ \hline 
      \multirow{4}{*}{\STAB{\rotatebox[origin=c]{90}{\tiny{HDBSCAN}}}}&       S   & 1  &  -- & --  & --  &  0 & 0    & 1     \\
 &       G   &       &   68 & --  & -- & 0 & 0 & 68 \\
 &       Q   &       &        & 3,578 & -- & 2 & 15 & 3,561 \\
 &       O   &       &       &       & 19 & 0 & 0 & 19 \\ \hline
        
     \multirow{3}{*}{\STAB{\rotatebox[origin=c]{90}{\tiny{SDSS}}}} &       S    &       &       &       &  & 2 & -- & --\\
 &       G    &       &       &       &  &       &  15 & --\\
 &       Q    &       &       &       &  &       &       &  3,649 \\ \hline
        
    \end{tabular}
}
    \caption{Confusion matrix between HDBSCAN and SDSS DR14 for the sample of 3,666 DR14Q quasars. The `O' refers to the post-consolidation outliers.}
    \label{tab:dr14q_cross_matched}
\end{table}

\begin{figure*}
\centering
\begin{tabular}{ccc}
\includegraphics[width=0.325\linewidth]{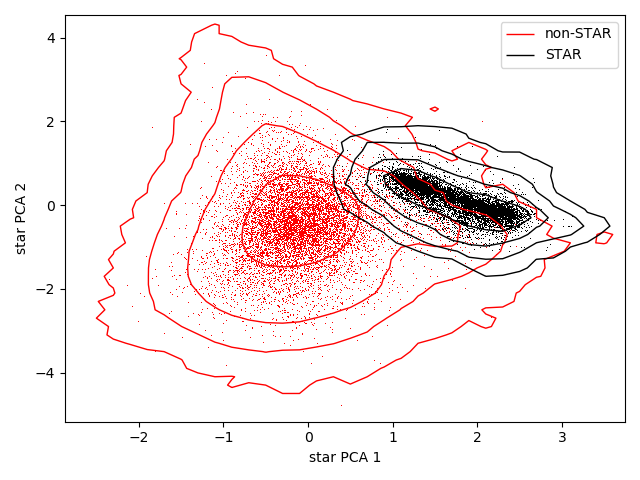} & 
\includegraphics[width=0.325\linewidth]{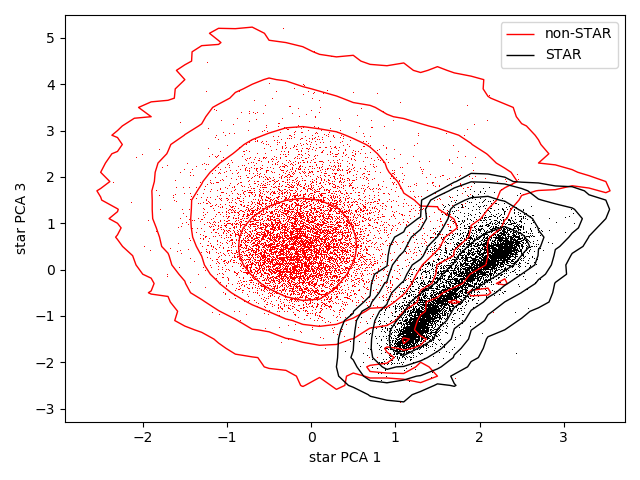} &
\includegraphics[width=0.325\linewidth]{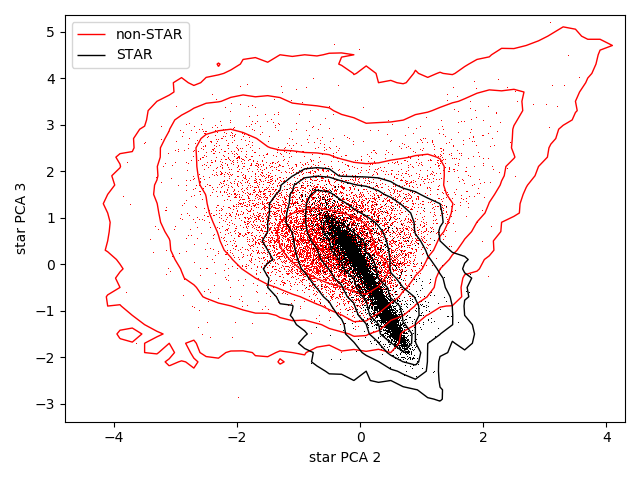} \\
\includegraphics[width=0.325\linewidth]{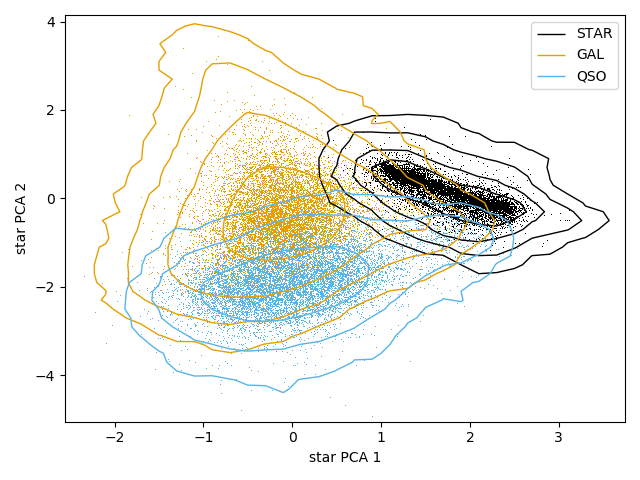}&
\includegraphics[width=0.325\linewidth]{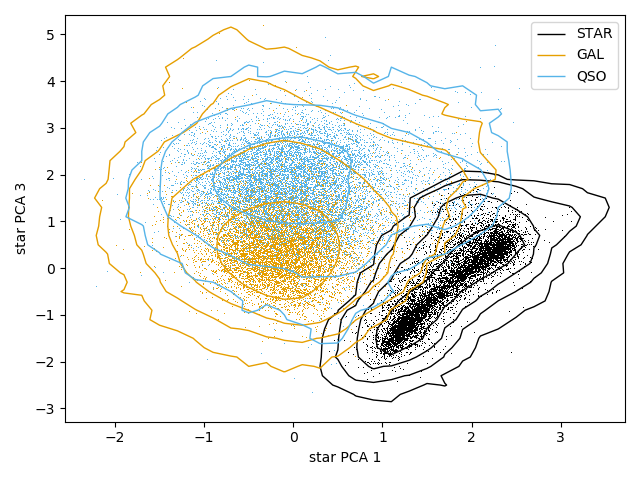}&
\includegraphics[width=0.325\linewidth]{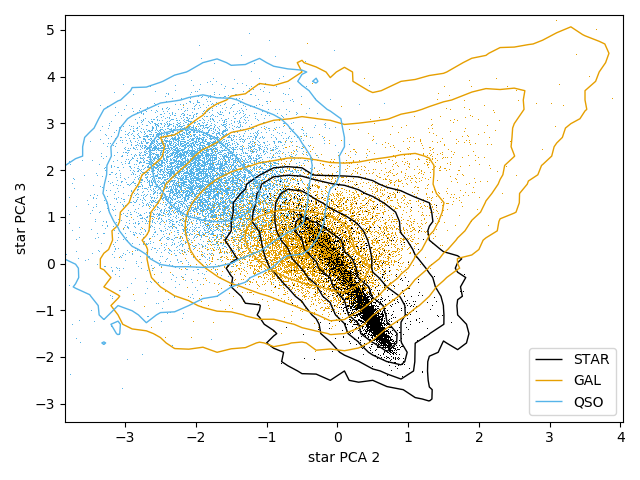} \\
\end{tabular}
\caption{KiDS-VW classification using the colour-only STAR classifier. Left: \textsc{hdbscan} STAR classifier output, right: post-consolidation labels. The contours mark iso-number locations of the full KiDS-VW sample. A subset of 10,000 points per class is shown for guidance. 
}
\label{fig:star_PCA_newcat}
\end{figure*}

\begin{figure*}
\centering
\begin{tabular}{ccc}
\includegraphics[width=0.325\linewidth]{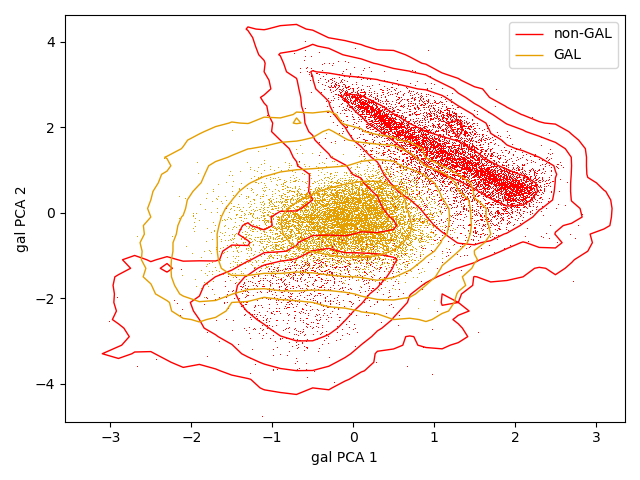} & 
\includegraphics[width=0.325\linewidth]{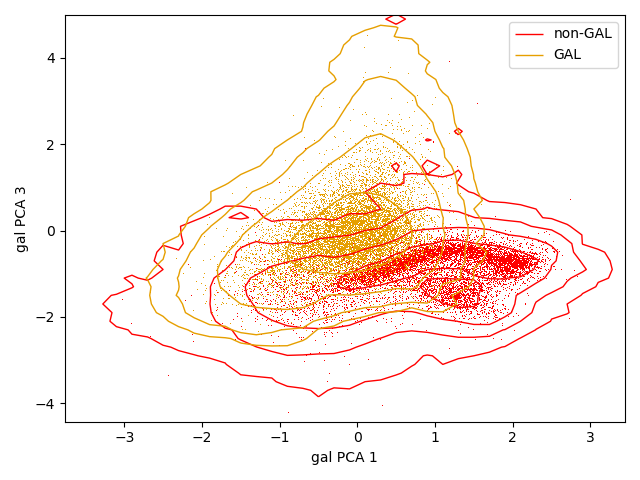} &
\includegraphics[width=0.325\linewidth]{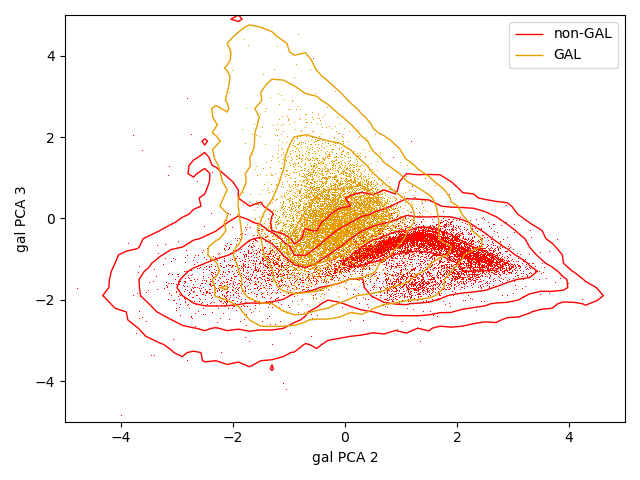} \\
\includegraphics[width=0.325\linewidth]{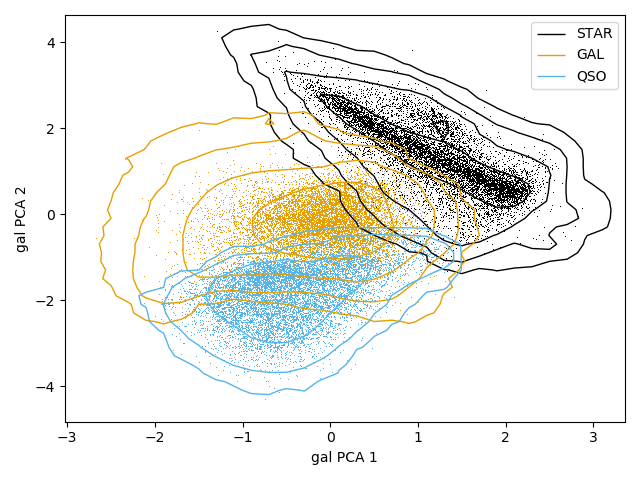}&
\includegraphics[width=0.325\linewidth]{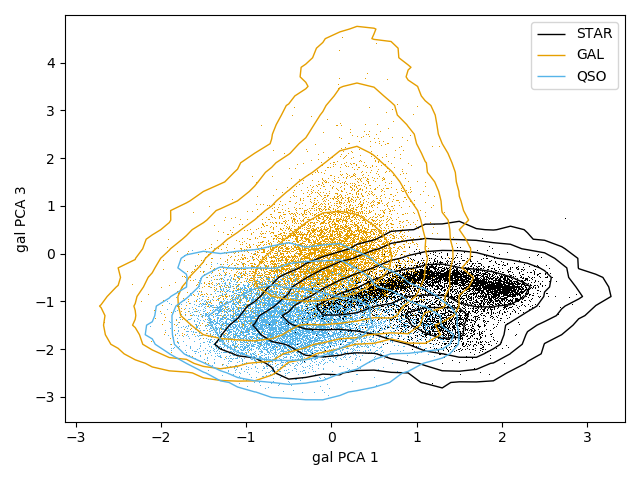}&
\includegraphics[width=0.325\linewidth]{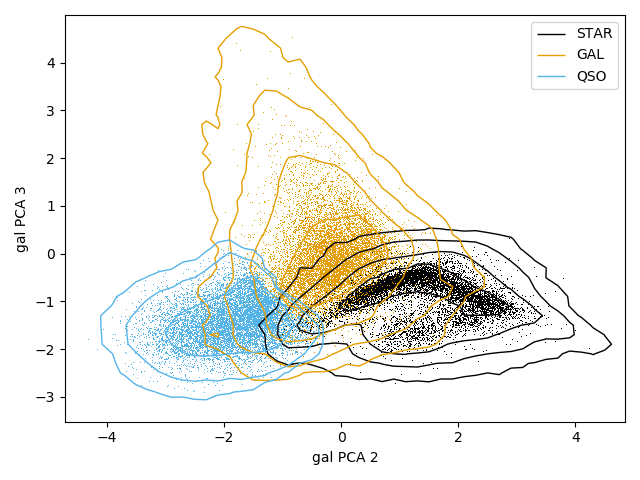} \\ 
\end{tabular}
\caption{Same as \ref{fig:star_PCA_newcat} for the GAL classifier.
}
\label{fig:gal_PCA_newcat}
\end{figure*}

\begin{figure*}
\centering
\begin{tabular}{ccc}
\includegraphics[width=0.325\linewidth]{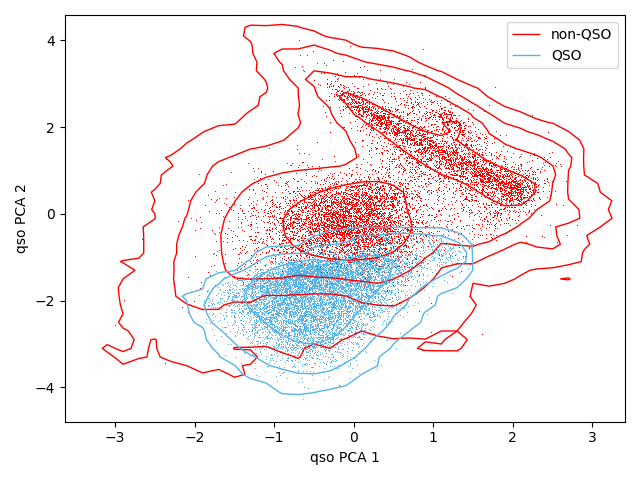} & 
\includegraphics[width=0.325\linewidth]{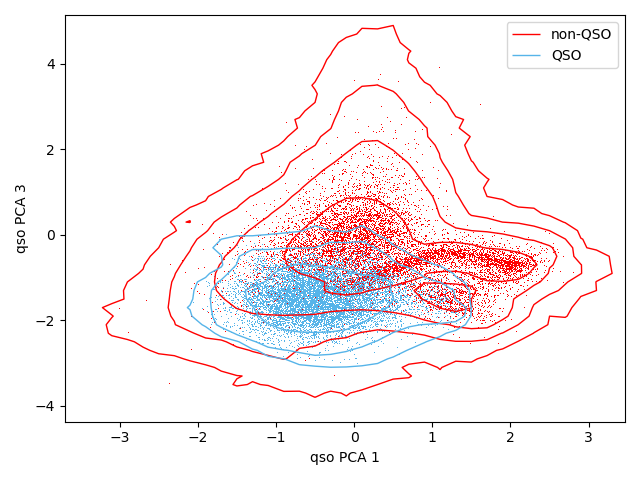} &
\includegraphics[width=0.325\linewidth]{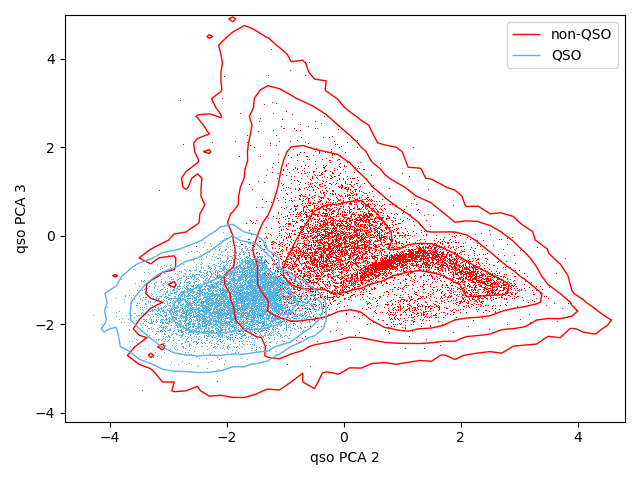} \\
\includegraphics[width=0.325\linewidth]{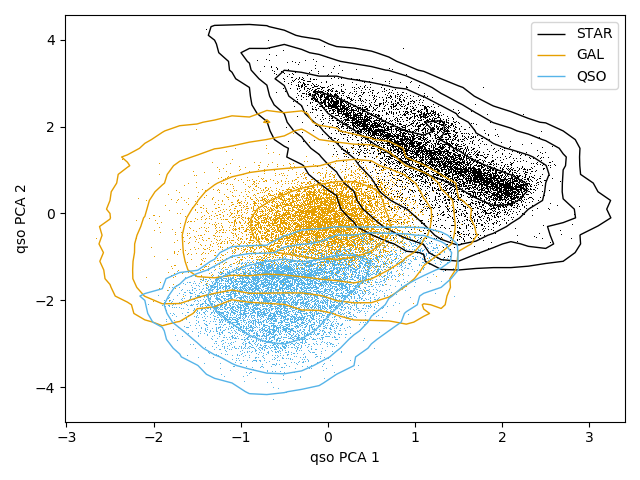}&
\includegraphics[width=0.325\linewidth]{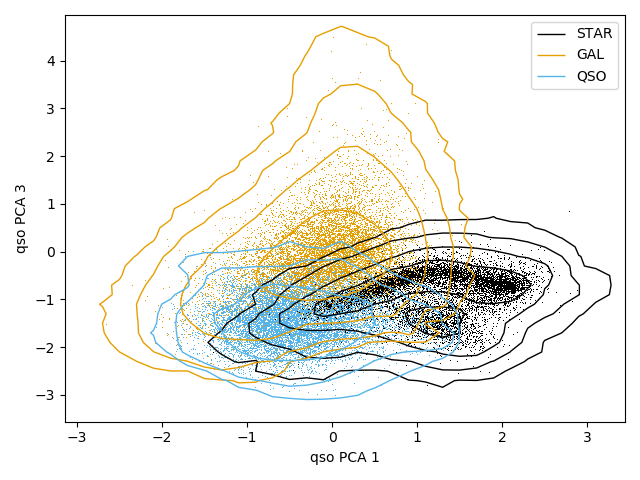}&
\includegraphics[width=0.325\linewidth]{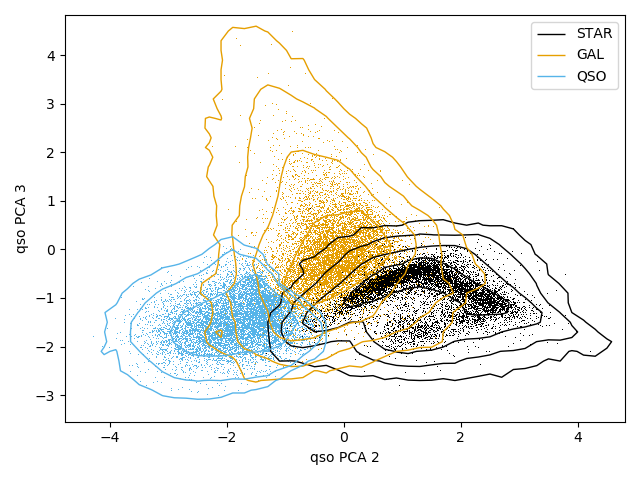} \\ 
\end{tabular}
\caption{Same as \ref{fig:star_PCA_newcat} for the QSO classifier.
}
\label{fig:qso_PCA_newcat}
\end{figure*}

\subsection{KiDS-VW quasar catalogue} \label{section.KiDS-VW_quasar_cat}
\begin{table}
    \centering
    \scalebox{0.85}{
    \begin{tabular}{cc|cccc|ccc}\hline
 &           & \multicolumn{4}{c|}{HDBSCAN} & \multicolumn{3}{c}{SDSS DR14} \\
 &          & S     & G & Q & O & S     & G     & Q   \\ 
 &          & 17.5\%  & 67.7\% & 13.8\% & 1.1\% & 18.3\%     & 65.8\%     & 15.9\%   \\ \hline 
      \multirow{4}{*}{\STAB{\rotatebox[origin=c]{90}{\tiny{HDBSCAN}}}}&       S   & 5,352  &  -- & --  & --  &                                5,249 & 97    & 6     \\
 &       G   &       &   20,659 & --  & -- & 170 & 19,843  & 646\\
 &       Q   &       &        & 4,196 & -- & 50 & 52 & 4,094 \\
 &       O   &       &       &       & 307 &  110 & 88 & 109 \\ \hline
        
     \multirow{3}{*}{\STAB{\rotatebox[origin=c]{90}{\tiny{SDSS}}}} &       S    &       &       &       &  & 5,579 & -- & -- \\
 &       G    &       &       &       &  &       & 20,080 & -- \\
 &       Q    &       &       &       &  &       &       & 4,855 \\ \hline
        
     \end{tabular}
    }
    \caption{Classification overlap among HDBSCAN and SDSS DR14 for the KiDS-VW/SDSS-DR14 matched sample of 30,514 objects, excluding any sources present in CPz. The `O' refers to the post-consolidation outliers.}
    \label{tab:sdss_cross_matched_not_in_CPz}
\end{table}

\begin{table}
    \centering
\scalebox{0.85}{
    \begin{tabular}{cc|cccc|ccc}\hline
 &           & \multicolumn{4}{c|}{HDBSCAN} & \multicolumn{3}{c}{SDSS DR14}\\
 &          & S     & G & Q & O & S     & G     & Q \\ 
  &          & 0.0\%  & 6.0\% & 91.9\% & 2.1\% & 0.12\%     & 0.42\%     & 99.5\% \\ \hline 
      \multirow{4}{*}{\STAB{\rotatebox[origin=c]{90}{\tiny{HDBSCAN}}}}&       S   & 0  &  -- & --  & --  &                                   0 & 0    & 0     \\
 &       G   &       &   259 & --  & -- &      0 & 8 & 251 \\
 &       Q   &       &        & 3,973 & -- & 3 & 10 & 3,960 \\
 &       O   &       &       &       & 89 &   2 & 0 & 87 \\ \hline
        
     \multirow{3}{*}{\STAB{\rotatebox[origin=c]{90}{\tiny{SDSS}}}} &       S    &       &       &       &  & 5 & -- & --\\
 &       G    &       &       &       &  &       &  18 & --\\
 &       Q    &       &       &       &  &       &       &  4,298 \\ \hline
        
    \end{tabular}
}
    \caption{Classification overlap among HDBSCAN, SDSS DR14 and DR14Q for the KiDS-VW/DR14Q matched sample of 4,321 DR14Q quasars, excluding any sources present in CPz. The `O' refers to the post-consolidation outliers. }
    \label{tab:dr14q_cross_matched_not_in_CPz}
\end{table}

\begin{figure*}
    \centering
    \begin{tabular}{cc}
    \includegraphics[width=\columnwidth]{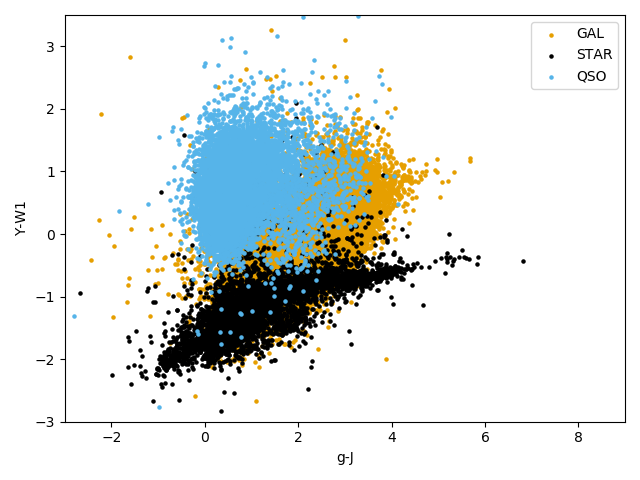} &
    \includegraphics[width=\columnwidth]{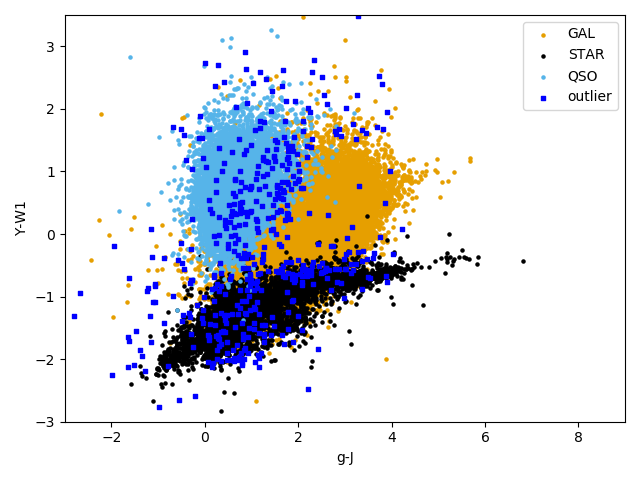} \\
    (a) & (b) \\
    \includegraphics[width=\columnwidth]{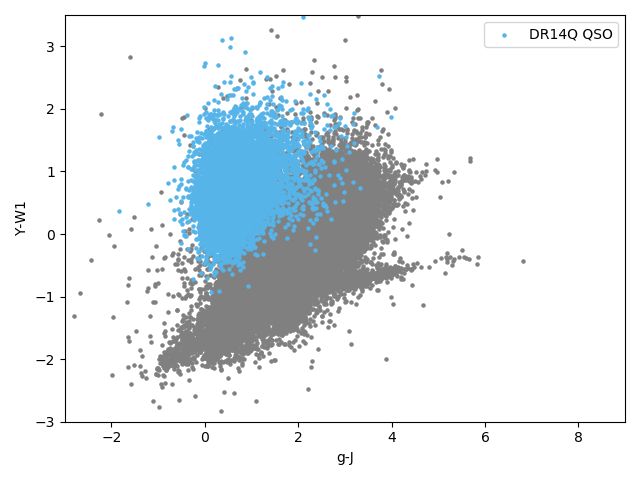} &
    \includegraphics[width=\columnwidth]{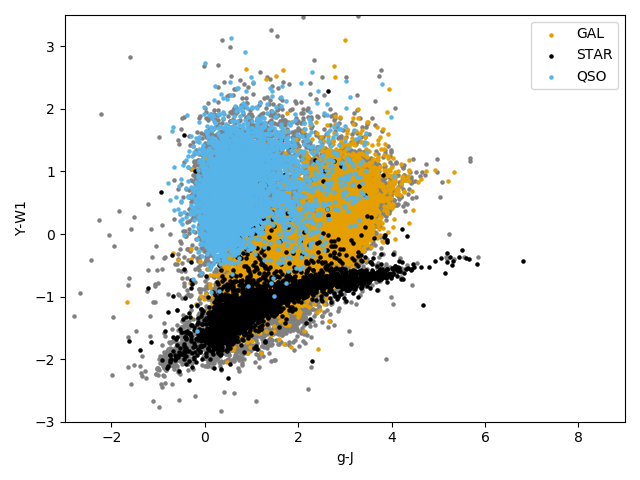} \\
    (c) & (d) \\
    \end{tabular}
    \caption{Classifications of all sources in SDSS DR14 matched to the KiDS-VW photometric catalogue. (a) SDSS DR14 spectroscopic label assignment. The SDSS-QSO class contains a mixture of black holes in various states (BALs, AGN, obscured systems etc) (b) SDSS DR14 points coloured by their \textsc{hdbscan} labels (using the `optimal' consolidation method). The outliers are post-consolidation outliers and are  defined as sources that could not be assigned unambiguously to a single object class by our \textsc{hdbscan} model in the PCA parameter space. (c) SDSS DR14 points, coloured in blue if in DR14Q, or grey if not in DR14Q. (d) SDSS DR14 points, coloured according to the labels of KIDS-DR3, or grey if not in KIDS-DR3. }
    \label{fig:SDSSDR14_classcomparison}
\end{figure*}

\begin{figure}
\begin{tabular}{c}
\includegraphics[width=\linewidth]{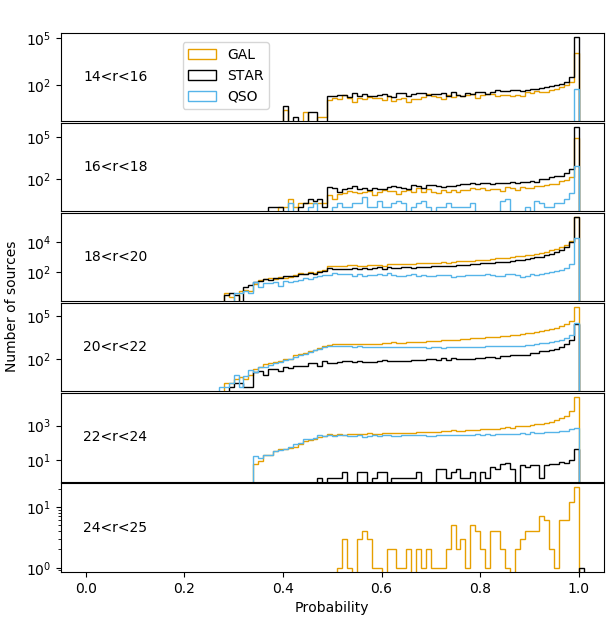}\\
\end{tabular}
\caption{The probability distributions for the KiDSVW sample for six different r magnitude bins are shown. The classification probability of each object is obtained using the `highest-probability' method (see \S \ref{subsection.consolidation}). The objects are in classes according to their `highest-probability' final label.}
\label{figure.av_prob_of_classification_for_highest_prob_consolidation_kidsvw}
\end{figure}

A positional match of 1'' radius between the KiDS-VW and SDSS DR14 yields 55,383. From this subset, we select the sources that are not already in the CPz sample, giving us 30,514 points. 
Figure \ref{fig:median_qso_spec} shows the median stacked SDSS DR14 spectra for the latter sample of \textsc{hdbscan} identified QSO in two redshift bins. The QSO identified with this method are protopytical unobscured QSO.

Table \ref{tab:highest_prob_label_comp_KiDS-VW} shows the performance metrics using the SDSS DR14 spectroscopic labels as the truth, for all three consolidation methods. We also estimated the uncertainties for the metric values (using the method described in \S \ref{section.uncertainties}, having used the `optimal' method for consolidation) for these 30,514 points, using the SDSS DR14 labels as the truth. We find the mean F1 scores for the star, galaxy and
QSO class to be 95.97$\pm$0.05, 97.26$\pm$0.03 and 89.49$\pm$0.15 (error given  is  the  standard  deviation).

In comparison to the performance metrics found for the CPz sample when just colours are used as attributes (shown in Table \ref{tab:finalmodelsetups_nohlr}) the F1 scores are lower for when training on the CPz sample and applying to the new points in KiDS-VW (see Table \ref{tab:highest_prob_label_comp_KiDS-VW}). This is expected, as the \textsc{hdbscan} was optimized for the original CPz dataset. However, the QSO F1 score barely decreases from the CPz sample to the application on the KiDS-VW sample (at least for those with SDSS spectroscopic labels). The precision for the prediction on the KiDS-VW sample is actually higher than that for the original CPz sample. 

For the 30,514 points in the overlap between the KiDS-VW and SDSS DR14 samples, we show the breakdown of the classifications between KiDS-VW and SDSS DR14 in Table \ref{tab:sdss_cross_matched_not_in_CPz}. Overall, the fraction of stars, galaxies and QSO is roughly comparable (stars: $\sim$18\%, galaxies: $\sim$66-68\%, QSO: $\sim$14-16\%). 

Figure \ref{fig:SDSSDR14_classcomparison} (a) shows the 55,383 SDSSDR14 sources matched to KiDS-VW, colour codes according the spectroscopic labels, while panel (b) shows the same objects colour-coded according to their \textsc{hdbscan} classification. The dark blue points are outliers as identified by the optimal consolidation method. The SDSSDR14 QSO labels sources seem to span a large area on this colour plot, extending into the star locus. Contrary, our method localizes the three populations very well, and the consolidation procedure identifies potential very problematic objects.

In addition, we present the AUC scores for all consolidation methods in Table \ref{tab:highest_prob_label_comp_KiDS-VW}. We find the AUC scores to always be 0.979 for star, 0.963 for galaxies, and between 0.920 and 0.927 for QSOs. In Figure \ref{figure.av_prob_of_classification_for_highest_prob_consolidation_kidsvw} we show the probability distributions of the ‘highest-probability’ consolidation method (see \S \ref{subsection.consolidation}) in different r magnitude bins for the KiDSVW sample.

\subsubsection{Comparison to DR14Q}

We also matched the 30,514 sources from the overlap between the KiDS-VW and SDSS DR14 samples with the \citet{Paris2018} DR14Q catalogue (4,321 matches) within 1'' match radius. In Table \ref{tab:dr14q_cross_matched_not_in_CPz} we show the breakdown of the classification of HDBSCAN and SDSS DR14 using the 4,321 matched quasars of the DR14Q quasar catalogue of \citet{Paris2018}. 

Figure \ref{fig:SDSSDR14_classcomparison} (c) shows the DR14Q sources that are in SDSSDR14 - KiDS-VW sample (cyan for QSO, grey otherwise). It is clear that the verification procedure of \cite{Paris2018} has removed a lot of interlopers, particularly sources with star-like colours.

\subsubsection{Comparison to the KiDS-DR3 QSO catalogue}

\begin{figure}
    \centering
    \includegraphics[width=\columnwidth]{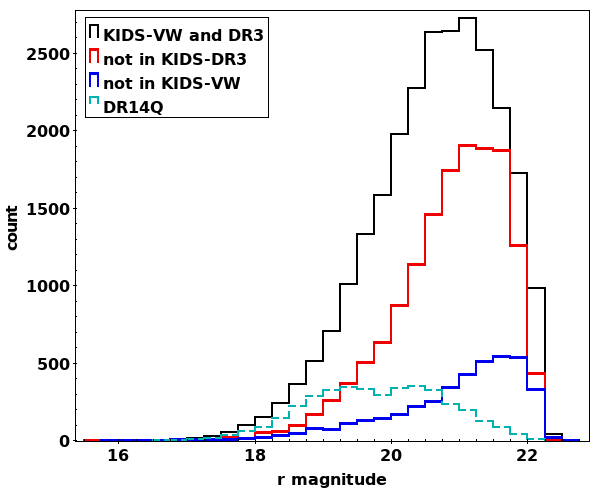}
    \caption{Magnitude distribution of the KiDS-VW and KIDS-DR3 quasar samples. Black line: quasars identified in both samples. Red line: quasars identified only in KiDS-VW. Blue line: quasars identified only in KIDS-DR3. Cyan line: DR14Q quasars matched to the overlap photometric catalogue of KiDS-VW and KIDS-DR3.}
    \label{fig:KIDS_DR4_vs_VW}
\end{figure}

\citet{Nakoneczny2019} used KIDS DR3 to search for quasars using $ugri$  colours and magnitudes, and the stellarity morphological index. They created a parent catalogue of $\mathrm{3.4\;10^6}$ sources and using SDSS DR14 spectroscopic labels, they identified a total of 190,000 quasar candidates ($r$<22).
We matched our KiDS-VW catalogue ($\mathrm{2.7\;10^6}$ sources) with the \citet{Nakoneczny2019} parent catalogue ($\mathrm{3.4\;10^6}$ sources) using a matching radius of 1'' and found 837,624 matches. Our catalogue contains less sources since it was built with the requirement to have complete photometric detections in the full wavelength range of the $u$ -- $W2$ bands. 

Out of the 837,624 sources in common, we examined the overlap of quasar positive classifications. We selected KIDS-DR3 quasars using a threshold of Pr[QSO]>0.7 suggested for optimized completeness in \citet{Nakoneczny2019}. Using the matched sample, we find that the KIDS-DR3 catalogue identifies 29,878 quasar candidates while the KiDS-VW sample identifies 40,621 quasars. The sample in common between the two samples is 25,784 sources. Figure \ref{fig:KIDS_DR4_vs_VW} shows the magnitude distribution of the quasar samples. The black lines shows the 25,784 sources in common between the two samples, the red lines shows the 14,837 quasars identified within the KiDS-VW sample but not in the KIDS-DR3, and the blue line shows the 4,094 sources identified within the KIDS-DR3 sample but not in the KiDS-VW. Clearly, the discrepancies between the two samples have a strong magnitude dependence, but they are however already eminent at $r$<20. For comparison, we matched the DR14Q quasar catalogues \citep{Paris2018} to the overlap of KiDS-VW and KIDS-DR3 catalogues finding 3,889 sources (cyan line in Fig. \ref{fig:KIDS_DR4_vs_VW}).

According to the feature importance ranking discussed in \citet{Nakoneczny2019}, the stellarity index carried the highest weight. Hence, we conclude that the $ugri$ colours where suffiencent to disentangle between stars and quasars leading to the creation of a pure quasar catalogue. On the other hand, the limited input attribute parameter space has impacted the completeness of the identified quasars. This is clear demonstration for the importance of near and mid-infrared photometry in quasar identification.

Finally, Figure \ref{fig:SDSSDR14_classcomparison} (d) shows the 55,383 sources in SDSSDR14 and the KiDS-VW sample, colour coded according to the KIDS-DR3 quasar classification of \citet{Nakoneczny2019}. Since this classification was based on the spectroscopic labels of SDSS DR14, it replicates the distribution of the sources in panel (a), including with the caveats of mislabelled sources that where addressed in \citet{Paris2018} shown in pabel (c).

Figure \ref{fig:SDSSDR14_classcomparison} reveals that our method serves a good alternative to spectroscopic source labeling.

\section{Conclusions} \label{section.conclusion}
We used Hierarchical Density-Based Spatial Clustering of Applications with Noise (\textsc{hdbscan}) for the classification of stars, galaxies, QSO using only photometric data. 

   \begin{enumerate}
      \item Exploring a range of input attributes for our classification model, we found that using PCA to reduce the number of attributes to a lower dimension results to optimal performance for \textsc{hdbscan}.
      
      \item We find that the best performance is achieved using a different model set ups for each of the STAR, GAL and QSO classes.
      
      \item The \textsc{hdbscan} classification benefits from the inclusion of morphological information (in our case HLR data).
      
      \item Our automatic classification is in very good agreement with the \citet{Paris2018} quasar catalogue, and at cases outperforms the automated classification of SDSS, with the only caveat  $z>2.5$ quasars with prominent Lyman Limit Systems.
      
      \item We applied our final model to the latest public version of KiDS, VIKING and ALLWISE catalogues (KiDS-VW) and publicly release the classifications as well as photometric redshift estimations.

   \end{enumerate}
   
      Our final model can classify stars, galaxies and QSOs with an F1 score of 98.64, 98.7 and 91.07 respectively when using only colour data, and 98.9, 98.9 and 93.13 when also using HLR information. Given the scalability of the application of the trained model, we motivate the use of our approach for current and upcoming data-rich surveys. 
      
\begin{acknowledgements}
  We thank the referee for their useful comments on this work.     
  SF acknowledges the financial support of the Swiss National Science Foundation. 
  Based on data products from observations made with ESO Telescopes at the La Silla or Paranal Observatories under ESO programme IDs 179.A-2006 and 179.A-2004.
Funding for SDSS-III has been provided by the Alfred P. Sloan Foundation, the Participating Institutions, the National Science Foundation, and the U.S. Department of Energy Office of Science. The SDSS-III web site is http://www.sdss3.org/.
SDSS-III is managed by the Astrophysical Research Consortium for the Participating Institutions of the SDSS-III Collaboration including the University of Arizona, the Brazilian Participation Group, Brookhaven National Laboratory, Carnegie Mellon University, University of Florida, the French Participation Group, the German Participation Group, Harvard University, the Instituto de Astrofisica de Canarias, the Michigan State/Notre Dame/JINA Participation Group, Johns Hopkins University, Lawrence Berkeley National Laboratory, Max Planck Institute for Astrophysics, Max Planck Institute for Extraterrestrial Physics, New Mexico State University, New York University, Ohio State University, Pennsylvania State University, University of Portsmouth, Princeton University, the Spanish Participation Group, University of Tokyo, University of Utah, Vanderbilt University, University of Virginia, University of Washington, and Yale University.
Based on data products from observations made with ESO Telescopes at the La Silla Paranal Observatory under programme IDs 177.A-3016, 177.A-3017 and 177.A-3018, and on data products produced by Target/OmegaCEN, INAF-OACN, INAF-OAPD and the KiDS production team, on behalf of the KiDS consortium. OmegaCEN and the KiDS production team acknowledge support by NOVA and NWO-M grants. Members of INAF-OAPD and INAF-OACN also acknowledge the support from the Department of Physics \& Astronomy of the University of Padova, and of the Department of Physics of Univ. Federico II (Naples).
This publication makes use of data products from the Wide-field Infrared Survey Explorer, which is a joint project of the University of California, Los Angeles, and the Jet Propulsion Laboratory/California Institute of Technology, funded by the National Aeronautics and Space Administration.
Based on observations obtained with MegaPrime/MegaCam, a joint project of CFHT and CEA/IRFU, at the Canada-France-Hawaii Telescope (CFHT) which is operated by the National Research Council (NRC) of Canada, the Institut National des Science de l'Univers of the Centre National de la Recherche Scientifique (CNRS) of France, and the University of Hawaii. This work is based in part on data products produced at Terapix available at the Canadian Astronomy Data Centre as part of the Canada-France-Hawaii Telescope Legacy Survey, a collaborative project of NRC and CNRS. 
 This research has made use of the ASPIC database, operated at CeSAM/LAM, Marseille, France. 
This paper uses data from the VIMOS Public Extragalactic Redshift Survey (VIPERS). VIPERS has been performed using the ESO Very Large Telescope, under the "Large Programme" 182.A-0886. The participating institutions and funding agencies are listed at http://vipers.inaf.it

GAMA is a joint European-Australasian project based around a spectroscopic campaign using the Anglo-Australian Telescope. The GAMA input catalogue is based on data taken from the Sloan Digital Sky Survey and the UKIRT Infrared Deep Sky Survey. Complementary imaging of the GAMA regions is being obtained by a number of independent survey programmes including GALEX MIS, VST KiDS, VISTA VIKING, WISE, Herschel-ATLAS, GMRT and ASKAP providing UV to radio coverage. GAMA is funded by the STFC (UK), the ARC (Australia), the AAO and the participating institutions. The GAMA website is http://www.gama-survey.org/ . 
This research uses data from the VIMOS VLT Deep Survey, obtained from the VVDS database operated by Cesam, Laboratoire d'Astrophysique de Marseille, France.
Funding for PRIMUS is provided by NSF (AST-0607701, AST-0908246, AST-0908442, AST-0908354) and NASA (Spitzer-1356708, 08-ADP08-0019, NNX09AC95G). 

\end{acknowledgements}

\begin{appendix}
\section{Appendix} \label{section.appendix}
\subsection{Practical Considerations}
For any readers interested in replicating this work, or applying \textsc{hdbscan} to other data in a similar fashion, we note the following practical considerations. 

To reproduce our results, the random\_state in the PCA (from \texttt{scikit-learn}) was set to 1. Importantly, for the application of the CPz trained model on the KiDS-VW data, both the normalization (we used StandardScaler from \texttt{scikit-learn}) and the PCA were the same that were applied to the CPz, and that had `learned' using the CPz data. One quirk that we found to be important is that the order of the attributes into the StandardScaler and PCA had to be the same each time, as if they were different, the output values were sometimes different by a tiny amount (of the order of 1e-12 at most), and \textsc{hdbscan} was extremely sensitive to the values of the attributes. Dealing with this was vital for reproducible results, and the solution was trivial: simply maintaining the order of the input attributes for each run (the order of the attributes given in Table \ref{tab:appendix_colours_conversion} is the required order to obtain the results reported in this paper).     

To obtain the results presented in the paper, the packages we used were as follows:
python 3.7.3; \texttt{scikit-learn} 0.21.2; \texttt{pandas} 0.24.2, \texttt{numpy} 1.16.4

\subsection{Description of Catalogues} \label{section.appendix.catalogue_descriptions}
We present two catalogues. The first catalogue is the CPz sample with predicted labels and other outputs for when we train the \textsc{hdbscan} model setup with both colour and HLR attributes, and predicted labels and other outputs for when we train the \textsc{hdbscan} model setup with just colour attributes. The second catalogue is the KiDS-VW catalogue with the labels and other outputs for when we trained the \textsc{hdbscan} model on the CPz sample using just colour data and then used this trained model to predict the labels for the KiDS-VW catalogue. The KiDS-VW catalogue also contains predicted photometric redshifts. For both catalogues we describe the columns here:

Missing values are denoted by: empty entries, N/A, nan, or -99. In the CPz catalogue, for the colour and HLR outputs (columns 67-83) all -99 values are for those without Y -- K HLR values (as their labels and other outputs could not be generated using that attribute setup due to having missing or unrealistic Y -- K HLR values).

\textit{CPz catalogue}: For this catalogue, any data in the u, g, r, i bands are from KiDS, any data in the Y, J, H, K bands are from VIKING, and the data for the W1 and W2 bands are from WISE. The catalogue is described in more detail in FP18, and briefly in this paper in \S \ref{section.dataset}.
\begin{itemize}
\item[--] Col 1: Spectroscopic redshift ID (same as column 1 from the CPz catalogue presented in FP18) 
\item[--] Col 2,3: Spectroscopic redshift coordinates (same as column 2,3 in the CPz catalogue presented in FP18) 
\item[--] Col 4: Spectroscopic redshift value (same as column 4 in the CPz catalogue presented in FP18) 
\item[--] Col 5: Spectroscopic redshift classification (same as column 5 in the CPz catalogue presented in FP18, with the change as described in \S \ref{section.dataset_speclabels} for AGN and UNKNOWN). 0=star, 1=galaxy, 3=QSO  
\item[--] Col 6-27: u, g, r, i, z, Y, J, H, K, W1, W2 total magnitudes, with associated errors  
\item[--] Col 28-45: u, g, r, i, z, Y, J, H, K 3'' aperture magnitudes, with associated errors  
\item[--] Col 46-49: Half light radius (in arcseconds) in Y, J, H K bands \\
\end{itemize}

Outputs for colour data as attributes (see \S \ref{subsection.consolidation} for information on these)
\begin{itemize}
\item[--] Col 50-58: PCA components for the STAR, GAL, QSO \textsc{hdbscan} binary classifiers for the `optimal' method setup
\item[--] Col 59: hdbscanclass\_optimal\_method\_colours labels (0=outlier, 1=star, 2=gal, 3=QSO) 
\item[--] Col 60: hdbscanclass\_alternative\_method\_colours  labels (0=outlier, 1=star, 2=gal, 3=QSO) 
\item[--] Col 61: double\_positives\_colours 
\item[--] Col 62-65: outlier\_probability\_colours, star\_probability\_colours, gal\_probability\_colours, QSO\_probability\_colours (probabilities are from the `highest-probability' consolidation method)
\item[--] Col 66: highest\_probability\_labels\_colours  labels (0=outlier, 1=star, 2=gal, 3=QSO) \\
\end{itemize}

Outputs for colour and HLR data as attributes (see \S \ref{subsection.consolidation} for information on these)
\begin{itemize}
\item[--] Col 67-75: PCA components for the STAR, GAL, QSO \textsc{hdbscan} binary classifiers for the `optimal' method setup
\item[--] Col 76: hdbscanclass\_optimal\_method\_colours+HLR labels (0=outlier, 1=star, 2=gal, 3=QSO) 
\item[--] Col 77: hdbscanclass\_alternative\_method\_colours+HLR labels (0=outlier, 1=star, 2=gal, 3=QSO) 
\item[--] Col 78: double\_positives\_colours+HLR 
\item[--] Col 79-82: outlier\_probability\_colours+HLR, star\_probability\_colours+HLR, gal\_probability\_colours+HLR, QSO\_probability\_colours+HLR (probabilities are from the `highest-probability' consolidation method)
\item[--] Col 83: highest\_probability\_labels\_colours+HLR (0=outlier, 1=star, 2=gal, 3=QSO)
\end{itemize}

\textit{KiDS-VW catalogue}: For this catalogue, any data in the u, g, r, i bands are from KiDS, and any data in the Y, J, H, K bands are from VIKING, and the data for the W1 and W2 bands are from WISE. The catalogue construction is described in detail in \S \ref{section.app_to_KiDS-VW}.  We note that all of the outputs are for just colour data as attributes for this catalogue.

\begin{itemize}
\item[--] Col 1: serialid\_kids\_u\_dr4v3 
\item[--] Col 2,3:  alpha\_j2000\_kids\_u\_dr4v3, delta\_j2000\_kids\_u\_dr4v3
\item[--] Col 4: CLASS\_sdssdr14 Spectroscopic redshift classification 
\item[--] Col 5: Z\_sdssdr14 Spectroscopic redshift value from SDSS DR14 
\item[--] Col 6: Z\_dr14q Spectroscopic redshift value from DR14Q - this column can be used as a flag for if a point is in the DR14Q catalogue - if it has a value, it is in the DR14Q catalogue.
\item[--] Col 7: id\_cpz Spectroscopic redshift ID for data points in the CPz sample (same as column 1 from the CPz catalogue presented in FP18) 
\item[--] Col 8-29: u, g, r, i, z, Y, J, H, K, W1, W2 total magnitudes, with associated errors  
\item[--] Col 30-47: u, g, r, i, z, Y, J, H, K 3'' aperture magnitudes, with associated errors  
\item[--] Col 48-56: PCA components for the STAR, GAL, QSO \textsc{hdbscan} binary classifiers for the `optimal' method setup \\
See \S \ref{subsection.consolidation} for information on columns 57-64 
\item[--] Col 57: hdbscanclass\_optimal\_method labels (0=outlier, 1=star, 2=gal, 3=QSO) 
\item[--] Col 58: hdbscanclass\_alternative\_method labels (0=outlier, 1=star, 2=gal, 3=QSO) 
\item[--] Col 59: double\_positives 
\item[--] Col 60-63: outlier\_hdbscan\_prob\_CL\_method, star\_hdbscan\_prob\_CL\_method, gal\_hdbscan\_prob\_CL\_method, QSO\_hdbscan\_prob\_CL\_method (probabilities are from the `highest-probability' consolidation method presented in \S \ref{subsection.consolidation})
\item[--] Col 64: highest\_probability\_labels (0=outlier, 1=star, 2=gal, 3=QSO) 
\item[--] Col 65: GAL\_PHOTOZ\_PREDICTOR (see \S \ref{subsection:photometric_redshifts}) photometric redshift predictions
\item[--] Col 66: QSO\_PHOTOZ\_PREDICTOR (see \S \ref{subsection:photometric_redshifts}) photometric redshift predictions 
\item[--] Col 67: Training/validation/test set labels for the photometric redshift predictions (see \S \ref{subsection:photometric_redshifts}) . If the source is in SDSS DR14 it has a value from 1-10. The training set has values 1-6, validation 7-8 and test 9-10. -99 values are for sources not in SDSSDR14.
\end{itemize}

\end{appendix}

\bibliographystyle{aa}

\begin{thebibliography}{65}
\expandafter\ifx\csname natexlab\endcsname\relax\def\natexlab#1{#1}\fi

\bibitem[{{Alam} {et~al.}(2015){Alam}, {Albareti}, {Allende Prieto}, {Anders},
  {Anderson}, {Anderton}, {Andrews}, {Armengaud}, {Aubourg}, {Bailey}, {Basu},
  {Bautista}, {Beaton}, {Beers}, {Bender}, {Berlind}, {Beutler}, {Bhardwaj},
  {Bird}, {Bizyaev}, {Blake}, {Blanton}, {Blomqvist}, {Bochanski}, {Bolton},
  {Bovy}, {Shelden Bradley}, {Brandt}, {Brauer}, {Brinkmann}, {Brown},
  {Brownstein}, {Burden}, {Burtin}, {Busca}, {Cai}, {Capozzi}, {Carnero
  Rosell}, {Carr}, {Carrera}, {Chambers}, {Chaplin}, {Chen}, {Chiappini},
  {Chojnowski}, {Chuang}, {Clerc}, {Comparat}, {Covey}, {Croft}, {Cuesta},
  {Cunha}, {da Costa}, {Da Rio}, {Davenport}, {Dawson}, {De Lee}, {Delubac},
  {Deshpande}, {Dhital}, {Dutra-Ferreira}, {Dwelly}, {Ealet}, {Ebelke},
  {Edmondson}, {Eisenstein}, {Ellsworth}, {Elsworth}, {Epstein}, {Eracleous},
  {Escoffier}, {Esposito}, {Evans}, {Fan}, {Fern{\'a}ndez-Alvar}, {Feuillet},
  {Filiz Ak}, {Finley}, {Finoguenov}, {Flaherty}, {Fleming}, {Font-Ribera},
  {Foster}, {Frinchaboy}, {Galbraith-Frew}, {Garc{\'\i}a},
  {Garc{\'\i}a-Hern{\'a}ndez}, {Garc{\'\i}a P{\'e}rez}, {Gaulme}, {Ge},
  {G{\'e}nova-Santos}, {Georgakakis}, {Ghezzi}, {Gillespie}, {Girardi},
  {Goddard}, {Gontcho}, {Gonz{\'a}lez Hern{\'a}ndez}, {Grebel}, {Green},
  {Grieb}, {Grieves}, {Gunn}, {Guo}, {Harding}, {Hasselquist}, {Hawley},
  {Hayden}, {Hearty}, {Hekker}, {Ho}, {Hogg}, {Holley-Bockelmann}, {Holtzman},
  {Honscheid}, {Huber}, {Huehnerhoff}, {Ivans}, {Jiang}, {Johnson},
  {Kinemuchi}, {Kirkby}, {Kitaura}, {Klaene}, {Knapp}, {Kneib}, {Koenig},
  {Lam}, {Lan}, {Lang}, {Laurent}, {Le Goff}, {Leauthaud}, {Lee}, {Lee},
  {Licquia}, {Liu}, {Long}, {L{\'o}pez-Corredoira}, {Lorenzo-Oliveira},
  {Lucatello}, {Lundgren}, {Lupton}, {Mack}, {Mahadevan}, {Maia}, {Majewski},
  {Malanushenko}, {Malanushenko}, {Manchado}, {Manera}, {Mao}, {Maraston},
  {Marchwinski}, {Margala}, {Martell}, {Martig}, {Masters}, {Mathur},
  {McBride}, {McGehee}, {McGreer}, {McMahon}, {M{\'e}nard}, {Menzel},
  {Merloni}, {M{\'e}sz{\'a}ros}, {Miller}, {Miralda-Escud{\'e}}, {Miyatake},
  {Montero-Dorta}, {More}, {Morganson}, {Morice-Atkinson}, {Morrison},
  {Mosser}, {Muna}, {Myers}, {Nand ra}, {Newman}, {Neyrinck}, {Nguyen},
  {Nichol}, {Nidever}, {Noterdaeme}, {Nuza}, {O'Connell}, {O'Connell},
  {O'Connell}, {Ogando}, {Olmstead}, {Oravetz}, {Oravetz}, {Osumi}, {Owen},
  {Padgett}, {Padmanabhan}, {Paegert}, {Palanque-Delabrouille}, {Pan},
  {Parejko}, {P{\^a}ris}, {Park}, {Pattarakijwanich}, {Pellejero-Ibanez},
  {Pepper}, {Percival}, {P{\'e}rez-Fournon}, {Ṕrez-Ra`fols}, {Petitjean},
  {Pieri}, {Pinsonneault}, {Porto de Mello}, {Prada}, {Prakash},
  {Price-Whelan}, {Protopapas}, {Raddick}, {Rahman}, {Reid}, {Rich}, {Rix},
  {Robin}, {Rockosi}, {Rodrigues}, {Rodr{\'\i}guez-Torres}, {Roe}, {Ross},
  {Ross}, {Rossi}, {Ruan}, {Rubi{\~n}o-Mart{\'\i}n}, {Rykoff},
  {Salazar-Albornoz}, {Salvato}, {Samushia}, {S{\'a}nchez}, {Santiago},
  {Sayres}, {Schiavon}, {Schlegel}, {Schmidt}, {Schneider}, {Schultheis},
  {Schwope}, {Sc{\'o}ccola}, {Scott}, {Sellgren}, {Seo}, {Serenelli}, {Shane},
  {Shen}, {Shetrone}, {Shu}, {Silva Aguirre}, {Sivarani}, {Skrutskie},
  {Slosar}, {Smith}, {Sobreira}, {Souto}, {Stassun}, {Steinmetz}, {Stello},
  {Strauss}, {Streblyanska}, {Suzuki}, {Swanson}, {Tan}, {Tayar}, {Terrien},
  {Thakar}, {Thomas}, {Thomas}, {Thompson}, {Tinker}, {Tojeiro}, {Troup},
  {Vargas-Maga{\~n}a}, {Vazquez}, {Verde}, {Viel}, {Vogt}, {Wake}, {Wang},
  {Weaver}, {Weinberg}, {Weiner}, {White}, {Wilson}, {Wisniewski},
  {Wood-Vasey}, {Ye`che}, {York}, {Zakamska}, {Zamora}, {Zasowski}, {Zehavi},
  {Zhao}, {Zheng}, {Zhou}, {Zhou}, {Zou}, \& {Zhu}}]{Alam2015}
{Alam}, S., {Albareti}, F.~D., {Allende Prieto}, C., {et~al.} 2015, \apjs, 219,
  12

\bibitem[{{Arnaboldi} {et~al.}(2007){Arnaboldi}, {Neeser}, {Parker}, {Rosati},
  {Lombardi}, {Dietrich}, \& {Hummel}}]{Arnaboldi2007}
{Arnaboldi}, M., {Neeser}, M.~J., {Parker}, L.~C., {et~al.} 2007, The
  Messenger, 127, 28

\bibitem[{{Assef} {et~al.}(2018){Assef}, {Stern}, {Noirot}, {Jun}, {Cutri}, \&
  {Eisenhardt}}]{Assef2018}
{Assef}, R.~J., {Stern}, D., {Noirot}, G., {et~al.} 2018, \apjs, 234, 23

\bibitem[{{Bai} {et~al.}(2019){Bai}, {Liu}, {Wang}, \& {Yang}}]{Bai2019}
{Bai}, Y., {Liu}, J., {Wang}, S., \& {Yang}, F. 2019, \aj, 157, 9

\bibitem[{{Baldwin} {et~al.}(1981){Baldwin}, {Phillips}, \&
  {Terlevich}}]{BPT1981}
{Baldwin}, J.~A., {Phillips}, M.~M., \& {Terlevich}, R. 1981, \pasp, 93, 5

\bibitem[{{Barchi} {et~al.}(2019){Barchi}, {de Carvalho}, {Rosa}, {Sautter},
  {Soares-Santos}, {Marques}, \& {Clua}}]{Barchi2019}
{Barchi}, P.~H., {de Carvalho}, R.~R., {Rosa}, R.~R., {et~al.} 2019, arXiv
  e-prints, arXiv:1901.07047

\bibitem[{{Bertin} \& {Arnouts}(1996)}]{Bertin1996}
{Bertin}, E. \& {Arnouts}, S. 1996, \aaps, 117, 393

\bibitem[{{Bovy} {et~al.}(2011){Bovy}, {Hennawi}, {Hogg}, {Myers},
  {Kirkpatrick}, {Schlegel}, {Ross}, {Sheldon}, {McGreer}, {Schneider}, \&
  {Weaver}}]{Bovy2011}
{Bovy}, J., {Hennawi}, J.~F., {Hogg}, D.~W., {et~al.} 2011, \apj, 729, 141

\bibitem[{Bradley(1997)}]{Bradley:1997a}
Bradley, A.~P. 1997, Pattern Recognition, 30, 1145

\bibitem[{Breiman(2001)}]{Breiman:2001a}
Breiman, L. 2001, Machine Learning, 45, 5

\bibitem[{Campello {et~al.}(2013)Campello, Moulavi, \& Sander}]{Campello:2013a}
Campello, R. J. G.~B., Moulavi, D., \& Sander, J. 2013, in Advances in
  Knowledge Discovery and Data Mining, ed. J.~Pei, V.~S. Tseng, L.~Cao,
  H.~Motoda, \& G.~Xu (Berlin, Heidelberg: Springer Berlin Heidelberg),
  160--172

\bibitem[{{Cardelli} {et~al.}(1989){Cardelli}, {Clayton}, \&
  {Mathis}}]{Cardelli1989}
{Cardelli}, J.~A., {Clayton}, G.~C., \& {Mathis}, J.~S. 1989, \apj, 345, 245

\bibitem[{{Coil} {et~al.}(2011){Coil}, {Blanton}, {Burles}, {Cool},
  {Eisenstein}, {Moustakas}, {Wong}, {Zhu}, {Aird}, {Bernstein}, {Bolton}, \&
  {Hogg}}]{Coil2011}
{Coil}, A.~L., {Blanton}, M.~R., {Burles}, S.~M., {et~al.} 2011, \apj, 741, 8

\bibitem[{{Cool} {et~al.}(2013){Cool}, {Moustakas}, {Blanton}, {Burles},
  {Coil}, {Eisenstein}, {Wong}, {Zhu}, {Aird}, {Bernstein}, {Bolton}, {Hogg},
  \& {Mendez}}]{Cool2013}
{Cool}, R.~J., {Moustakas}, J., {Blanton}, M.~R., {et~al.} 2013, \apj, 767, 118

\bibitem[{{Daddi} {et~al.}(2004){Daddi}, {Cimatti}, {Renzini}, {Fontana},
  {Mignoli}, {Pozzetti}, {Tozzi}, \& {Zamorani}}]{Daddi2004}
{Daddi}, E., {Cimatti}, A., {Renzini}, A., {et~al.} 2004, \apj, 617, 746

\bibitem[{{de Jong} {et~al.}(2015){de Jong}, {Verdoes Kleijn}, {Boxhoorn},
  {Buddelmeijer}, {Capaccioli}, {Getman}, {Grado}, {Helmich}, {Huang},
  {Irisarri}, {Kuijken}, {La Barbera}, {McFarland}, {Napolitano}, {Radovich},
  {Sikkema}, {Valentijn}, {Begeman}, {Brescia}, {Cavuoti}, {Choi}, {Cordes},
  {Covone}, {Dall'Ora}, {Hildebrandt}, {Longo}, {Nakajima}, {Paolillo},
  {Puddu}, {Rifatto}, {Tortora}, {van Uitert}, {Buddendiek},
  {Harnois-D{\'e}raps}, {Erben}, {Eriksen}, {Heymans}, {Hoekstra}, {Joachimi},
  {Kitching}, {Klaes}, {Koopmans}, {K{\"o}hlinger}, {Roy}, {Sif{\'o}n},
  {Schneider}, {Sutherland}, {Viola}, \& {Vriend}}]{deJong2015}
{de Jong}, J. T.~A., {Verdoes Kleijn}, G.~A., {Boxhoorn}, D.~R., {et~al.} 2015,
  \aap, 582, A62

\bibitem[{{Dubath} {et~al.}(2017){Dubath}, {Apostolakos}, {Bonchi}, {Belikov},
  {Brescia}, {Cavuoti}, {Capak}, {Coupon}, {Dabin}, {Degaudenzi}, {Desai},
  {Dubath}, {Fontana}, {Fotopoulou}, {Frailis}, {Galametz}, {Hoar}, {Holliman},
  {Hoyle}, {Hudelot}, {Ilbert}, {Kuemmel}, {Melchior}, {Mellier}, {Mohr},
  {Morisset}, {Paltani}, {Pello}, {Pilo}, {Polenta}, {Poncet}, {Saglia},
  {Salvato}, {Sauvage}, {Schefer}, {Serrano}, {Soldati}, {Tramacere},
  {Williams}, \& {Zacchei}}]{Dubath2017}
{Dubath}, P., {Apostolakos}, N., {Bonchi}, A., {et~al.} 2017, in IAU Symposium,
  Vol. 325, Astroinformatics, ed. M.~{Brescia}, S.~G. {Djorgovski}, E.~D.
  {Feigelson}, G.~{Longo}, \& S.~{Cavuoti}, 73--82

\bibitem[{Ester {et~al.}(1996)Ester, Kriegel, Sander, \& Xu}]{Ester:1996a}
Ester, M., Kriegel, H.-P., Sander, J., \& Xu, X. 1996, in Proceedings of the
  Second International Conference on Knowledge Discovery and Data Mining,
  KDD'96 (AAAI Press), 226--231

\bibitem[{{Fotopoulou} \& {Paltani}(2018)}]{Fotopoulou:2018a}
{Fotopoulou}, S. \& {Paltani}, S. 2018, \aap, 619, A14

\bibitem[{{Francis} \& {Wills}(1999)}]{Francis:1999a}
{Francis}, P.~J. \& {Wills}, B.~J. 1999, in Astronomical Society of the Pacific
  Conference Series, Vol. 162, Quasars and Cosmology, ed. G.~{Ferland} \&
  J.~{Baldwin}, 363

\bibitem[{{Garilli} {et~al.}(2014){Garilli}, {Guzzo}, {Scodeggio},
  {Bolzonella}, {Abbas}, {Adami}, {Arnouts}, {Bel}, {Bottini}, {Branchini},
  {Cappi}, {Coupon}, {Cucciati}, {Davidzon}, {De Lucia}, {de la Torre},
  {Franzetti}, {Fritz}, {Fumana}, {Granett}, {Ilbert}, {Iovino}, {Krywult}, {Le
  Brun}, {Le F{\`e}vre}, {Maccagni}, {Ma{\l}ek}, {Marulli}, {McCracken},
  {Paioro}, {Polletta}, {Pollo}, {Schlagenhaufer}, {Tasca}, {Tojeiro},
  {Vergani}, {Zamorani}, {Zanichelli}, {Burden}, {Di Porto}, {Marchetti},
  {Marinoni}, {Mellier}, {Moscardini}, {Nichol}, {Peacock}, {Percival},
  {Phleps}, \& {Wolk}}]{Garilli2014}
{Garilli}, B., {Guzzo}, L., {Scodeggio}, M., {et~al.} 2014, \aap, 562, A23

\bibitem[{{Geach}(2012)}]{Geach2012}
{Geach}, J.~E. 2012, \mnras, 419, 2633

\bibitem[{{Gieseke} {et~al.}(2017){Gieseke}, {Bloemen}, {van den Bogaard},
  {Heskes}, {Kindler}, {Scalzo}, {Ribeiro}, {van Roestel}, {Groot}, {Yuan},
  {M{\"o}ller}, \& {Tucker}}]{Gieseke2017}
{Gieseke}, F., {Bloemen}, S., {van den Bogaard}, C., {et~al.} 2017, \mnras,
  472, 3101

\bibitem[{{Hemmati} {et~al.}(2019){Hemmati}, {Capak}, {Pourrahmani}, {Nayyeri},
  {Stern}, {Mobasher}, {Darvish}, {Davidzon}, {Ilbert}, {Masters}, \&
  {Shahidi}}]{Hemmati2019ApJ...881L..14H}
{Hemmati}, S., {Capak}, P., {Pourrahmani}, M., {et~al.} 2019, \apjl, 881, L14

\bibitem[{{Hubble}(1926)}]{Hubble1926}
{Hubble}, E.~P. 1926, \apj, 64, 321

\bibitem[{{Hudelot} {et~al.}(2012){Hudelot}, {Cuillandre}, {Withington},
  {Goranova}, {McCracken}, {Magnard}, {Mellier}, {Regnault}, {Betoule},
  {Aussel}, {Kavelaars}, {Fernique}, {Bonnarel}, {Ochsenbein}, \&
  {Ilbert}}]{Hudelot2012}
{Hudelot}, P., {Cuillandre}, J.~C., {Withington}, K., {et~al.} 2012, VizieR
  Online Data Catalog, II/317

\bibitem[{{Huertas-Company} {et~al.}(2015){Huertas-Company}, {Gravet},
  {Cabrera-Vives}, {P{\'e}rez-Gonz{\'a}lez}, {Kartaltepe}, {Barro}, {Bernardi},
  {Mei}, {Shankar}, {Dimauro}, {Bell}, {Kocevski}, {Koo}, {Faber}, \&
  {Mcintosh}}]{Huertas-Company2015}
{Huertas-Company}, M., {Gravet}, R., {Cabrera-Vives}, G., {et~al.} 2015, \apjs,
  221, 8

\bibitem[{{Jin} {et~al.}(2019){Jin}, {Zhang}, {Zhang}, {Zhao}, {Wu}, \&
  {Fan}}]{Jin2019}
{Jin}, X., {Zhang}, Y., {Zhang}, J., {et~al.} 2019, \mnras, 485, 4539

\bibitem[{{Jones} {et~al.}(2009){Jones}, {Read}, {Saunders}, {Colless},
  {Jarrett}, {Parker}, {Fairall}, {Mauch}, {Sadler}, {Watson}, {Burton},
  {Campbell}, {Cass}, {Croom}, {Dawe}, {Fiegert}, {Frankcombe}, {Hartley},
  {Huchra}, {James}, {Kirby}, {Lahav}, {Lucey}, {Mamon}, {Moore}, {Peterson},
  {Prior}, {Proust}, {Russell}, {Safouris}, {Wakamatsu}, {Westra}, \&
  {Williams}}]{Jones2009}
{Jones}, D.~H., {Read}, M.~A., {Saunders}, W., {et~al.} 2009, \mnras, 399, 683

\bibitem[{{Jones} {et~al.}(2004){Jones}, {Saunders}, {Colless}, {Read},
  {Parker}, {Watson}, {Campbell}, {Burkey}, {Mauch}, {Moore}, {Hartley},
  {Cass}, {James}, {Russell}, {Fiegert}, {Dawe}, {Huchra}, {Jarrett}, {Lahav},
  {Lucey}, {Mamon}, {Proust}, {Sadler}, \& {Wakamatsu}}]{Jones2004}
{Jones}, D.~H., {Saunders}, W., {Colless}, M., {et~al.} 2004, \mnras, 355, 747

\bibitem[{{Kuntzer} {et~al.}(2016){Kuntzer}, {Tewes}, \&
  {Courbin}}]{Kuntzer2016}
{Kuntzer}, T., {Tewes}, M., \& {Courbin}, F. 2016, \aap, 591, A54

\bibitem[{{Lahav} {et~al.}(1995){Lahav}, {Naim}, {Buta}, {Corwin}, {de
  Vaucouleurs}, {Dressler}, {Huchra}, {van den Bergh}, {Raychaudhury}, {Sodre},
  \& {Storrie-Lombardi}}]{Lahav1995}
{Lahav}, O., {Naim}, A., {Buta}, R.~J., {et~al.} 1995, Science, 267, 859

\bibitem[{{Le F{\`e}vre} {et~al.}(2013){Le F{\`e}vre}, {Cassata}, {Cucciati},
  {Garilli}, {Ilbert}, {Le Brun}, {Maccagni}, {Moreau}, {Scodeggio}, {Tresse},
  {Zamorani}, {Adami}, {Arnouts}, {Bardelli}, {Bolzonella}, {Bondi},
  {Bongiorno}, {Bottini}, {Cappi}, {Charlot}, {Ciliegi}, {Contini}, {de la
  Torre}, {Foucaud}, {Franzetti}, {Gavignaud}, {Guzzo}, {Iovino}, {Lemaux},
  {L{\'o}pez-Sanjuan}, {McCracken}, {Marano}, {Marinoni}, {Mazure}, {Mellier},
  {Merighi}, {Merluzzi}, {Paltani}, {Pell{\`o}}, {Pollo}, {Pozzetti},
  {Scaramella}, {Tasca}, {Vergani}, {Vettolani}, {Zanichelli}, \&
  {Zucca}}]{LeFevre2013}
{Le F{\`e}vre}, O., {Cassata}, P., {Cucciati}, O., {et~al.} 2013, \aap, 559,
  A14

\bibitem[{{Lintott} {et~al.}(2011){Lintott}, {Schawinski}, {Bamford}, {Slosar},
  {Land}, {Thomas}, {Edmondson}, {Masters}, {Nichol}, {Raddick}, {Szalay},
  {Andreescu}, {Murray}, \& {Vandenberg}}]{Lintott:2011a}
{Lintott}, C., {Schawinski}, K., {Bamford}, S., {et~al.} 2011, \mnras, 410, 166

\bibitem[{{Lintott} {et~al.}(2008){Lintott}, {Schawinski}, {Slosar}, {Land},
  {Bamford}, {Thomas}, {Raddick}, {Nichol}, {Szalay}, {Andreescu}, {Murray}, \&
  {Vandenberg}}]{Lintott:2008a}
{Lintott}, C.~J., {Schawinski}, K., {Slosar}, A., {et~al.} 2008, \mnras, 389,
  1179

\bibitem[{{Liske} {et~al.}(2015){Liske}, {Baldry}, {Driver}, {Tuffs},
  {Alpaslan}, {Andrae}, {Brough}, {Cluver}, {Grootes}, {Gunawardhana},
  {Kelvin}, {Loveday}, {Robotham}, {Taylor}, {Bamford}, {Bland-Hawthorn},
  {Brown}, {Drinkwater}, {Hopkins}, {Meyer}, {Norberg}, {Peacock}, {Agius},
  {Andrews}, {Bauer}, {Ching}, {Colless}, {Conselice}, {Croom}, {Davies}, {De
  Propris}, {Dunne}, {Eardley}, {Ellis}, {Foster}, {Frenk}, {H{\"a}u{\ss}ler},
  {Holwerda}, {Howlett}, {Ibarra}, {Jarvis}, {Jones}, {Kafle}, {Lacey},
  {Lange}, {Lara-L{\'o}pez}, {L{\'o}pez-S{\'a}nchez}, {Maddox}, {Madore},
  {McNaught-Roberts}, {Moffett}, {Nichol}, {Owers}, {Palamara}, {Penny},
  {Phillipps}, {Pimbblet}, {Popescu}, {Prescott}, {Proctor}, {Sadler},
  {Sansom}, {Seibert}, {Sharp}, {Sutherland}, {V{\'a}zquez-Mata}, {van Kampen},
  {Wilkins}, {Williams}, \& {Wright}}]{Liske2015}
{Liske}, J., {Baldry}, I.~K., {Driver}, S.~P., {et~al.} 2015, \mnras, 452, 2087

\bibitem[{{Mainzer} {et~al.}(2011){Mainzer}, {Bauer}, {Grav}, {Masiero},
  {Cutri}, {Dailey}, {Eisenhardt}, {McMillan}, {Wright}, {Walker}, {Jedicke},
  {Spahr}, {Tholen}, {Alles}, {Beck}, {Brand enburg}, {Conrow}, {Evans},
  {Fowler}, {Jarrett}, {Marsh}, {Masci}, {McCallon}, {Wheelock}, {Wittman},
  {Wyatt}, {DeBaun}, {Elliott}, {Elsbury}, {Gautier}, {Gomillion}, {Leisawitz},
  {Maleszewski}, {Micheli}, \& {Wilkins}}]{Mainzer2011}
{Mainzer}, A., {Bauer}, J., {Grav}, T., {et~al.} 2011, \apj, 731, 53

\bibitem[{{Masters} {et~al.}(2017){Masters}, {Stern}, {Cohen}, {Capak},
  {Rhodes}, {Castander}, \& {Paltani}}]{Masters2017ApJ...841..111M}
{Masters}, D.~C., {Stern}, D.~K., {Cohen}, J.~G., {et~al.} 2017, \apj, 841, 111

\bibitem[{{Masters} {et~al.}(2019){Masters}, {Stern}, {Cohen}, {Capak},
  {Stanford}, {Hernitschek}, {Galametz}, {Davidzon}, {Rhodes}, {Sand ers},
  {Mobasher}, {Castander}, {Pruett}, \&
  {Fotopoulou}}]{Masters2019ApJ...877...81M}
{Masters}, D.~C., {Stern}, D.~K., {Cohen}, J.~G., {et~al.} 2019, \apj, 877, 81

\bibitem[{{McGurk} {et~al.}(2010){McGurk}, {Kimball}, \&
  {Ivezi{\'c}}}]{McGurk:2010a}
{McGurk}, R.~C., {Kimball}, A.~E., \& {Ivezi{\'c}}, {\v{Z}}. 2010, \aj, 139,
  1261

\bibitem[{McInnes {et~al.}(2017)McInnes, Healy, \& Astels}]{McInnes:2017a}
McInnes, L., Healy, J., \& Astels, S. 2017, The Journal of Open Source
  Software, 2

\bibitem[{{Nakoneczny} {et~al.}(2019){Nakoneczny}, {Bilicki}, {Solarz},
  {Pollo}, {Maddox}, {Spiniello}, {Brescia}, \& {Napolitano}}]{Nakoneczny2019}
{Nakoneczny}, S., {Bilicki}, M., {Solarz}, A., {et~al.} 2019, Astronomy and
  Astrophysics, 624, A13

\bibitem[{{Odewahn} {et~al.}(1992){Odewahn}, {Stockwell}, {Pennington},
  {Humphreys}, \& {Zumach}}]{Odewahn1992}
{Odewahn}, S.~C., {Stockwell}, E.~B., {Pennington}, R.~L., {Humphreys}, R.~M.,
  \& {Zumach}, W.~A. 1992, \aj, 103, 318

\bibitem[{{Paraficz} {et~al.}(2016){Paraficz}, {Courbin}, {Tramacere},
  {Joseph}, {Metcalf}, {Kneib}, {Dubath}, {Droz}, {Filleul}, {Ringeisen}, \&
  {Sch{\"a}fer}}]{Paraficz:2016a}
{Paraficz}, D., {Courbin}, F., {Tramacere}, A., {et~al.} 2016, \aap, 592, A75

\bibitem[{{P{\^a}ris} {et~al.}(2018){P{\^a}ris}, {Petitjean}, {Aubourg},
  {Myers}, {Streblyanska}, {Lyke}, {Anderson}, {Armengaud}, {Bautista},
  {Blanton}, {Blomqvist}, {Brinkmann}, {Brownstein}, {Brand t}, {Burtin},
  {Dawson}, {de la Torre}, {Georgakakis}, {Gil-Mar{\'\i}n}, {Green}, {Hall},
  {Kneib}, {LaMassa}, {Le Goff}, {MacLeod}, {Mariappan}, {McGreer}, {Merloni},
  {Noterdaeme}, {Palanque-Delabrouille}, {Percival}, {Ross}, {Rossi},
  {Schneider}, {Seo}, {Tojeiro}, {Weaver}, {Weijmans}, {Y{\`e}che}, {Zarrouk},
  \& {Zhao}}]{Paris2018}
{P{\^a}ris}, I., {Petitjean}, P., {Aubourg}, {\'E}., {et~al.} 2018, \aap, 613,
  A51

\bibitem[{{P{\^a}ris} {et~al.}(2017){P{\^a}ris}, {Petitjean}, {Ross}, {Myers},
  {Aubourg}, {Streblyanska}, {Bailey}, {Armengaud}, {Palanque-Delabrouille},
  {Y{\`e}che}, {Hamann}, {Strauss}, {Albareti}, {Bovy}, {Bizyaev}, {Niel
  Brandt}, {Brusa}, {Buchner}, {Comparat}, {Croft}, {Dwelly}, {Fan},
  {Font-Ribera}, {Ge}, {Georgakakis}, {Hall}, {Jiang}, {Kinemuchi},
  {Malanushenko}, {Malanushenko}, {McMahon}, {Menzel}, {Merloni}, {Nandra},
  {Noterdaeme}, {Oravetz}, {Pan}, {Pieri}, {Prada}, {Salvato}, {Schlegel},
  {Schneider}, {Simmons}, {Viel}, {Weinberg}, \& {Zhu}}]{Paris2017}
{P{\^a}ris}, I., {Petitjean}, P., {Ross}, N.~P., {et~al.} 2017, \aap, 597, A79

\bibitem[{{Patel} {et~al.}(2012){Patel}, {Holden}, {Kelson}, {Franx}, {van der
  Wel}, \& {Illingworth}}]{Patel2012}
{Patel}, S.~G., {Holden}, B.~P., {Kelson}, D.~D., {et~al.} 2012, \apj, 748, L27

\bibitem[{Pedregosa {et~al.}(2011)Pedregosa, Varoquaux, Gramfort, Michel,
  Thirion, Grisel, Blondel, Prettenhofer, Weiss, Dubourg, Vanderplas, Passos,
  Cournapeau, Brucher, Perrot, \& Duchesnay}]{Pedregosa:2011a}
Pedregosa, F., Varoquaux, G., Gramfort, A., {et~al.} 2011, Journal of Machine
  Learning Research, 12, 2825

\bibitem[{{Richards} {et~al.}(2001){Richards}, {Fan}, {Schneider}, {Vanden
  Berk}, {Strauss}, {York}, {Anderson}, {Anderson}, {Annis}, {Bahcall},
  {Bernardi}, {Briggs}, {Brinkmann}, {Brunner}, {Burles}, {Carey}, {Castand
  er}, {Connolly}, {Crocker}, {Csabai}, {Doi}, {Finkbeiner}, {Friedman},
  {Frieman}, {Fukugita}, {Gunn}, {Hindsley}, {Ivezi{\'c}}, {Kent}, {Knapp},
  {Lamb}, {Leger}, {Long}, {Loveday}, {Lupton}, {McKay}, {Meiksin}, {Merrelli},
  {Munn}, {Newberg}, {Newcomb}, {Nichol}, {Owen}, {Pier}, {Pope}, {Richmond},
  {Rockosi}, {Schlegel}, {Siegmund}, {Smee}, {Snir}, {Stoughton}, {Stubbs},
  {SubbaRao}, {Szalay}, {Szokoly}, {Tremonti}, {Uomoto}, {Waddell}, {Yanny}, \&
  {Zheng}}]{Richards2001}
{Richards}, G.~T., {Fan}, X., {Schneider}, D.~P., {et~al.} 2001, \aj, 121, 2308

\bibitem[{{Salvato} {et~al.}(2009){Salvato}, {Hasinger}, {Ilbert}, {Zamorani},
  {Brusa}, {Scoville}, {Rau}, {Capak}, {Arnouts}, {Aussel}, {Bolzonella},
  {Buongiorno}, {Cappelluti}, {Caputi}, {Civano}, {Cook}, {Elvis}, {Gilli},
  {Jahnke}, {Kartaltepe}, {Impey}, {Lamareille}, {Le Floc'h}, {Lilly},
  {Mainieri}, {McCarthy}, {McCracken}, {Mignoli}, {Mobasher}, {Murayama},
  {Sasaki}, {Sanders}, {Schiminovich}, {Shioya}, {Shopbell}, {Silverman},
  {Smol{\v{c}}i{\'c}}, {Surace}, {Taniguchi}, {Thompson}, {Trump}, {Urry}, \&
  {Zamojski}}]{Salvato2009}
{Salvato}, M., {Hasinger}, G., {Ilbert}, O., {et~al.} 2009, \apj, 690, 1250

\bibitem[{{Salvato} {et~al.}(2011){Salvato}, {Ilbert}, {Hasinger}, {Rau},
  {Civano}, {Zamorani}, {Brusa}, {Elvis}, {Vignali}, {Aussel}, {Comastri},
  {Fiore}, {Le Floc'h}, {Mainieri}, {Bardelli}, {Bolzonella}, {Bongiorno},
  {Capak}, {Caputi}, {Cappelluti}, {Carollo}, {Contini}, {Garilli}, {Iovino},
  {Fotopoulou}, {Fruscione}, {Gilli}, {Halliday}, {Kneib}, {Kakazu},
  {Kartaltepe}, {Koekemoer}, {Kovac}, {Ideue}, {Ikeda}, {Impey}, {Le Fevre},
  {Lamareille}, {Lanzuisi}, {Le Borgne}, {Le Brun}, {Lilly}, {Maier},
  {Manohar}, {Masters}, {McCracken}, {Messias}, {Mignoli}, {Mobasher}, {Nagao},
  {Pello}, {Puccetti}, {Perez-Montero}, {Renzini}, {Sargent}, {Sanders},
  {Scodeggio}, {Scoville}, {Shopbell}, {Silvermann}, {Taniguchi}, {Tasca},
  {Tresse}, {Trump}, \& {Zucca}}]{Salvato2011}
{Salvato}, M., {Ilbert}, O., {Hasinger}, G., {et~al.} 2011, \apj, 742, 61

\bibitem[{{Schlegel} {et~al.}(1998){Schlegel}, {Finkbeiner}, \&
  {Davis}}]{Schlegel1998}
{Schlegel}, D.~J., {Finkbeiner}, D.~P., \& {Davis}, M. 1998, \apj, 500, 525

\bibitem[{{Schmidt}(1963)}]{Schmidt1963}
{Schmidt}, M. 1963, \nat, 197, 1040

\bibitem[{{Schneider} {et~al.}(2010){Schneider}, {Richards}, {Hall}, {Strauss},
  {Anderson}, {Boroson}, {Ross}, {Shen}, {Brandt}, {Fan}, {Inada}, {Jester},
  {Knapp}, {Krawczyk}, {Thakar}, {Vanden Berk}, {Voges}, {Yanny}, {York},
  {Bahcall}, {Bizyaev}, {Blanton}, {Brewington}, {Brinkmann}, {Eisenstein},
  {Frieman}, {Fukugita}, {Gray}, {Gunn}, {Hibon}, {Ivezi{\'c}}, {Kent}, {Kron},
  {Lee}, {Lupton}, {Malanushenko}, {Malanushenko}, {Oravetz}, {Pan}, {Pier},
  {Price}, {Saxe}, {Schlegel}, {Simmons}, {Snedden}, {SubbaRao}, {Szalay}, \&
  {Weinberg}}]{Schneider2010}
{Schneider}, D.~P., {Richards}, G.~T., {Hall}, P.~B., {et~al.} 2010, \aj, 139,
  2360

\bibitem[{{Shlens}(2014)}]{Shlens:2014a}
{Shlens}, J. 2014, arXiv e-prints, arXiv:1404.1100

\bibitem[{{Stern} {et~al.}(2012){Stern}, {Assef}, {Benford}, {Blain}, {Cutri},
  {Dey}, {Eisenhardt}, {Griffith}, {Jarrett}, {Lake}, {Masci}, {Petty},
  {Stanford}, {Tsai}, {Wright}, {Yan}, {Harrison}, \& {Madsen}}]{Stern2012}
{Stern}, D., {Assef}, R.~J., {Benford}, D.~J., {et~al.} 2012, \apj, 753, 30

\bibitem[{{Stern} {et~al.}(2005){Stern}, {Eisenhardt}, {Gorjian}, {Kochanek},
  {Caldwell}, {Eisenstein}, {Brodwin}, {Brown}, {Cool}, {Dey}, {Green},
  {Jannuzi}, {Murray}, {Pahre}, \& {Willner}}]{Stern2005}
{Stern}, D., {Eisenhardt}, P., {Gorjian}, V., {et~al.} 2005, \apj, 631, 163

\bibitem[{{Storrie-Lombardi} {et~al.}(1992){Storrie-Lombardi}, {Lahav},
  {Sodre}, \& {Storrie-Lombardi}}]{Storrie-Lombardi1992}
{Storrie-Lombardi}, M.~C., {Lahav}, O., {Sodre}, L., J., \& {Storrie-Lombardi},
  L.~J. 1992, \mnras, 259, 8P

\bibitem[{{Torres} {et~al.}(2019){Torres}, {Cantero}, {Rebassa-Mansergas},
  {Skorobogatov}, {Jim{\'e}nez-Esteban}, \& {Solano}}]{Torres2019}
{Torres}, S., {Cantero}, C., {Rebassa-Mansergas}, A., {et~al.} 2019, \mnras,
  485, 5573

\bibitem[{{Vafaei Sadr} {et~al.}(2019){Vafaei Sadr}, {Vos}, {Bassett},
  {Hosenie}, {Oozeer}, \& {Lochner}}]{VafaeiSadr2019}
{Vafaei Sadr}, A., {Vos}, E.~E., {Bassett}, B.~A., {et~al.} 2019, \mnras, 484,
  2793

\bibitem[{{Vasconcellos} {et~al.}(2011){Vasconcellos}, {de Carvalho}, {Gal},
  {LaBarbera}, {Capelato}, {Frago Campos Velho}, {Trevisan}, \&
  {Ruiz}}]{Vasconcellos:2011a}
{Vasconcellos}, E.~C., {de Carvalho}, R.~R., {Gal}, R.~R., {et~al.} 2011, \aj,
  141, 189

\bibitem[{{Weir} {et~al.}(1995){Weir}, {Fayyad}, \& {Djorgovski}}]{Weir1995}
{Weir}, N., {Fayyad}, U.~M., \& {Djorgovski}, S. 1995, \aj, 109, 2401

\bibitem[{{Willett} {et~al.}(2013){Willett}, {Lintott}, {Bamford}, {Masters},
  {Simmons}, {Casteels}, {Edmondson}, {Fortson}, {Kaviraj}, {Keel}, {Melvin},
  {Nichol}, {Raddick}, {Schawinski}, {Simpson}, {Skibba}, {Smith}, \&
  {Thomas}}]{Willett:2013a}
{Willett}, K.~W., {Lintott}, C.~J., {Bamford}, S.~P., {et~al.} 2013, \mnras,
  435, 2835

\bibitem[{{Wright} {et~al.}(2010){Wright}, {Eisenhardt}, {Mainzer}, {Ressler},
  {Cutri}, {Jarrett}, {Kirkpatrick}, {Padgett}, {McMillan}, {Skrutskie},
  {Stanford}, {Cohen}, {Walker}, {Mather}, {Leisawitz}, {Gautier}, {McLean},
  {Benford}, {Lonsdale}, {Blain}, {Mendez}, {Irace}, {Duval}, {Liu}, {Royer},
  {Heinrichsen}, {Howard}, {Shannon}, {Kendall}, {Walsh}, {Larsen}, {Cardon},
  {Schick}, {Schwalm}, {Abid}, {Fabinsky}, {Naes}, \& {Tsai}}]{Wright2010}
{Wright}, E.~L., {Eisenhardt}, P. R.~M., {Mainzer}, A.~K., {et~al.} 2010, \aj,
  140, 1868

\bibitem[{{York} {et~al.}(2000){York}, {Adelman}, {Anderson}, {Anderson},
  {Annis}, {Bahcall}, {Bakken}, {Barkhouser}, {Bastian}, {Berman}, {Boroski},
  {Bracker}, {Briegel}, {Briggs}, {Brinkmann}, {Brunner}, {Burles}, {Carey},
  {Carr}, {Castander}, {Chen}, {Colestock}, {Connolly}, {Crocker}, {Csabai},
  {Czarapata}, {Davis}, {Doi}, {Dombeck}, {Eisenstein}, {Ellman}, {Elms},
  {Evans}, {Fan}, {Federwitz}, {Fiscelli}, {Friedman}, {Frieman}, {Fukugita},
  {Gillespie}, {Gunn}, {Gurbani}, {de Haas}, {Haldeman}, {Harris}, {Hayes},
  {Heckman}, {Hennessy}, {Hindsley}, {Holm}, {Holmgren}, {Huang}, {Hull},
  {Husby}, {Ichikawa}, {Ichikawa}, {Ivezi{\'c}}, {Kent}, {Kim}, {Kinney},
  {Klaene}, {Kleinman}, {Kleinman}, {Knapp}, {Korienek}, {Kron}, {Kunszt},
  {Lamb}, {Lee}, {Leger}, {Limmongkol}, {Lindenmeyer}, {Long}, {Loomis},
  {Loveday}, {Lucinio}, {Lupton}, {MacKinnon}, {Mannery}, {Mantsch}, {Margon},
  {McGehee}, {McKay}, {Meiksin}, {Merelli}, {Monet}, {Munn}, {Narayanan},
  {Nash}, {Neilsen}, {Neswold}, {Newberg}, {Nichol}, {Nicinski}, {Nonino},
  {Okada}, {Okamura}, {Ostriker}, {Owen}, {Pauls}, {Peoples}, {Peterson},
  {Petravick}, {Pier}, {Pope}, {Pordes}, {Prosapio}, {Rechenmacher}, {Quinn},
  {Richards}, {Richmond}, {Rivetta}, {Rockosi}, {Ruthmansdorfer}, {Sand ford},
  {Schlegel}, {Schneider}, {Sekiguchi}, {Sergey}, {Shimasaku}, {Siegmund},
  {Smee}, {Smith}, {Snedden}, {Stone}, {Stoughton}, {Strauss}, {Stubbs},
  {SubbaRao}, {Szalay}, {Szapudi}, {Szokoly}, {Thakar}, {Tremonti}, {Tucker},
  {Uomoto}, {Vanden Berk}, {Vogeley}, {Waddell}, {Wang}, {Watanabe},
  {Weinberg}, {Yanny}, {Yasuda}, \& {SDSS Collaboration}}]{York2000}
{York}, D.~G., {Adelman}, J., {Anderson}, John~E., J., {et~al.} 2000, \aj, 120,
  1579

\end{thebibliography}

\end{document}